\documentclass[a4paper,fleqn,usenatbib]{mnras}

\usepackage{newtxtext,newtxmath}

\usepackage[T1]{fontenc}
\usepackage{ae,aecompl}

\usepackage{graphicx}	
\usepackage{amsmath}	
\usepackage{amssymb}	

\title[The SAMI Galaxy Survey: Global stellar populations]{The SAMI Galaxy Survey: Global stellar populations on the size--mass plane}

\author[N. Scott et al.]{
Nicholas Scott$^{1,2}$\thanks{E-mail: nscott@physics.usyd.edu.au},
S. Brough$^{3,4}$
Scott M. Croom$^{1,2}$,
Roger L. Davies$^{5}$,
\newauthor Jesse van de Sande$^{1}$,
 J. T. Allen$^{1,2}$,
Joss Bland-Hawthorn$^{1}$,
Julia J. Bryant$^{1,2,3}$,
\newauthor Luca Cortese$^{6}$,
Francesco D'Eugenio$^{2,7}$,
Christoph Federrath$^{7}$,
Ignacio Ferreras$^{8}$,
\newauthor Michael Goodwin$^{3}$,
Brent Groves$^{7}$,
Iraklis Konstantopoulos$^{3,9}$,
Jon S. Lawrence$^{3}$,
\newauthor Anne  M. Medling$^{7,10,11}$,
Amanda J. Moffett$^{12}$,
Matt S. Owers$^{3,13}$,
Samuel Richards$^{14}$,
\newauthor A. S. G. Robotham$^{6}$,
Chiara Tonini$^{15}$
and Sukyoung K. Yi$^{16}$
\\
$^{1}$ Sydney Institute for Astronomy, School of Physics, University of Sydney, NSW 2006, Australia.\\ 
$^{2}$ ARC Centre of Excellence for All-Sky Astrophysics (CAASTRO).\\
$^{3}$ Australian Astronomical Observatory, 105 Delhi Road, North Ryde, NSW 2113, Australia.\\
$^{4}$ School of Physics University of New South Wales NSW 2052 Australia. \\
$^{5}$ Sub-Dept. of Astrophysics, Department of Physics, University of Oxford, Denys Wilkinson Building, Keble Rd., Oxford, OX1 3RH, UK. \\
$^{6}$ International Centre for Radio Astronomy Research, University of Western Australia, 35 Stirling Highway, Crawley WA 6009, Australia. \\
$^{7}$ Research School for Astronomy \& Astrophysics Australian National University Canberra, ACT 2611, Australia. \\
$^{8}$ Mullard Space Science Laboratory University College London Holmbury St Mary, Dorking, Surrey, RH5 6NT United Kingdom. \\
$^{9}$ Atlassian, 341 George Street, Sydney, NSW 2000. \\
$^{10}$ Cahill Center for Astronomy and Astrophysics California Institute of Technology, MS 249-17 Pasadena, CA 91125, USA. \\
$^{11}$ Hubble Fellow \\
$^{12}$ Department of Physics and Astronomy, Vanderbilt University, PMB \#401807 2401 Vanderbilt Place, Nashville, TN 37240, USA \\
$^{13}$ Department of Physics and Astronomy, Macquarie University, NSW 2109, Australia. \\
$^{14}$ SOFIA Operations Center, USRA, NASA Armstrong Flight Research Center, 2825 East Avenue P, Palmdale, CA 93550, USA \\
$^{15}$ Melbourne University, School of Physics Parkville, 3010 Australia. \\
$^{16}$ Department of Astronomy and Yonsei University Observatory, Yonsei University, Seoul 03722, Republic of Korea. \\
}

\date{Accepted XXX. Received YYY; in original form ZZZ}
\pubyear{2017}

\begin{document}
\label{firstpage}
\pagerange{\pageref{firstpage}--\pageref{lastpage}}
\maketitle

\begin{abstract}
We present an analysis of the global stellar populations of galaxies in the SAMI Galaxy Survey. Our sample consists of 1319 galaxies spanning four orders of magnitude in stellar mass and includes all morphologies and environments. We derive luminosity-weighted, single stellar population equivalent stellar ages, metallicities and alpha enhancements from spectra integrated within one effective radius apertures. Variations in galaxy size explain the majority of the scatter in the age--mass and metallicity--mass relations. Stellar populations vary systematically in the plane of galaxy size and stellar mass, such that galaxies with high stellar surface mass density are older, more metal-rich and alpha-enhanced than less dense galaxies. Galaxies with high surface mass densities have a very narrow range of metallicities, however, at fixed mass, the spread in metallicity increases substantially with increasing galaxy size (decreasing density). We identify residual correlations with morphology and environment. At fixed mass {\it and} size, galaxies with late-type morphologies, small bulges and low S\'{e}rsic $n$ are younger than early-type, high $n$, high bulge-to-total galaxies. Age and metallicity both show small residual correlations with environment; at fixed mass {\it and} size, galaxies in denser environments or more massive halos are older and somewhat more metal rich than those in less dense environments. We connect these trends to evolutionary tracks within the size--mass plane.
\end{abstract}



\section{Introduction}
\label{sec:intro}

A galaxy's stellar population contains a fossil record of its entire formation history. The mass fraction of stars with a given age fully describes the star formation history of a galaxy, though does little to constrain when and how the galaxy was assembled. In contrast the metallicity of a given sub-population of stars is thought to be strongly correlated with the mass of the halo in which those stars formed, and therefore with the assembly history of a galaxy. The detailed abundance pattern of individual elements or groups of elements is also tightly correlated with assembly history, and is widely used to identify the remnants of dwarf galaxies in the halo of the Milky Way \citep[][and references therein]{Freeman:2002}.

Extracting this history from observational data is a non-trivial task, particularly for galaxies outside the local group where individual stars cannot be resolved. Galaxy colours have been used as a crude tool to distinguish different stellar populations for many decades \citep[see][for a review of early work on galaxy stellar populations]{Tinsley:1980}, but single colours suffer from a strong degeneracy between age, metallicity and reddening due to dust. The strongest constraints on galaxy stellar populations come from optical and near-infra red (NIR) spectroscopy. The optical region contains a wealth of spectral absorption lines that have varying sensitivities to the age, metallicity, and elemental abundances of a stellar population. By using a combination of age- and metallicity-sensitive features, the age-metallicity degeneracy can be broken \citep{Worthey:1994b} and the luminosity-weighted mean age, metallicity and abundance pattern of a galaxy can be well constrained.

The Lick absorption line strength system is a widely-used approach that utilises optical absorption features to measure the properties of a galaxy's stellar population \citep{Worthey:1994a}. The Lick system defines a set of absorption line indices using i) a central bandpass, ii) two side bandpasses used to define a pseudocontinuum level and iii) a spectral resolution at which the index should be measured. For a full description of the Lick absorption line indices used in this work see \citet{Trager:1998}. These absorption line indices can then be inverted into measurements of the luminosity-weighted single-burst stellar population (SSP) equivalent age, metallicity ([Z/H]) and alpha-element abundance ([$\alpha$/Fe]) through the use of stellar population models \citep[e.g.][]{Worthey:1994b,Schiavon:2007,Thomas:2010}.

Significant progress has been made on measuring galaxy stellar population parameters over the last two decades. The Lick index system was first applied to early-type galaxies by \citet{Faber:1973}. Using the same approach \citet{Gorgas:1993} and \citet{Trager:1998} found a substantial spread in mean age within the early-type population. In contrast, \citet{Kuntschner:2000} found a large spread in metallicity for early-type galaxies in the Fornax cluster. As observational techniques improved and large spectroscopic samples became available, clear trends emerged between stellar population parameters and structural galaxy parameters such as luminosity, stellar mass and velocity dispersion. These trends are in the sense that more luminous, massive galaxies tend to be older, more metal rich and more alpha-enhanced than lower mass galaxies \citep[e.g.][]{Bernardi:2005,Thomas:2005,delaRosa:2011}. Recent work has extended these trends from giant early-types towards the regime of dwarf galaxies \citep{Guerou:2015,Rys:2015}.

While the absorption line index technique has been widely applied to early-type galaxies, the late-type population is still relatively unexplored. This is largely due to the effects of on-going star formation in these galaxies, giving more extended, complex star formation histories and contaminating many of the absorption line features through emission from ionised gas. Despite these challenges, \citet{Peletier:2007}, \citet{Ganda:2007}, \citet{Sanchez-Blazquez:2014} and others have used high S/N spectra, combined with a careful subtraction of emission lines, to examine the populations of late-type galaxies. These studies found that the correlations between SSP parameters and mass (or velocity dispersion) found in early-type galaxies can generally be extended to late-type galaxies, though the scatter in the late-type population is significantly larger than in early-type galaxies. When mass-weighted parameters are used this scatter decreases significantly \citep{Ganda:2007}, suggesting the bursty nature of star formation in late-types is responsible for the increased scatter. \citet{Gallazzi:2005}, \citet[the CALIFA survey]{Gonzalez-Delgado:2015} and \citet[the MaNGA survey]{Goddard:2017a} used large samples with broad ranges in mass and morphology to show that the trends described above extend across the full range of morphological types E -- Sd and across a range in mass from $\sim 10^9 $M$_\odot$ to $\sim 10^{11} $M$_\odot$.

In addition to mass (or velocity dispersion) and morphology, a third parameter has been investigated as driving variations in galaxy stellar population parameters: environment. Because galaxy mass and environment are correlated, large samples spanning a substantial range in both mass and environment are required to disentangle the effects of the two variables. Despite samples of many thousands of galaxies the role of environment is still unclear. Most studies indicate older luminosity-weighted ages in dense environments, but with disagreement over whether metallicity is enhanced in lower-density environments \citep[e.g.][]{Thomas:2005,Sanchez-Blazquez:2006c} or alpha abundance enhanced in high-density environments \citep[e.g.][]{McDermid:2015}. Overall, the influence of environment, once the correlation with mass has been controlled for, is surprisingly weak \citep{Rogers:2010,Thomas:2010,LaBarbera:2014,McDermid:2015}. Most recently, \citet{Goddard:2017b,Zheng:2017} examined the effect of environment on stellar population gradients in the MaNGA survey, again finding at most a weak dependence.

Despite the large body of work outlined above, the dependence of galaxy stellar populations on galaxy mass, morphology and environment remains unclear. No single study has simultaneously explored the full parameter space with a large enough sample to break the known inter-dependencies between e.g. mass and environment. In addition, the largest existing studies are predominantly based on fibre spectroscopy, which suffers from uncorrected aperture effects that can modify the underlying correlations \citep[see Appendix B of][for an example]{McDermid:2015}. Integral Field Spectroscopy (IFS) can remove this aperture bias, allowing the effective aperture to be scaled to match the apparent diameter of each galaxy.

\begin{figure*}
\includegraphics[width=1.7in]{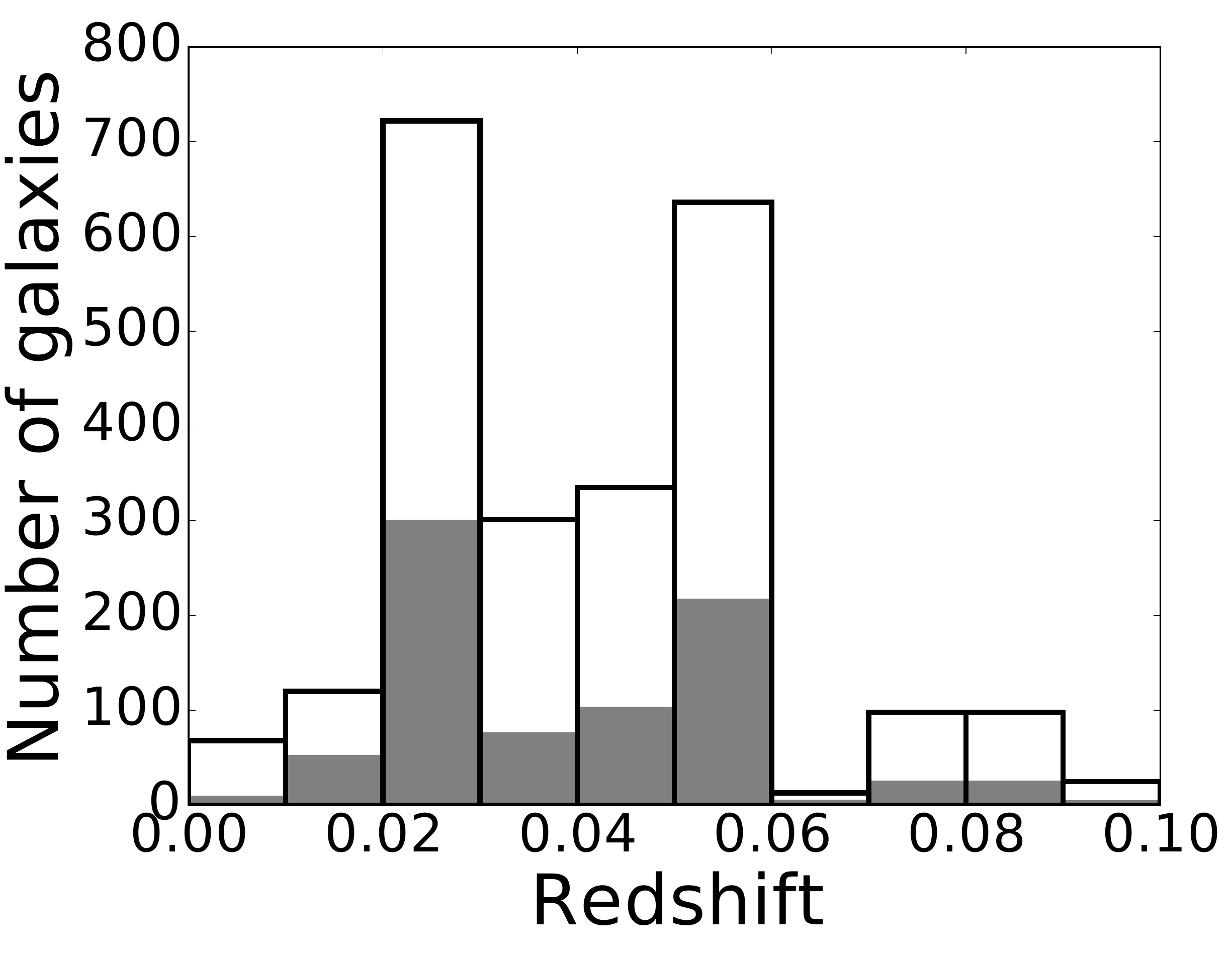}
\includegraphics[width=1.7in]{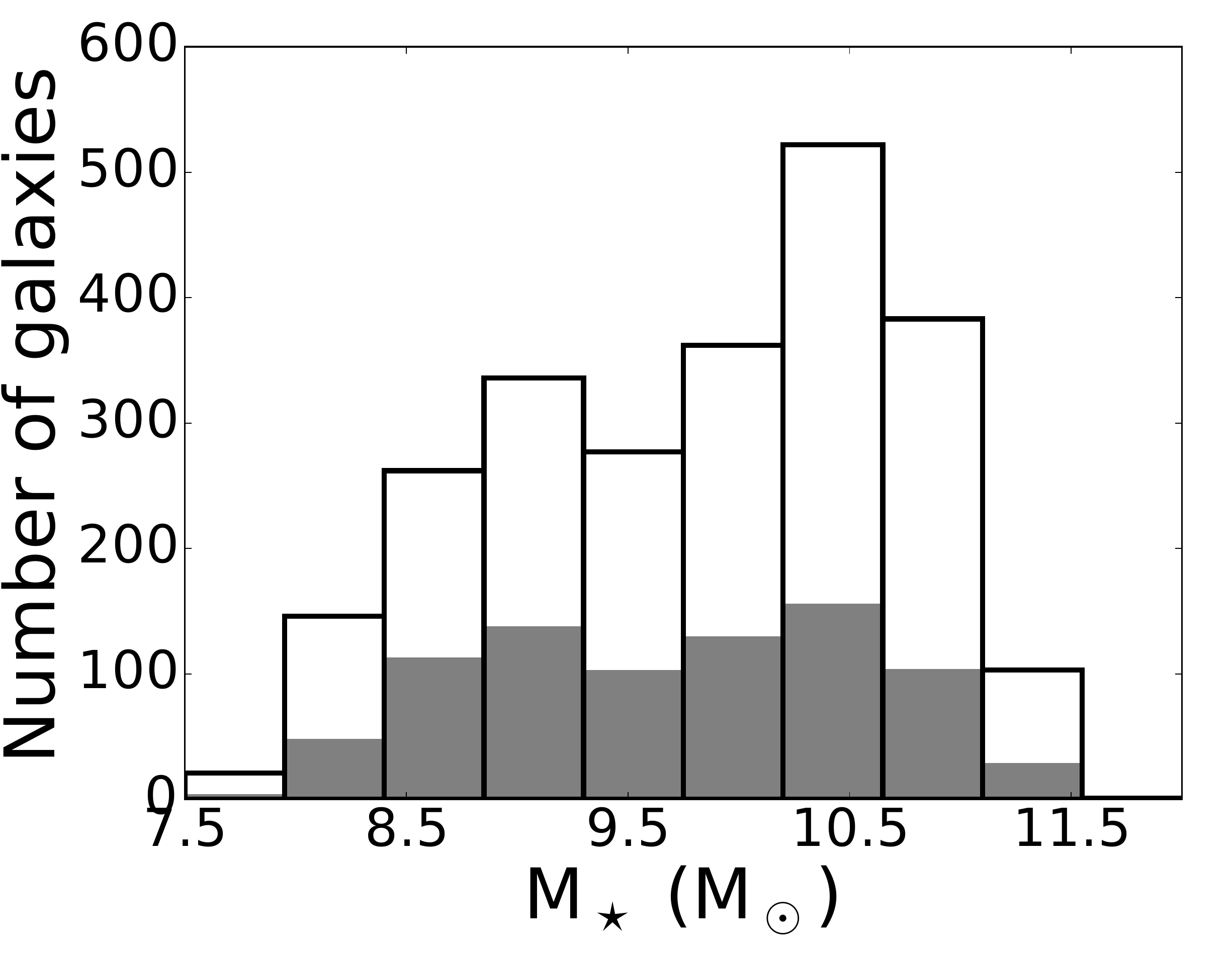}
\includegraphics[width=1.7in]{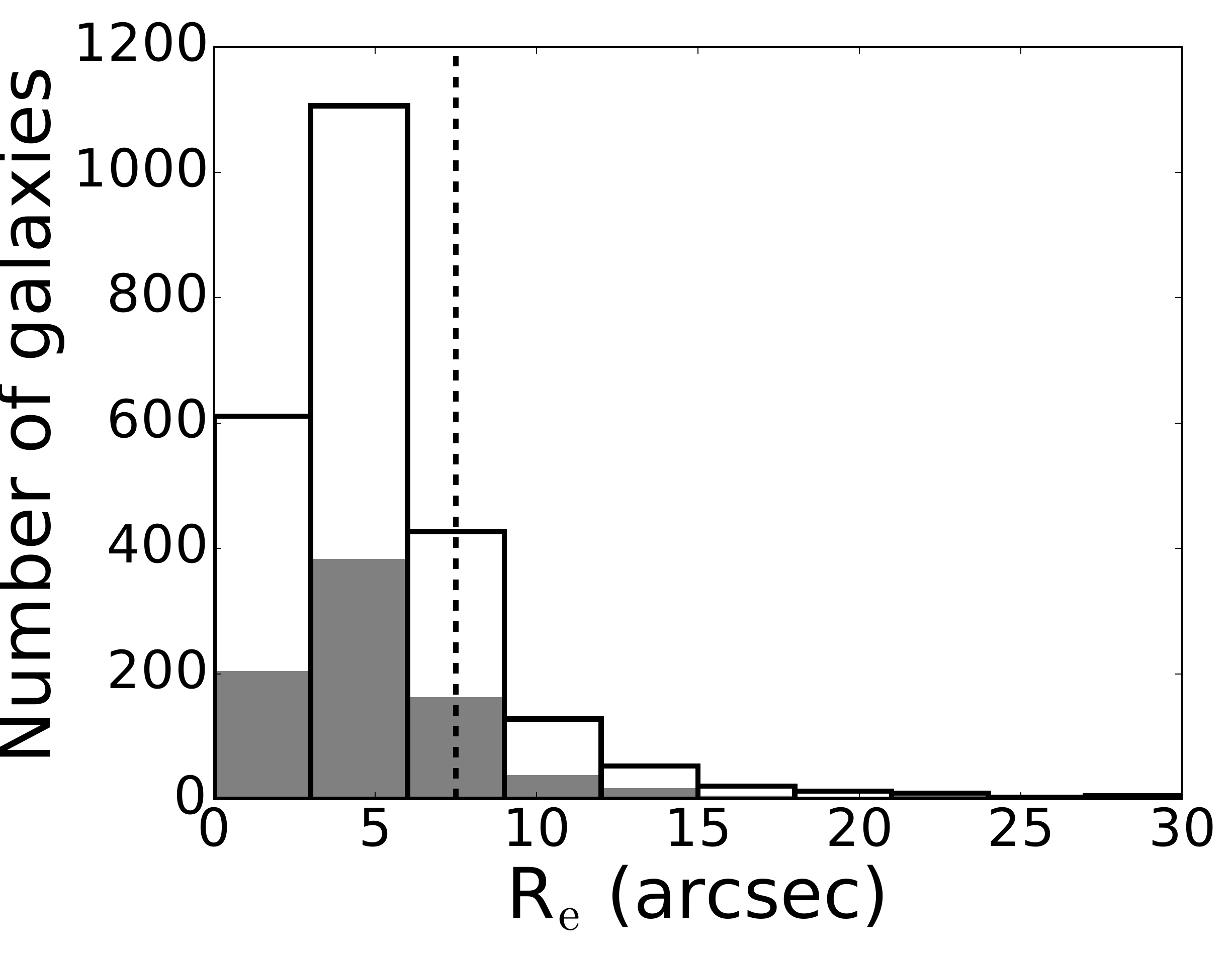}
\includegraphics[width=1.7in]{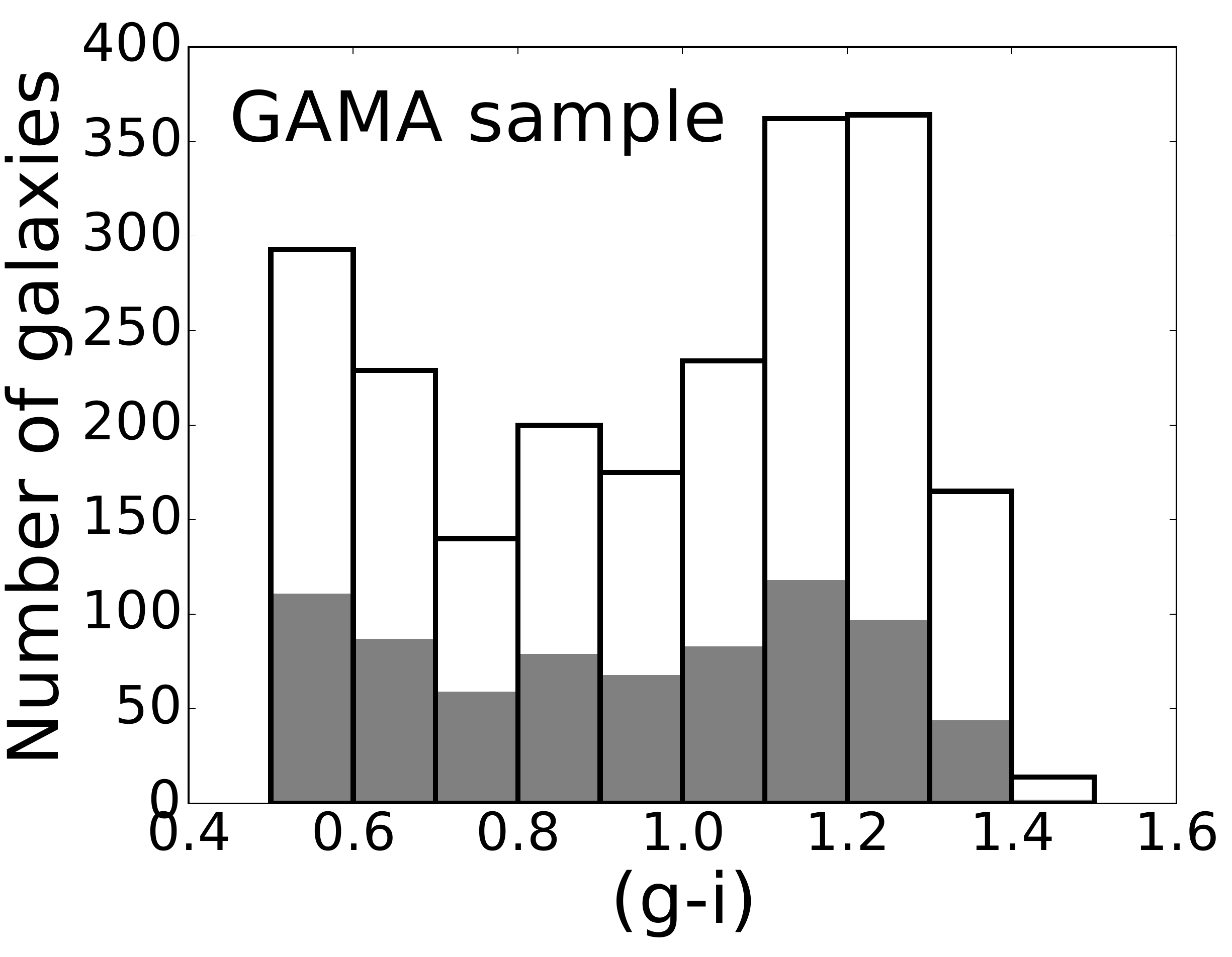}

\includegraphics[width=1.7in]{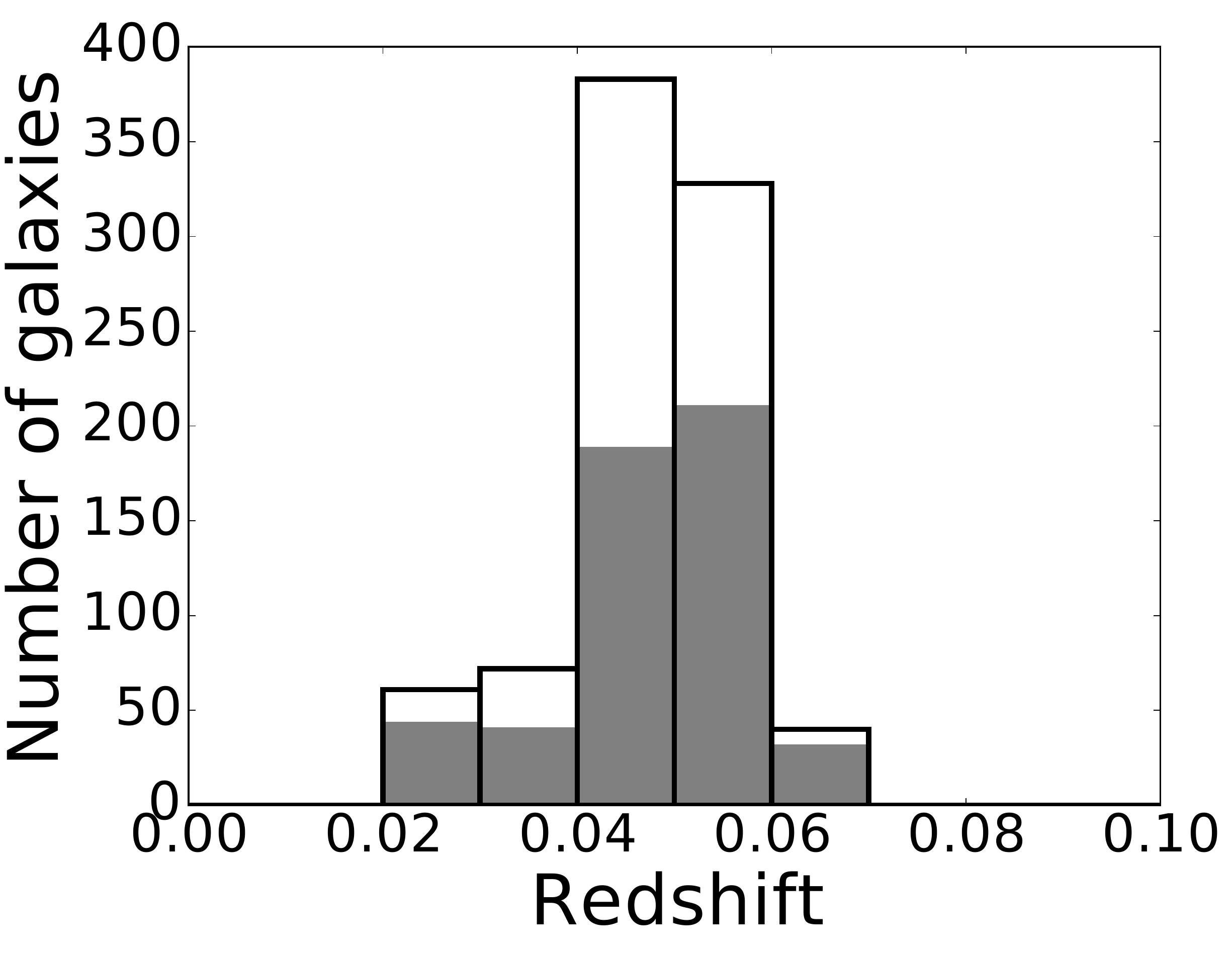}
\includegraphics[width=1.7in]{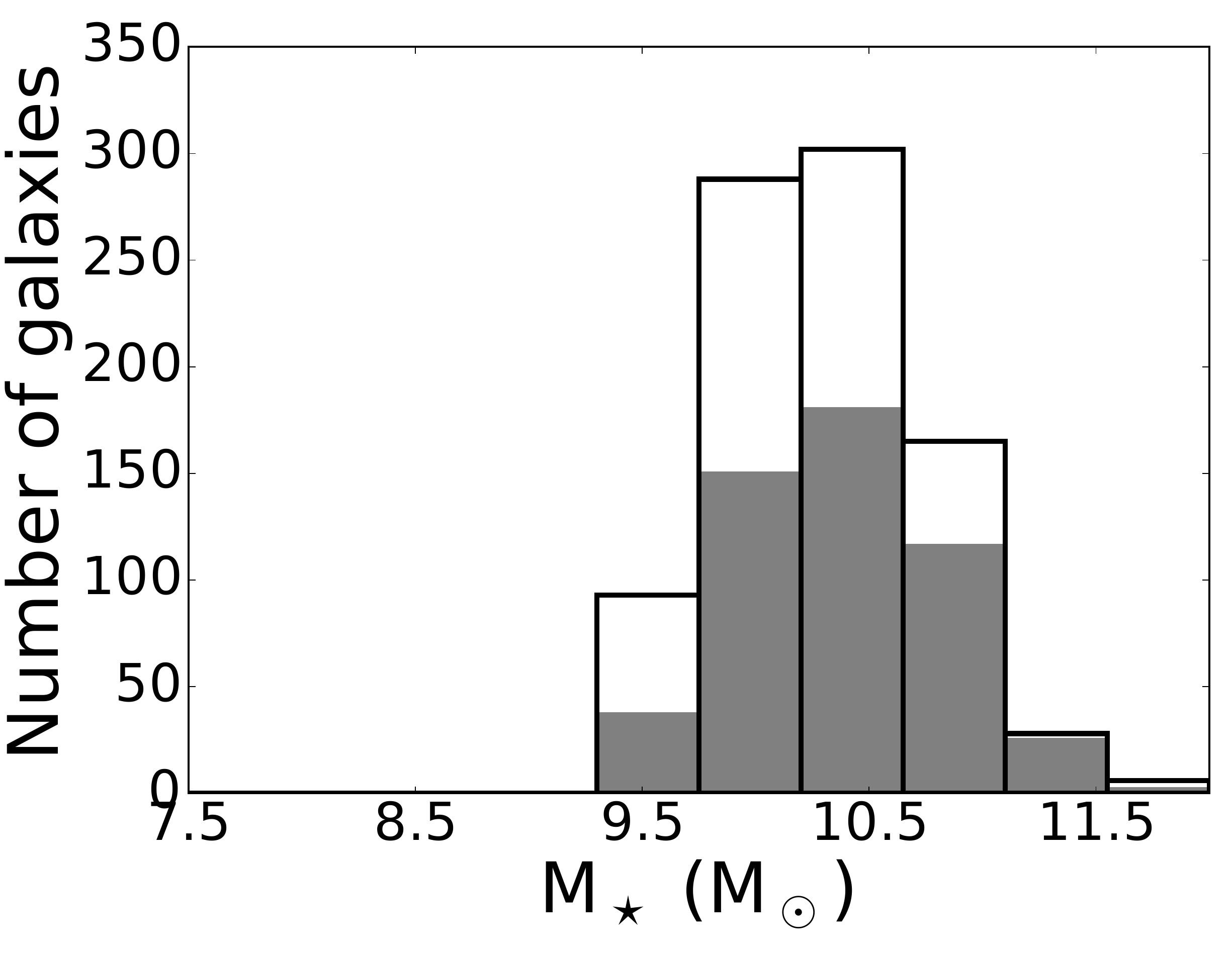}
\includegraphics[width=1.7in]{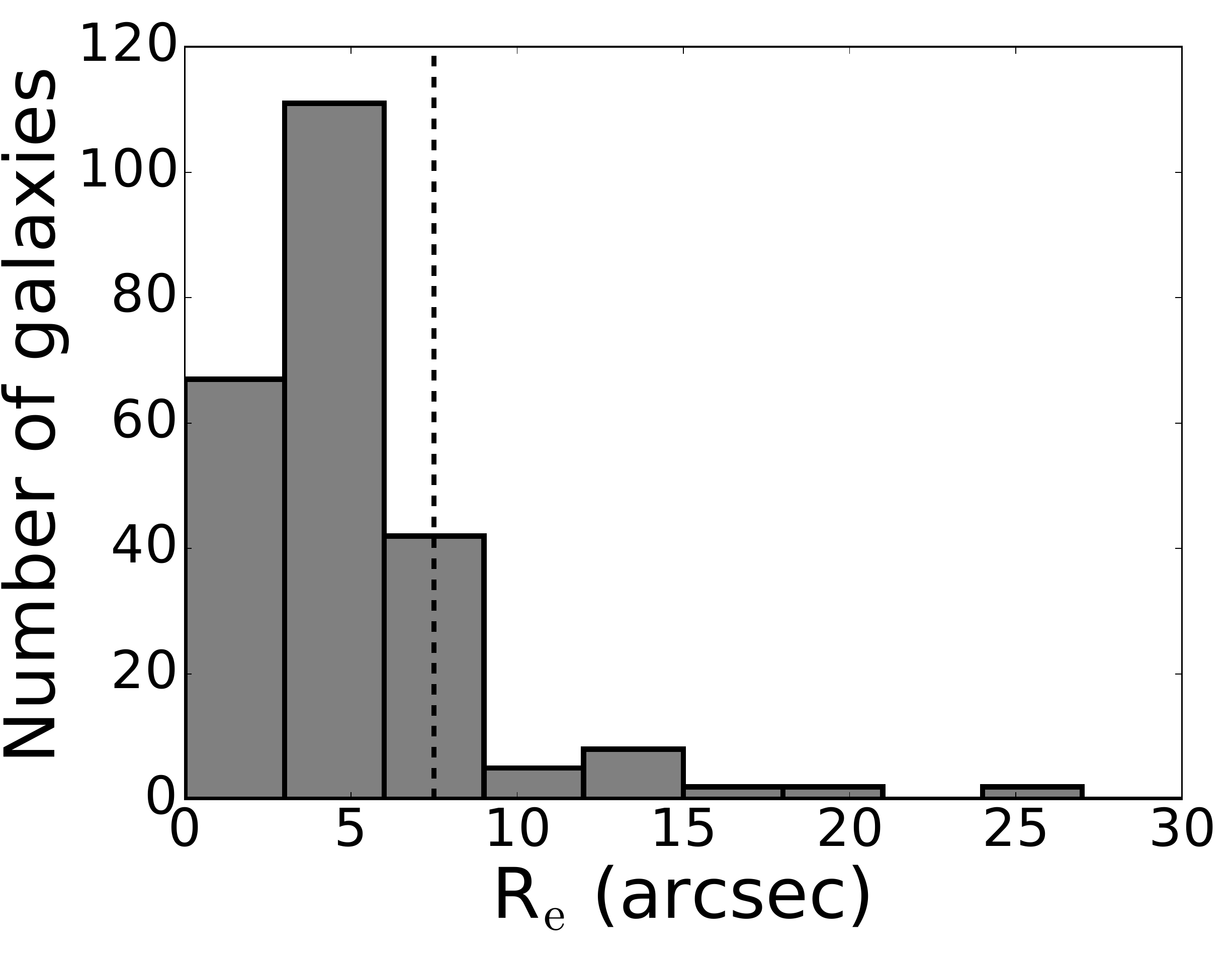}
\includegraphics[width=1.7in]{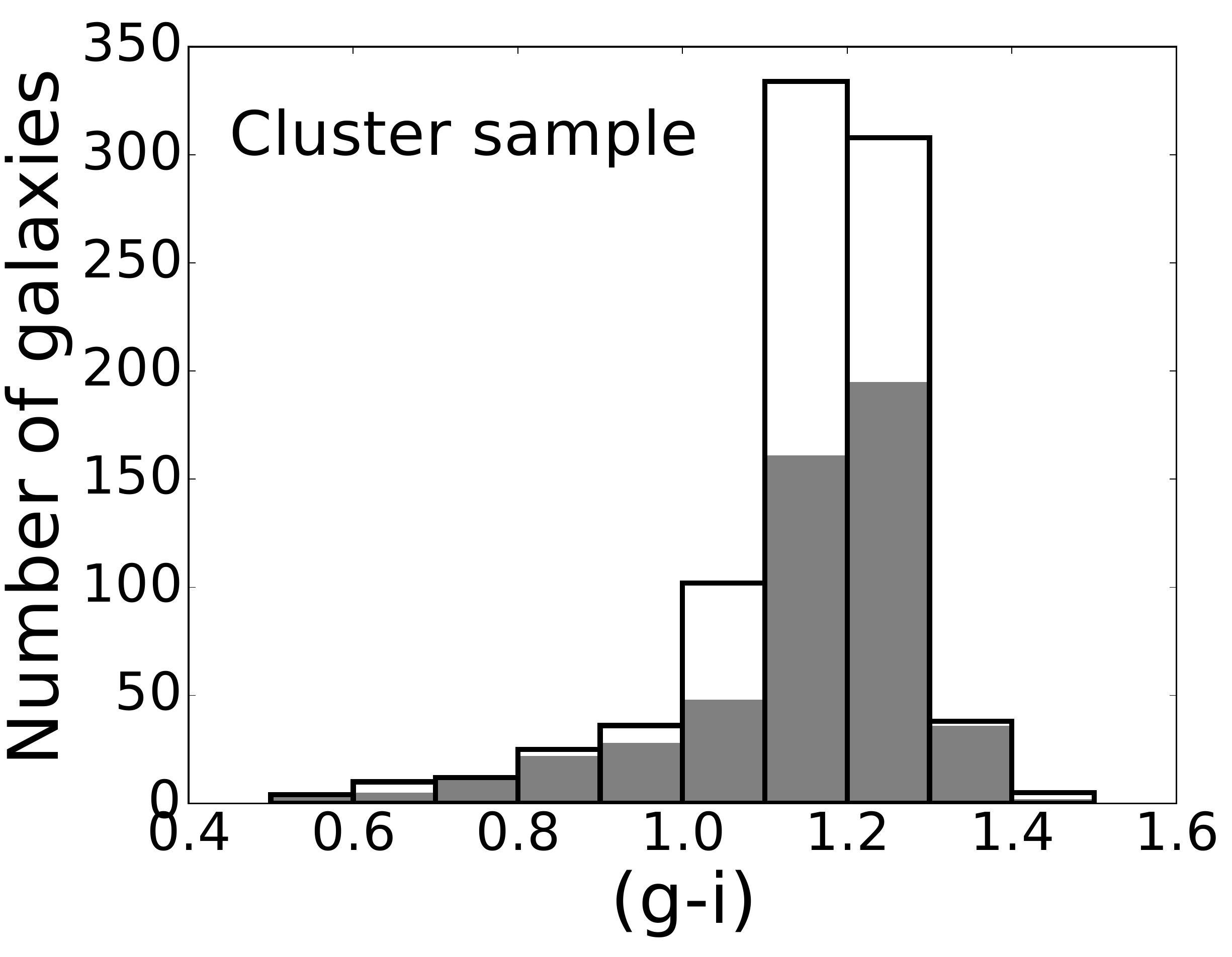}

\caption{Properties of the full SAMI Galaxy Survey sample (unfilled histograms), and the representative subset used in this work (solid histograms). The upper row shows galaxies from the GAMA regions, with the lower row showing the cluster galaxies. Left panel: histogram of galaxy number versus redshift. The effect of convolution of large-scale structure with the survey selection function is apparent in the peaks in galaxy number at z = 0.025 and 0.055. Centre left panel: histogram of galaxy number vs. stellar mass. Centre right panel: histogram of galaxy number vs. R$_e$ (in arcseconds). The SAMI hexabundle field of view of 7.5 arcseconds in radius (indicated by the vertical dashed line) is well matched to the radii of the sample. A typical target has coverage to 2.6 R$_e$, with less than 5 per cent of targets having coverage of less than 1 R$_e$. Only 20 galaxies have R$_e$ smaller than 1 arcsecond, therefore 98 per cent of galaxies are well sampled within 1 R$_e$ by SAMI's 0.5 arcsecond spaxels. For cluster galaxies, only the observed R$_e$ distribution is shown, because not all galaxies from the full cluster sample have measured radii. Right panel: histogram of galaxy number vs. ({\it g}-{\it i}) colour.}
\label{fig:sample}
\end{figure*}

Here we present results from a stellar population study of galaxies drawn from the SAMI Galaxy Survey (SGS). The SGS is a large ($\sim 3600$ galaxies) optical IFS survey of low-redshift galaxies spanning a broad range in mass, morphology and environment, allowing us to disentangle the controlling variables of galaxy stellar populations largely free from aperture bias. Using the large SAMI aperture, we extend the study of stellar populations in galaxies to lower stellar masses ($M_* < 10^{9.5}$ M$_\odot$) than has typically been explored in previous work. In Section \ref{sec:sample} we describe the subset of the SGS sample used in this study, and the sources of additional supporting data including galaxy sizes, masses, environments and morphologies. In Section \ref{sec:method} we describe the determination of the mean, luminosity-weighted SSP-equivalent age, metallicity and alpha-abundance for the sample. We present our key results --- the partial and residual dependence of the SSP parameters on mass, size, environment and morphology--- in Sections \ref{sec:results1} and \ref{sec:results2}. Finally, we discuss our results in the context of modern galaxy evolution theory and simulations in Section \ref{sec:discussion}, before concluding in Section \ref{sec:conclusion}.

\section{Sample and Data}
\label{sec:sample}

The SAMI Galaxy Survey (SGS) is a large integral-field survey of $\sim3600$ galaxies using the Sydney-AAO Multi-Object Integral Field Spectrograph (SAMI) on the 3.9m Anglo-Australian Telescope. The SAMI instrument \citep{Croom:2012} consists of 13 `hexabundles' --- fibre-based integral field units with an individual field of view of 15 arcseconds in diameter \citep{BlandHawthorn:2011,Bryant:2012,Bryant:2014} --- that can be allocated across a 1 degree field of view. In its typical observing mode SAMI is used to observe twelve science targets simultaneously, with the thirteenth hexabundle allocated to a calibration star. A further twenty six individual fibres are dedicated to blank sky observations across the 1 degree field. This multiplexing allows deep ($\sim 3.5$ hours) spectroscopic observations of large samples of galaxies to be efficiently obtained, while achieving accurate sky subtraction and absolute flux calibration.

The 3600 galaxies of the SGS consist of a field and group sample of $\sim$ 3000 galaxies drawn from the Galaxy And Mass Assembly (GAMA) survey \citep{Driver:2011} and an additional sample of more than 600 galaxies drawn from eight low-redshift clusters. The SGS sample is fully described in \citet{Bryant:2015}, with additional details of the cluster sample in \citet{Owers:2017}. The sample spans a range in redshift from z = 0.004 to z = 0.095, a range in stellar mass from $M_* \sim 10^7$ to $10^{12} $M$_\odot$, a range in effective radii, R$_e$ well-matched to the SAMI hexabundle radius and the full range of morphologies from irregulars through spirals to elliptical galaxies. The GAMA sample consists of four volume-limited samples, with the lower bound on M$_*$ increasing in steps as the maximum redshift, and hence volume, of each sample increases. The same mass selection limits were applied to the cluster galaxies based on the cluster redshift, with the minimum M$_*$ for cluster galaxies being $10^{9.5}$ M$_\odot$ \citep[see][for details.]{Owers:2017}

\subsection{Sample}

As the SGS is currently in progress, this study makes use of a subset of the full SGS sample, hereafter simply `the sample'. We use the 1380 galaxies of the SAMI v0.9.1 internal data release, consisting of all galaxies observed up to December 2015 that satisfied the survey criteria for seeing, transparency and signal-to-noise ratio (S/N). For this work we impose an additional criterion of a minimum S/N of 10 per \AA\ within a 1 R$_e$ aperture, to ensure reliable absorption line index measurements. This cut further reduces the sample to 1319 galaxies, with the majority of rejected galaxies having stellar mass M$_* < 10^{8.5}$ M$_\odot$. This sample includes 517 cluster galaxies, with the remaining 826 galaxies drawn from the GAMA regions. While the selection criteria of the GAMA and cluster samples are the same, the two samples probe very different volumes. Because clusters are specifically targeted, cluster galaxies are overrepresented in the combined sample relative to an `average' region of space of the same volume. Therefore, when deriving population average properties we treat the cluster and GAMA samples separately, unless specifically controlling for the effects of environment. In Figure \ref{fig:sample} we show the mass, redshift, R$_e$ and ({\it g}-{\it i}) colour distributions of the present GAMA (upper row) and cluster (lower row) samples with respect to the complete survey selection.

The revised redshift and mass range of the final sample are: z = 0.004 -- 0.092 and $M_* = 10^{7.6} -10^{11.6} $M$_\odot$ (with the majority of galaxies in the range z = 0.02 -- 0.06 and $M_* \simeq 10^{8.5} - 10^{11.0} $M$_\odot$), spanning essentially the same range as the full survey. The sample includes the full range of morphological types present in the SGS, and galaxies spanning the full range of environments from relatively isolated field galaxies through objects in small groups to central cluster galaxies. The sample used in this work is unbiased relative to the full SGS sample. The only modest limitation is in the reduced number of galaxies relative to the full survey that mildly increases the statistical uncertainties by a factor $\sim \sqrt{3}$, compared to what could be achieved with the final SGS sample.

\subsection{SAMI data}

SAMI data consist of three-dimensional data cubes; two spatial dimensions (as projected on the sky) and a third spectral dimension. Two data cubes are produced for each galaxy target, a blue cube with wavelength coverage from 3750 \AA\ to 5750 \AA\ and spectral resolution $R \sim 1800$ (2.65 \AA\ full-width half-maximum) and a red cube with wavelength coverage from 6300 \AA\ to 7400 \AA\ and spectral resolution $R \sim 4300$ (1.61 \AA\ full-width half-maximum) \citep{vdSande:2017}. In this work we make use of the blue data only, because the SAMI red spectral region contains few prominent absorption features useful for constraining stellar population models.

SAMI observations are reduced using the dedicated {\sc sami} Python package\footnote{Available from https://bitbucket.org/james\_t\_allen/sami-package/}, which is described in detail in \citet{Allen:2015}. Data reduction consists of the usual stages of bias subtraction, flat fielding, cosmic ray removal, fibre extraction and absolute flux calibration. Further details of this process, and various quality metrics can be found in \citet{Allen:2015}. The internal v0.91 data release was processed in an identical fashion to the SAMI 1st Public Data Release (DR1), which is described in Green et al. (submitted), though includes more galaxies than DR1. The reduced fibre spectra are drizzled onto a regularly sampled grid following \citet{Sharp:2015}, after correcting for atmospheric dispersion. Final SAMI data cubes have a spatial sampling of 0.5 arc seconds, with a median seeing of 2.1 arc seconds \citep{Allen:2015}. 

\subsection{Ancilliary data}

In addition to SAMI spectroscopic data this study uses a number of additional measurements: effective radii (R$_e$),  stellar masses (M$_*$), stellar velocities and effective velocity dispersions ($\sigma_e$), visual morphological classifications, bulge-to-total ratios (B/T), S\'{e}rsic indices $n$, halo masses (M$_\mathrm{halo}$), and local environmental densities ($\Sigma_5$). Circularised effective radii are taken from the Multi Gaussian Expansion model \citep{Emsellem:1994} photometric fits of D'Eugenio et al. (in prep.) Stellar masses are taken from the SGS sample catalogue \citep{Bryant:2015}, and were originally derived from GAMA photometry in \citet{Taylor:2011}, with cluster galaxy masses derived in \citet{Owers:2017} using the same approach. 

Stellar kinematic parameters (including $\sigma_e$, the ratio of the velocity to the velocity dispersion, $(V/\sigma)_e$ and the specific stellar angular momentum, $\lambda_R$) were measured from the SAMI data cubes and are taken from van de Sande et al. (in prep.). Effective velocity dispersions, $\sigma_e$ were measured from luminosity-weighted aperture spectra extracted from 1 R$_e$ circular apertures where possible, otherwise they were measured from the closest possible aperture in radius and corrected to 1 R$_e$ measurements using the approach described in Sande et al. (in prep.). In \citet{Cortese:2016}, SAMI galaxies were classified into the four morphological groups (ellipticals, E, lenticulars, S0, early-type spirals, Sa/Sb and late-type spirals and irregulars, Sc/Sd/Irr) based on visual inspection of colour images by $\sim 10$ SAMI team members. Galaxies are given a morphological parameter between 0 (for ellipticals) and 3 (for late-type spirals), with non-integer values for galaxies where the classification was split between two morphological types. 

B/T and S\'{e}rsic $n$ were determined using the photometric fitting code {\sc Profit} \citep{Robotham:2017} by Moffett et al. (in prep.). The profiling was done using VST KiDS {\it r}-band imaging \citep{deJong:2013}, using both a single S\'{e}rsic and fixed exponential plus free S\'{e}rsic bulge profile to all SAMI survey targets. For galaxies where the rms deviation of the single S\'{e}rsic model, normalised by the degrees of freedom, was smaller than that of the exponential plus S\'{e}rsic model we assign a B/T of either 1 or 0 to the galaxy based on the single S\'{e}rsic n (B/T = 1 for galaxies with n $> 2$, and B/T = 0 for galaxies with n $\leq 2$). For galaxies where the exponential plus S\'{e}rsic model provided a better fit we define B/T as the luminosity ratio of the high S\'{e}rsic n component to the low S\'{e}rsic n component. If both components had n $\leq 1.5$ we assign B/T to be 0. S\'{e}rsic n was taken from the single S\'{e}rsic fits. Halo masses for the cluster galaxies were taken from \citet{Owers:2017}, and from \citet{Robotham:2011} for the GAMA galaxies. Local environmental measures for all galaxies were provided by \citep{Brough:2017}, based on data from the GAMA survey \citep{Driver:2011} or from \citet{Owers:2017} for the cluster galaxies.

\section{Measuring stellar population parameters}
\label{sec:method}

In this work we focus on the global stellar population properties of galaxies. We form spectra from circular apertures within a radius R$_e$  from the SAMI data cubes, and from these spectra derive Lick absorption line strength indices and stellar population parameters that are representative of the whole galaxy. To extract the aperture spectra, we first identify the centre of each galaxy in the SAMI cubes by collapsing each cube in wavelength space to create a high S/N, two-dimensional image of each galaxy. We use the {\sc Python} function, {\it find\_galaxy.py}\footnote{Available as part of the {\sc mge\_fit\_sectors} package from http://purl.org/cappellari/software}, to identify the luminosity-weighted centroid of each galaxy. For the majority of galaxies this was deliberately chosen to be the central spaxel of the cube, (24,24) during the drizzling process, but for highly disturbed objects or cubes containing multiple galaxies other locations may be chosen. We then identify all spaxels whose centre lies within a projected distance $d \leq R_e$ and sum the spectra for those spaxels to create a single aperture spectrum. We calculate the variance of each aperture spectrum in a similar fashion, by summing the variance spectra of individual spaxels, with the additional step of using the {\sc weight} and {\sc covar} extensions of each data cube to account for the covariance between spaxels \citep[see][for an example of this procedure]{Sharp:2015}. The median and maximum S/N for the R$_e$ aperture spectra are 60 and 187 per \AA\ respectively, measured over the full SAMI blue spectral range.

\subsection{Lick index measurements}

For each galaxy we measure a set of 20 Lick indices, using the index definitions of \citet{Worthey:1997} and \citet{Trager:1998}. The Lick indices we measure are:
\begin{itemize}
\item The Balmer indices: H$\delta_A$, H$\delta_F$, H$\gamma_A$, H$\gamma_F$, H$\beta$
\item The iron-dominated indices: Fe4383, Fe4531, Fe5015, Fe5270, Fe5335, Fe5406
\item The molecular indices: CN1, CN2, Mg1, Mg2
\item The remaining indices in the SAMI blue wavelength range: Ca4227, G4300, Ca4455, C4668, Mgb
\end{itemize}
Each index is defined by a central passband, two side passbands that define a pseudocontinuum level and a spectral resolution. 

\begin{figure*}
\includegraphics[width=2.25in]{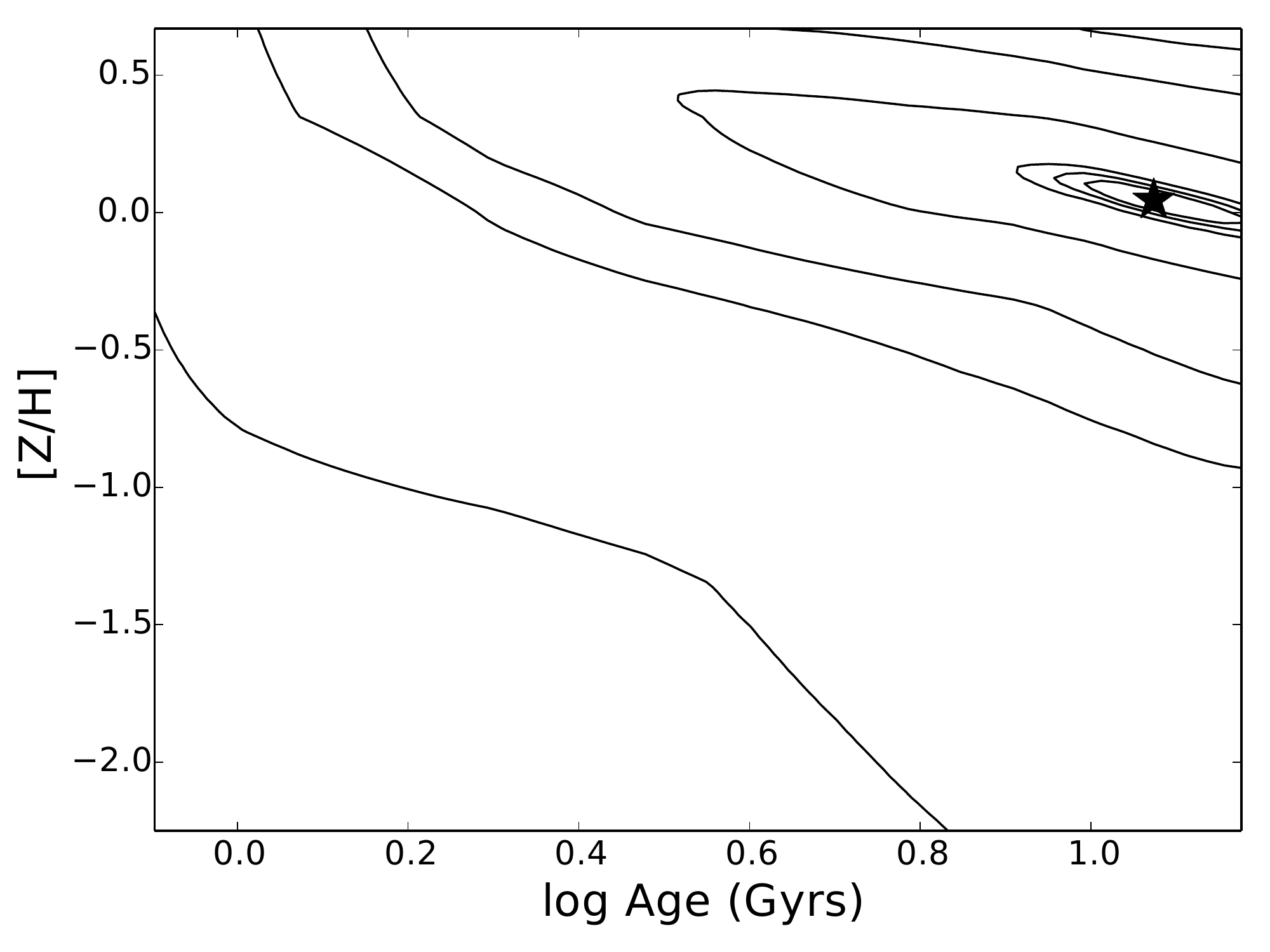}
\includegraphics[width=2.25in]{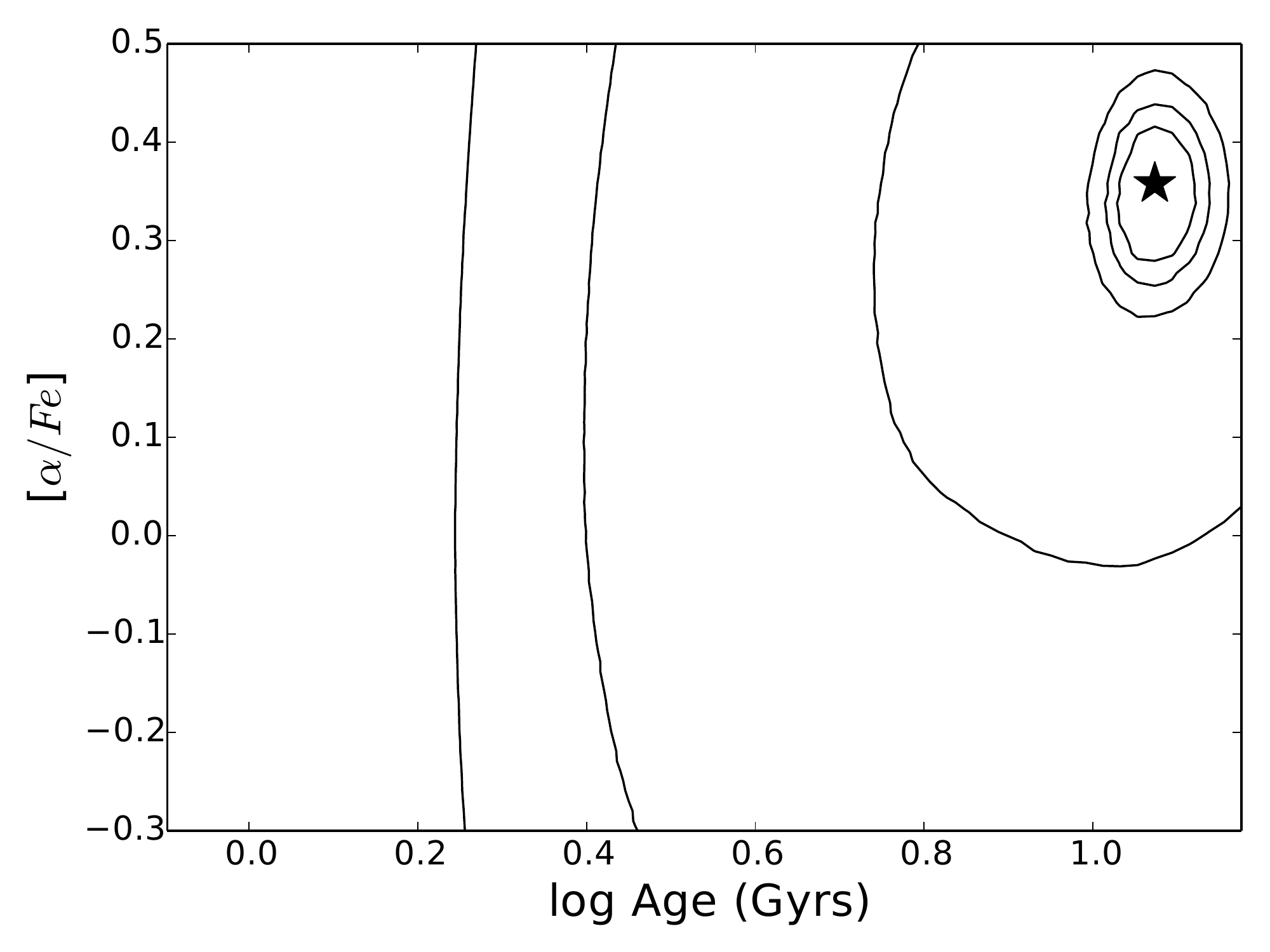}
\includegraphics[width=2.25in]{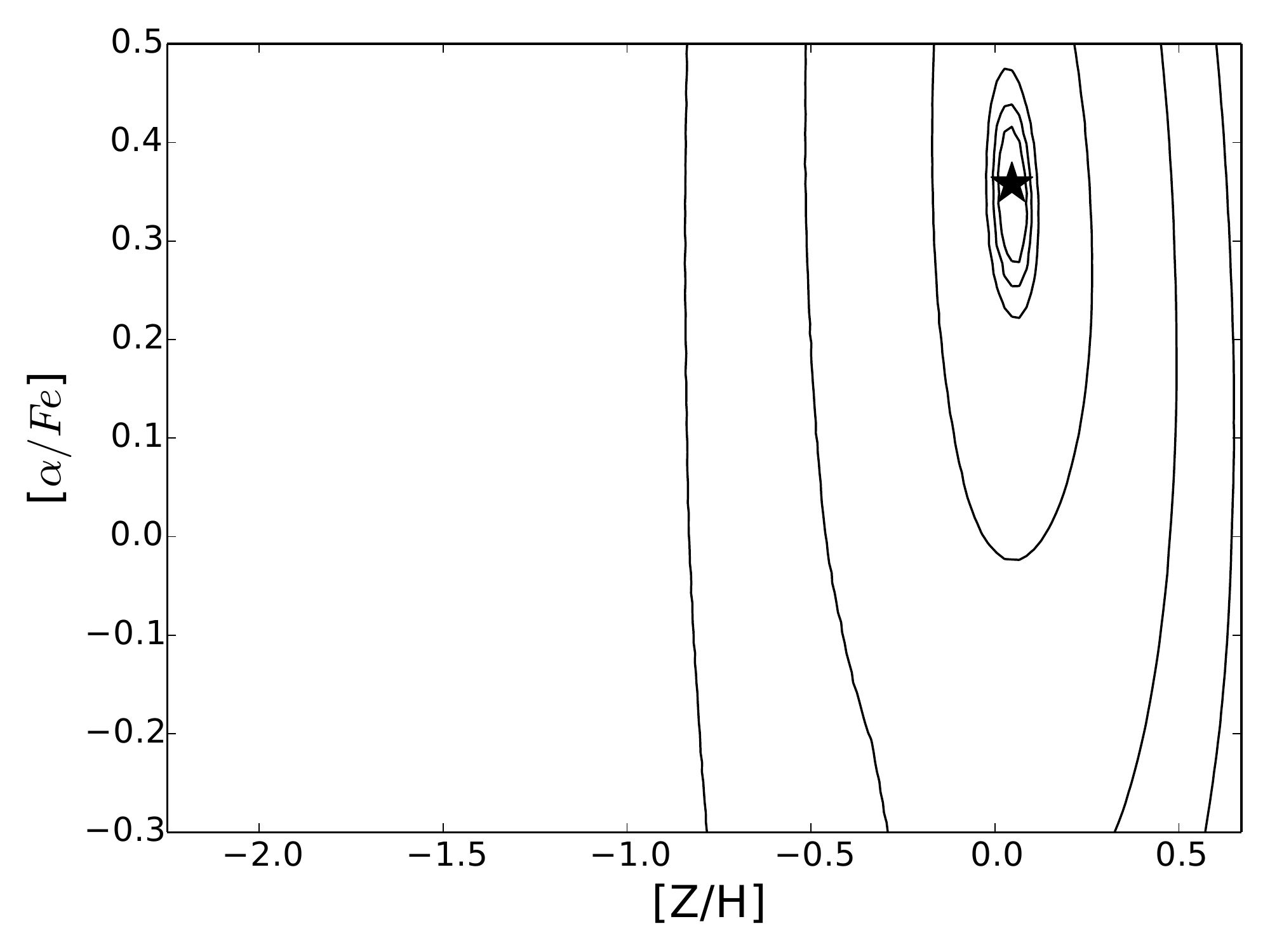}
\caption{$\chi^2$ distribution of the SSP parameters from the TMJ model for an example galaxy. The parameters were derived from the R$_e$ aperture spectrum of the galaxy with SAMI ID 93880000068. This is a high S/N galaxy, chosen to best illustrate the fitting technique, rather than be representative of a typical galaxy in the sample. The best-fitting values of age, [Z/H] and [$\alpha$/Fe] are indicated with the black star. The inner contours indicate the $\chi^2$ level corresponding to 1, 2 and 3 $\sigma$ uncertainties on the parameters, with the outer contours showing order-of-magnitude increases in $\chi^2$. Left panel: $\chi^2$ contours for [Z/H] vs log age. Centre panel: [$\alpha$/Fe] vs log age. Right panel: [$\alpha$/Fe] vs [Z/H].}
\label{fig:ssp_chi2}
\end{figure*}

Before measuring the indices, we perform some initial processing on the aperture spectra. Emission from star formation can significantly affect the depth of some absorption lines, particularly the Balmer lines, and this emission must be removed before measuring the absorption line strengths. We correct for emission by comparing the observed spectra to a set of template spectra that are unaffected by emission, identifying pixels that differ significantly between the observed and best fitting linear combination of template spectra, and replacing affected pixels with the values of the emission-free template spectra. We found this method to provide more physical absorption line strengths than simply subtracting emission line fits derived using the {\sc LZIFU} software \citet{Ho:2016}. In addition, because of the design of the Lick spectrograph, each index is measured at a different spectral resolution, and, to ensure consistency with the Lick system, all spectra must be broadened to the relevant Lick resolution before measuring the absorption line strength. The observed spectrum is broadened by a Gaussian, after accounting for the instrumental broadening and the intrinsic broadening due to a galaxy's velocity dispersion. These steps are described in detail in the following paragraphs.

We begin by fitting each spectrum with a set of template spectra using the penalised Pixel Fitting (pPXF) code of \citet{Cappellari:2004}. The input template spectra consist of 30 stellar templates chosen from the MILES spectral library \citep{Sanchez-Blazquez:2006b,Falcon-Barroso:2011}. We perform a fit to each input spectrum using pPXF three times, fitting for the first two moments of the line-of-sight velocity distribution, $v$ and $\sigma$, and including a 10$^\mathrm{th}$ order additive polynomial in each case. In the initial fit we assume a uniform noise spectrum, and use this to determine an initial approximate match to the observed spectrum. We use the absolute residuals between the initial fit and the observed spectrum to determine a new estimate of the noise in the spectrum. The new noise spectrum is clipped then smoothed using a broad Gaussian window to produce a smooth noise spectrum, unaffected by large spikes in the residuals due to emission lines or bad pixels. 

The smooth noise spectrum is then used as input to pPXF, with the original observed spectrum, to perform a second fit to the data. In this iteration we use the {\sc clean} keyword in pPXF to iteratively reject bad pixels from the fit. This rejection process successfully identifies regions of bad pixels or emission, even when the emission is weak or coincident with an absorption feature. These regions of bad pixels are expanded by +25 \% to include the faint wings of emission lines that may still suffer from contamination. Finally, using the smooth noise spectrum and the set of good pixels identified in the second iteration we perform a third fit to the data, the output of which is a measured $v$ and $\sigma$ and a best-fitting template spectrum for each input spectrum. Pixels in the input spectrum identified as bad in the second pPXF fit are flagged and replaced with corresponding pixels in the best-fitting template spectrum. This procedure successfully corrects for any bad pixels that were not accounted for in the data reduction process and for emission-line infill of absorption features or contamination of their side passbands. We found this iterative identification then replacement method to provide more physical absorption line strengths than simply subtracting emission line fits, particularly in the presence of weak emission, in the wings of emission lines, or where the continuum S/N was low.

\begin{figure*}
\includegraphics[width=2.25in]{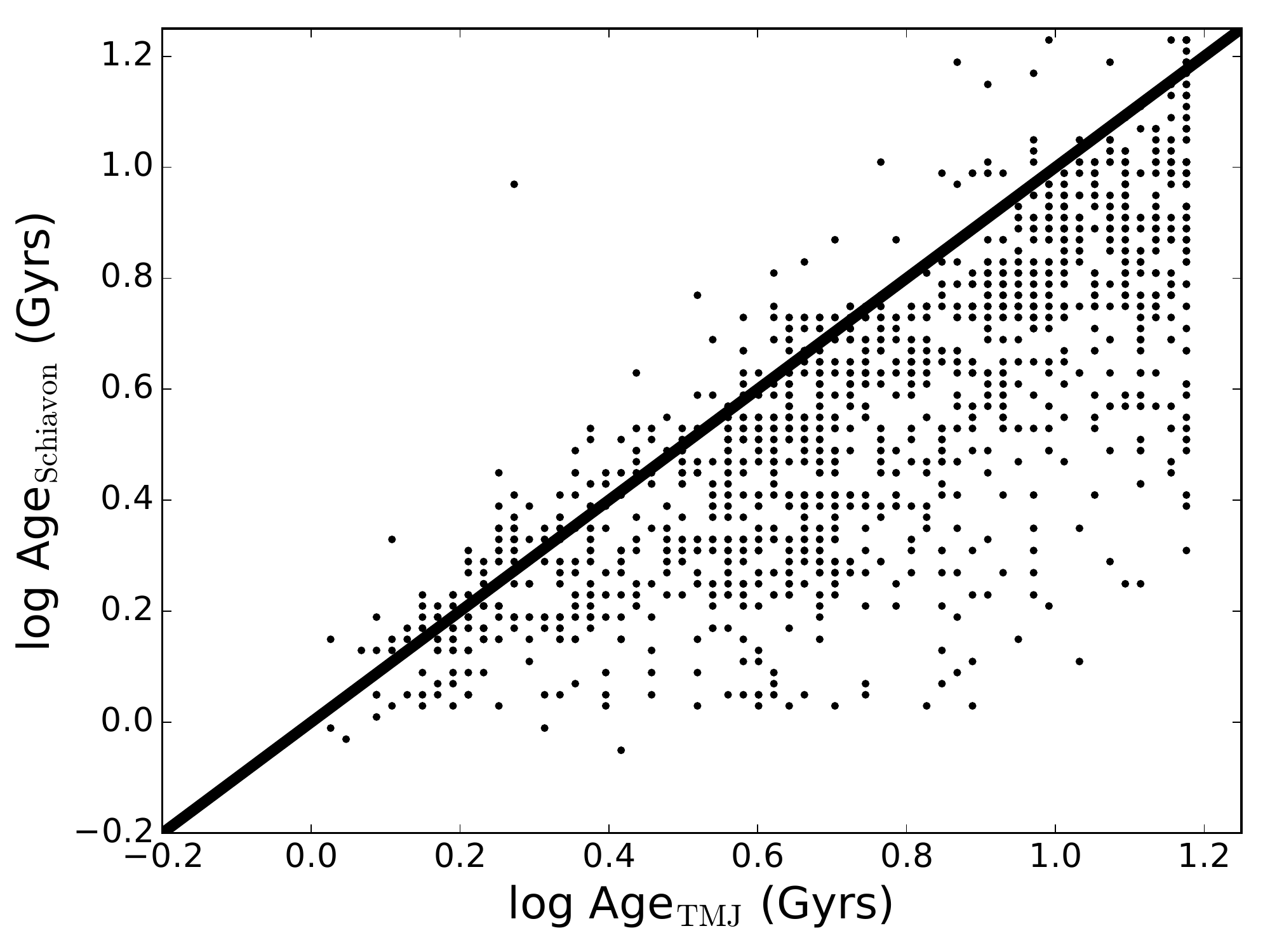}
\includegraphics[width=2.25in]{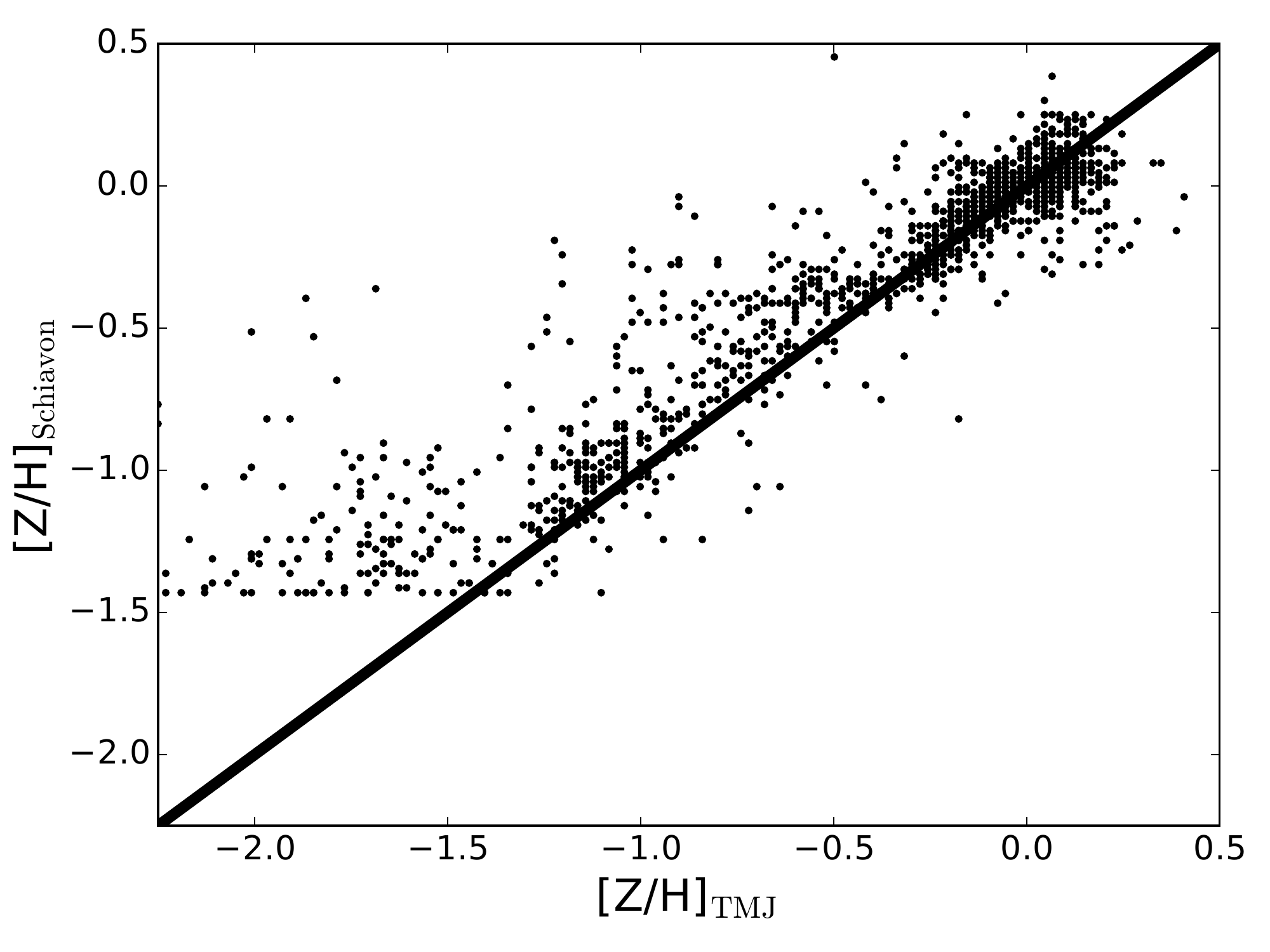}
\includegraphics[width=2.25in]{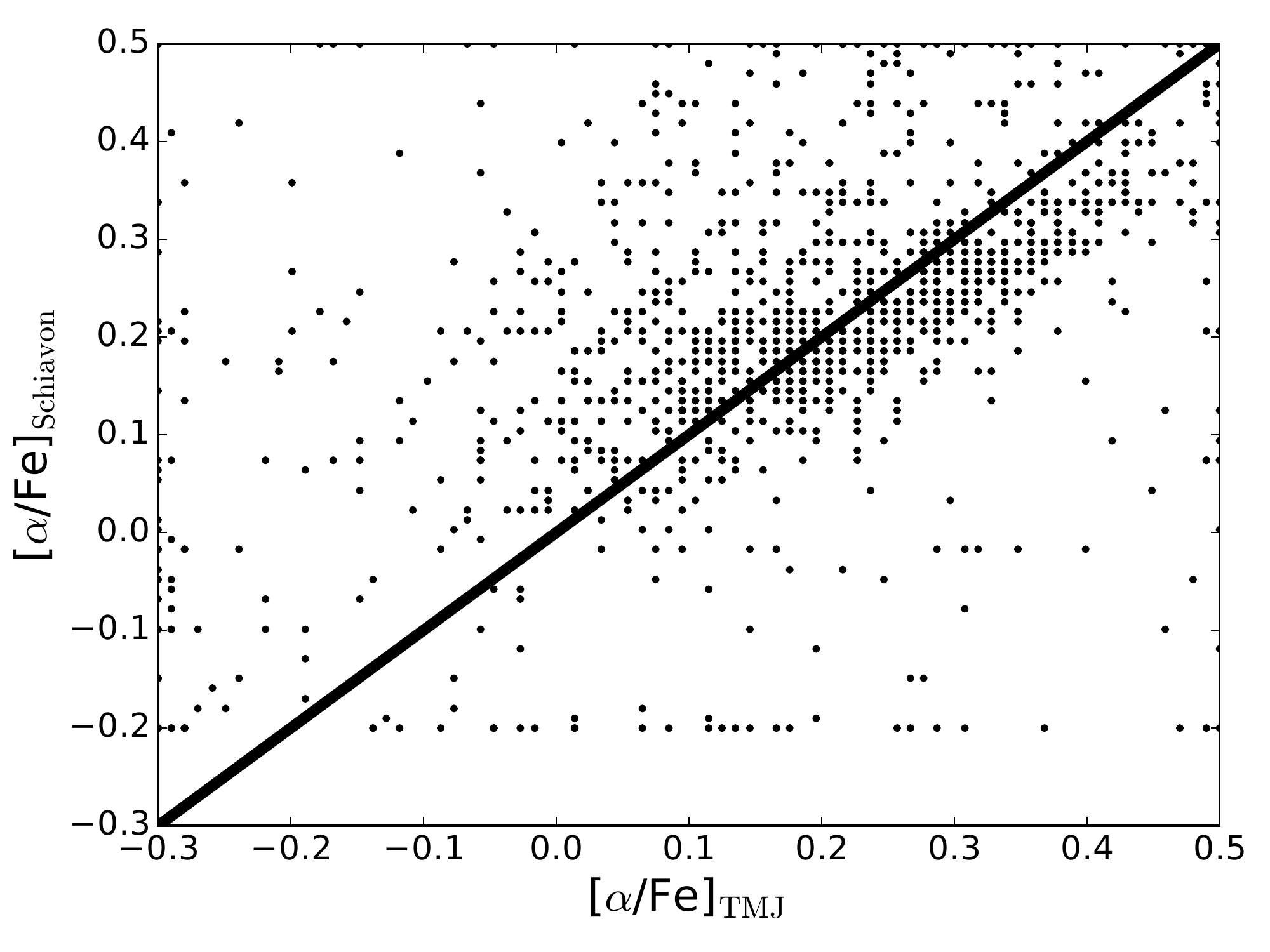}
\caption{Comparison of SSP equivalent age, metallicity and alpha-abundance determined from SAMI aperture spectra using the models of \citet[$x$-axes]{Thomas:2010} and \citet[$y$-axes]{Schiavon:2007}. The one-to-one relation in each panel is indicated with the solid black line. For [$\alpha$/Fe] we show only galaxies with S/N $> 20$ where the abundance measurements are reliable. In general, the agreement between the two models is good, though we discuss this in detail in the text.}
\label{fig:tmj_sch_comp}
\end{figure*}

Each Lick index was measured on the emission corrected observed spectrum, after shifting the spectrum to the appropriate rest-frame wavelength scale based on the measured velocity. Before measuring the absorption line indices, the observed spectrum was broadened to the corresponding resolution of the index in the Lick system \citep{Worthey:1997,Trager:1998}. For each index, the spectrum was then convolved with a Gaussian, such that the total broadening of the spectrum matches that of the Lick system for the given index, i.e. $\sigma_\mathrm{instrument}^2 + \sigma_\mathrm{galaxy}^2 + \sigma_\mathrm{applied}^2 = \sigma_\mathrm{Lick}^2$), where $\sigma_\mathrm{instrument}$ is the instrumental broadening, $\sigma_\mathrm{galaxy}$ is the velocity broadening due to each galaxies intrinsic velocity dispersion, $\sigma_\mathrm{applied}$ is the additional broadening applied to each spectrum to match the Lick resolution and $\sigma_\mathrm{Lick}$ is the broadening of each index in the Lick system. For the $\sim 8$ per cent of galaxies where the observed broadening is already greater than the desired Lick resolution no additional broadening is applied. Instead a correction is made to the measured index following \citet{Schiavon:2007}, to account for absorption line flux that may be shifted outside the central passband. This process was repeated for each index, resulting in all Lick indices being measured on spectra with resolution matching the Lick resolution for that index. While this is not the standard approach implemented in e.g. \citet{Kuntschner:2006}, where the effects of instrumental resolution and galaxy intrinsic dispersion are corrected for separately, it is equivalent in that it results in Lick indices measured at the Lick resolution on (effectively) zero velocity dispersion objects. Our correction is exactly as implemented in the {\it lick\_ew.pro} routine of \citet{Graves:2008}, part of their {\sc ez\_ages} package, which has been widely used to measure stellar population properties of galaxies and star clusters from Lick indices.

Uncertainties on all indices are estimated using a Monte Carlo procedure, where noise is randomly added to the best-fitting template spectrum and the indices remeasured for 100 different realisations. The standard deviation of the measured values is adopted as the uncertainty on each index, and the uncertainties obtained in this way are consistent with expectations from the S/N of the input spectra. We did not measure higher order moments of the line-of-sight velocity dispersion for this sample, as such measurements are highly uncertain at the low velocity dispersions and S/N of many of our galaxies \citep{vdSande:2017}. According to \citet{Kuntschner:2004}, this may introduce an additional uncertainty in the index measurements of up to 5 per cent in the most extreme cases, however for the majority of our galaxies this effect is negligible.

\subsection{Single Stellar Population (SSP) equivalent parameters}

To convert the observed Lick indices into SSP-equivalent age, metallicity and alpha-abundance we utilise the stellar population models of \citet[hereafter S07]{Schiavon:2007} and \citet[hereafter TMJ]{Thomas:2010}. These models predict Lick indices (as measured from flux-calibrated spectra at the appropriate Lick resolution) as a function of log age, [Z/H] and [$\alpha$/Fe]. We interpolate the standard models onto a finer grid, with a resolution of 0.02 in log age and [Z/H], and 0.01 in [$\alpha$/Fe]. We use a $\chi^2$ minimisation approach to determine the SSP that best reproduces the measured Lick indices, following a method first used by \citet{Proctor:2004}. This approach minimises the residuals between the observed indices and the predicted model indices, weighted by the measured uncertainty on each index. Where the observed value of an index lies more than $1 \sigma$ outside the range covered by the model it is excluded from the fit to avoid biasing the best-fitting solution. If fewer than five indices are included in the fit (or the fit lacks at least one Balmer index and one Fe index) then no solution is returned. 

For the TMJ model we utilise all 20 measured Lick indices. For the S07 model, we found that the model could not simultaneously reproduce the behaviour of several of the measured indices. In addition, the TMJ and S07 models give dramatically different behaviour for the dependencies of some abundance-sensitive indices with respect to [$\alpha$/Fe]. For example, G4300 and C4668 increase with increasing [$\alpha$/Fe] in the TMJ model, but show the opposite behaviour in the S07 model (with the TMJ predictions being consistent with the observed trends in the data). The S07 model also includes fewer indices overall. When using the S07 model, we restricted our fit to the following 11 indices: H$\delta_A$, H$\delta_F$, H$\gamma_A$, H$\gamma_F$, H$\beta$, Mgb, Mg$_2$, Fe4383, Fe5015, Fe5270 and Fe5335.

For each aperture spectrum we generate a $\chi^2$ distribution in the three-dimensional space of age, [Z/H] and [$\alpha$/Fe], an example of which is shown in Figure \ref{fig:ssp_chi2}. We adopt the values of age, [Z/H] and [$\alpha$/Fe] that result in the minimum $\chi^2$, and uncertainties on the three SSP parameters are determined from the $\chi^2$ distribution. We restrict the [$\alpha$/Fe] measurements to aperture spectra with S/N $> 20$, due to the increased uncertainty associated with this parameter. Typical uncertainties are $\pm 0.15$ in log age, $\pm 0.17$ in [Z/H] and $\pm 0.14$ in [$\alpha$/Fe]. The conversion of continuous index measurements into SSP values uses discretely sampled models, which introduces a degree of discretisation noise into the SSP values. This is particularly evident in the age values, where small changes in the index-SSP relation causes an excess of galaxies at certain ages. However, a comparison of ages between the two SSP models (which exhibit discretisation at different values) shows good agreement between ages, suggesting the discretisation noise does not significantly affect the derived SSP equivalent parameters.

\subsubsection{Choice of SSP model}
\label{sec:ssp_model_choice}

\begin{figure*}
\includegraphics[width=2.25in]{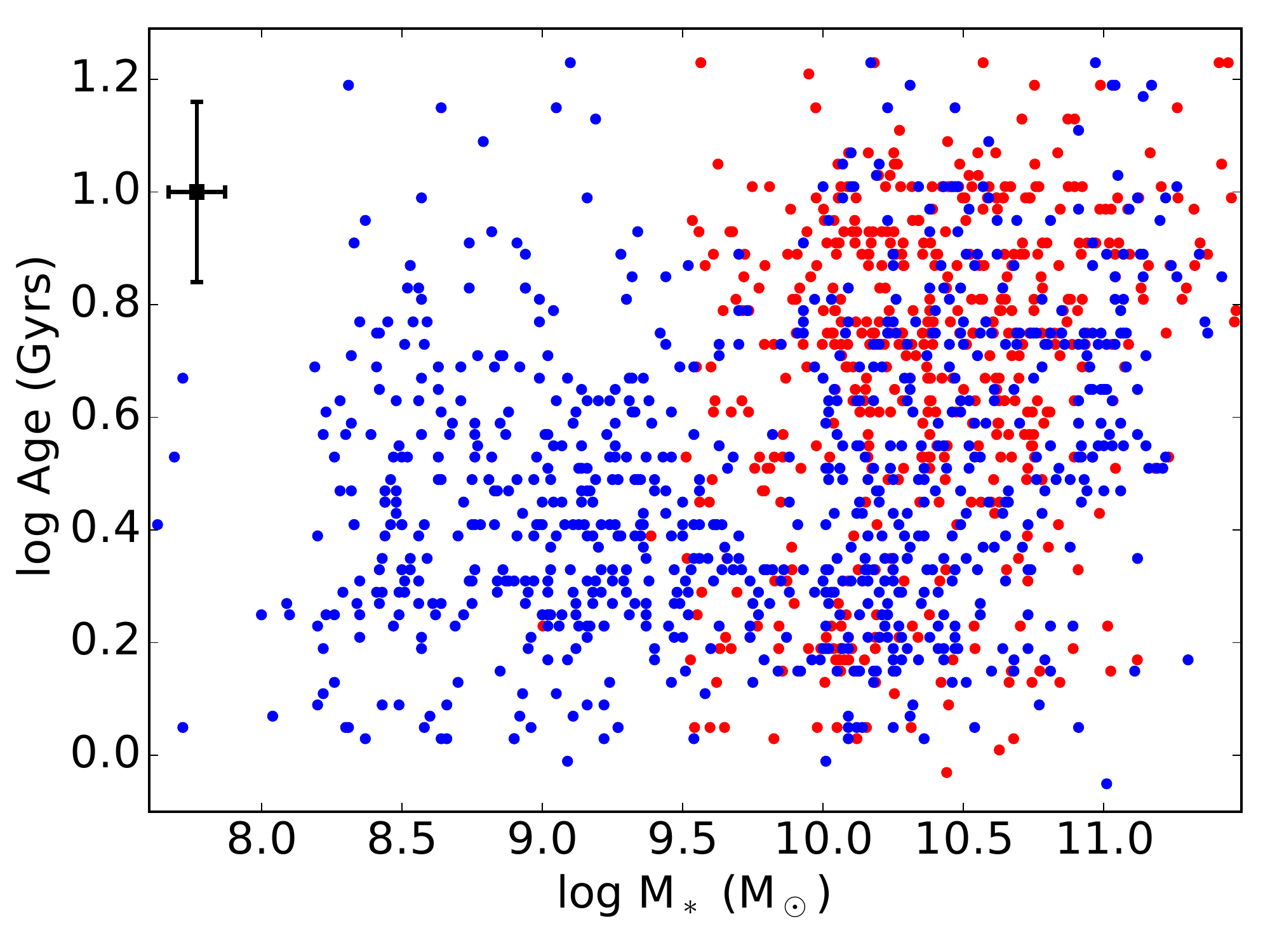}
\includegraphics[width=2.25in]{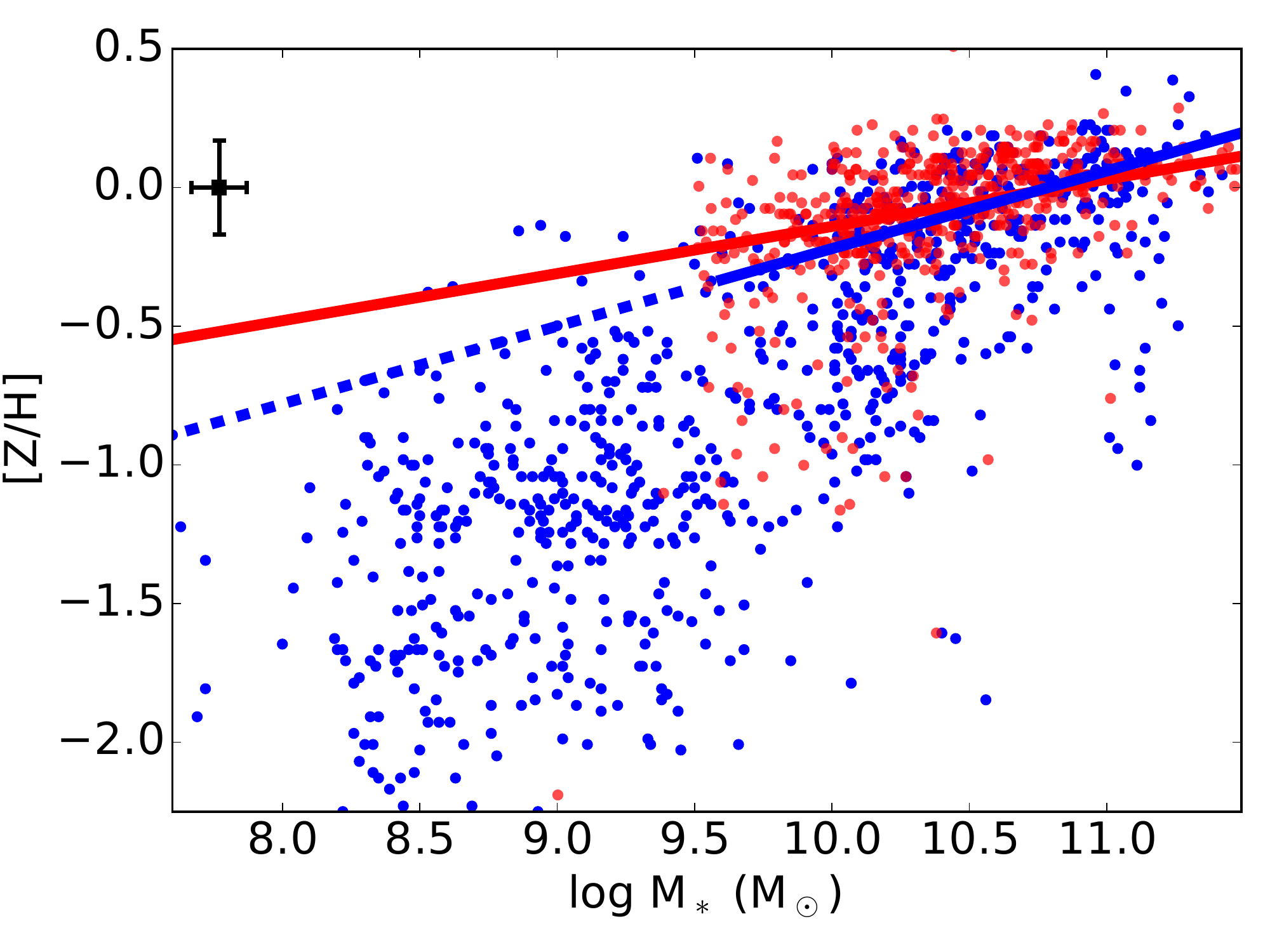}
\includegraphics[width=2.25in]{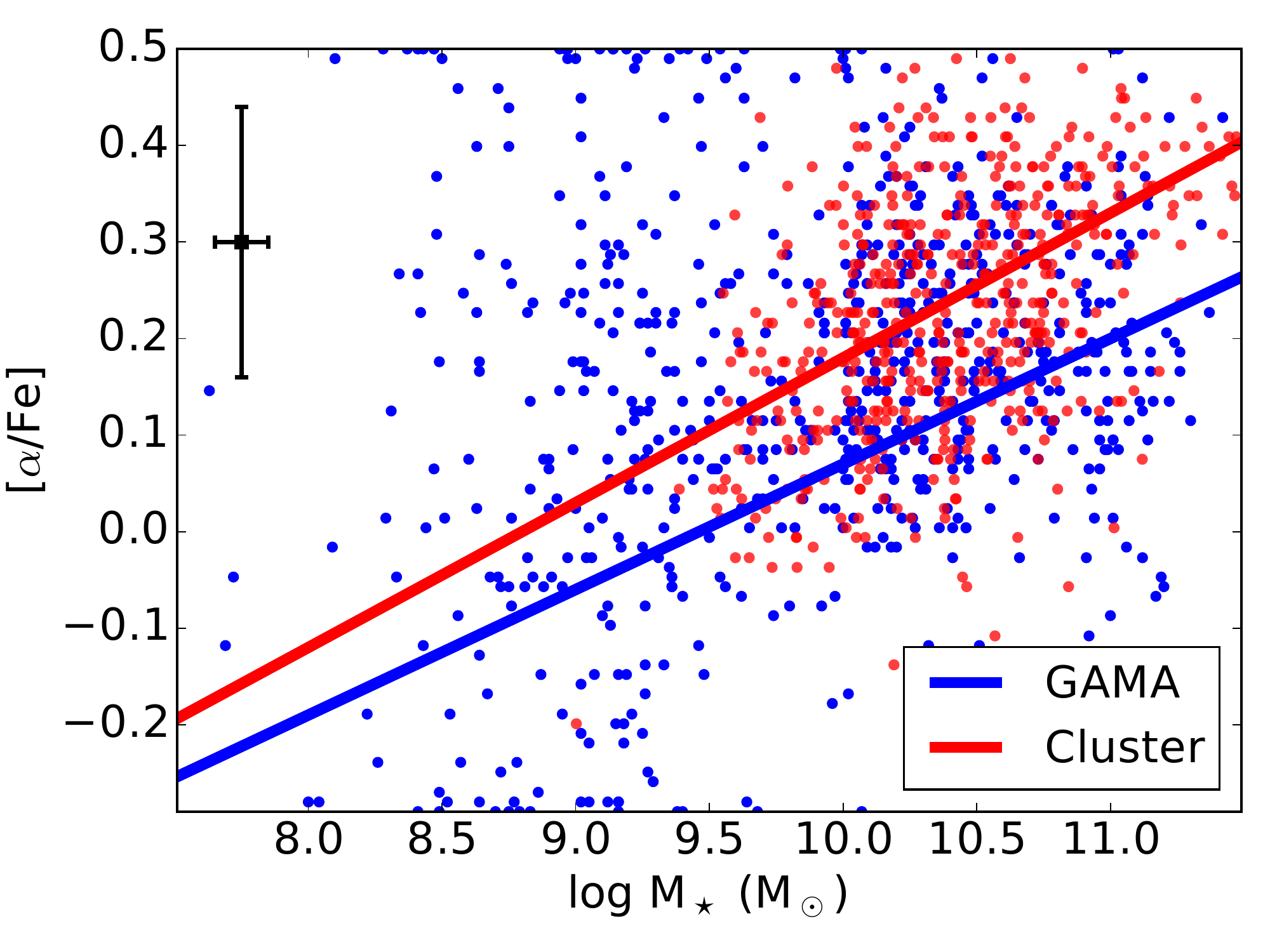}
\caption{Dependence of the SSP parameters on stellar mass, M$_*$. Red points indicate cluster galaxies and blue points those drawn from the group and field GAMA sample. Solid lines indicate linear fits to the data, whose parameters are given in Table \ref{tab:ssp_relations}. For [Z/H], the fit range was restricted to galaxies with M$_* > 10^{9.5}$ M$_\odot$, the extent of the solid line. The dashed lines indicate the extrapolation of these fits to lower masses. Left panel: age (in Gyrs) vs. M$_*$. Centre panel: [Z/H] vs. M$_*$. Right panel: $[\alpha$/Fe] vs. M$_*$. Representative error bars for each quantity are shown in the upper left of each panel. The variation of age and [Z/H] with M$_*$ shows a bimodal behaviour, with a transition at M$_* \sim 10^{10} $M$_\odot$. The dependence of [$\alpha$/Fe] on M$_*$ does not show clear bimodal behaviour.}
\label{fig:ssp_mass}
\end{figure*}

\begin{figure*}
\includegraphics[width=2.25in]{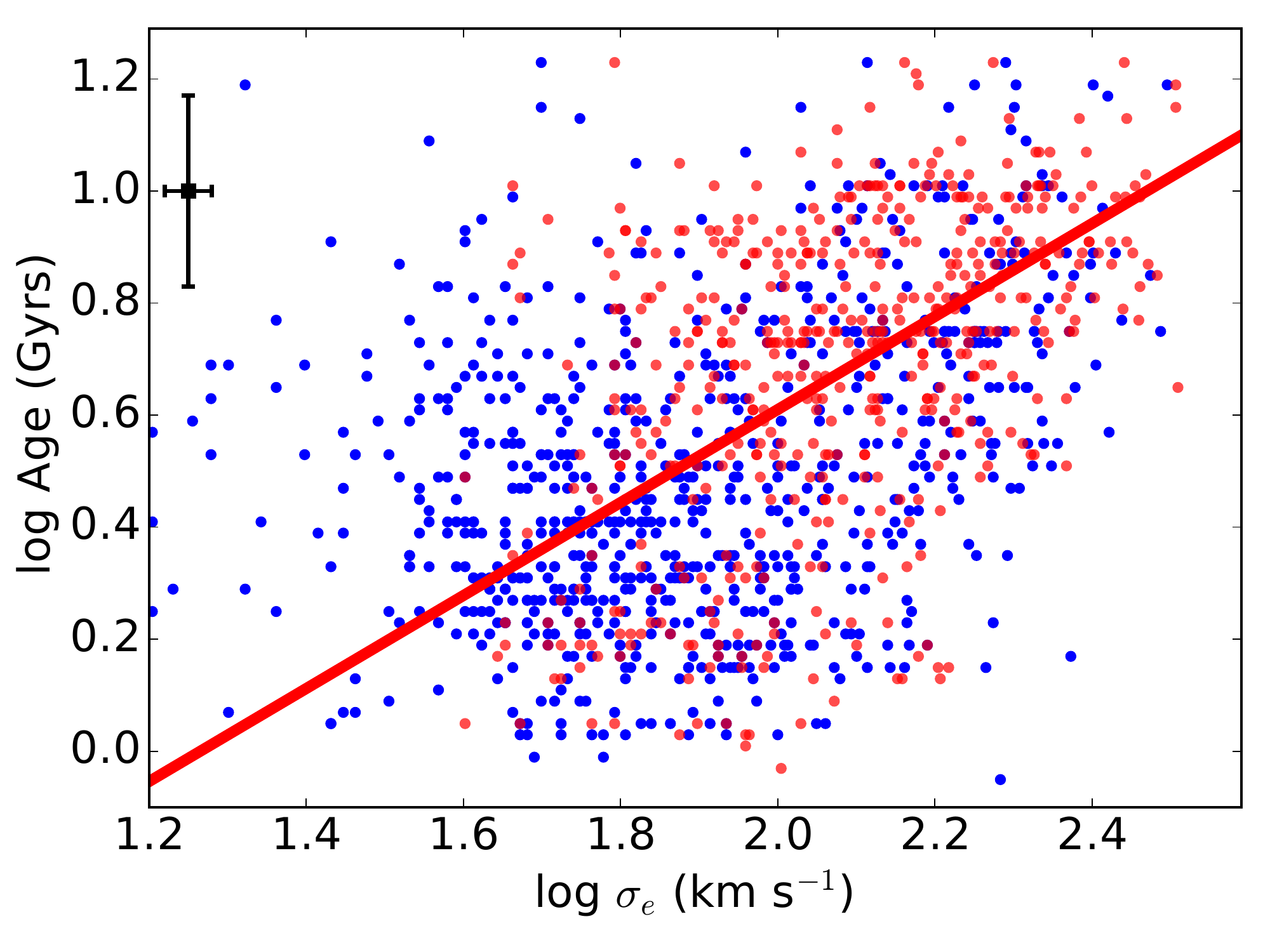}
\includegraphics[width=2.25in]{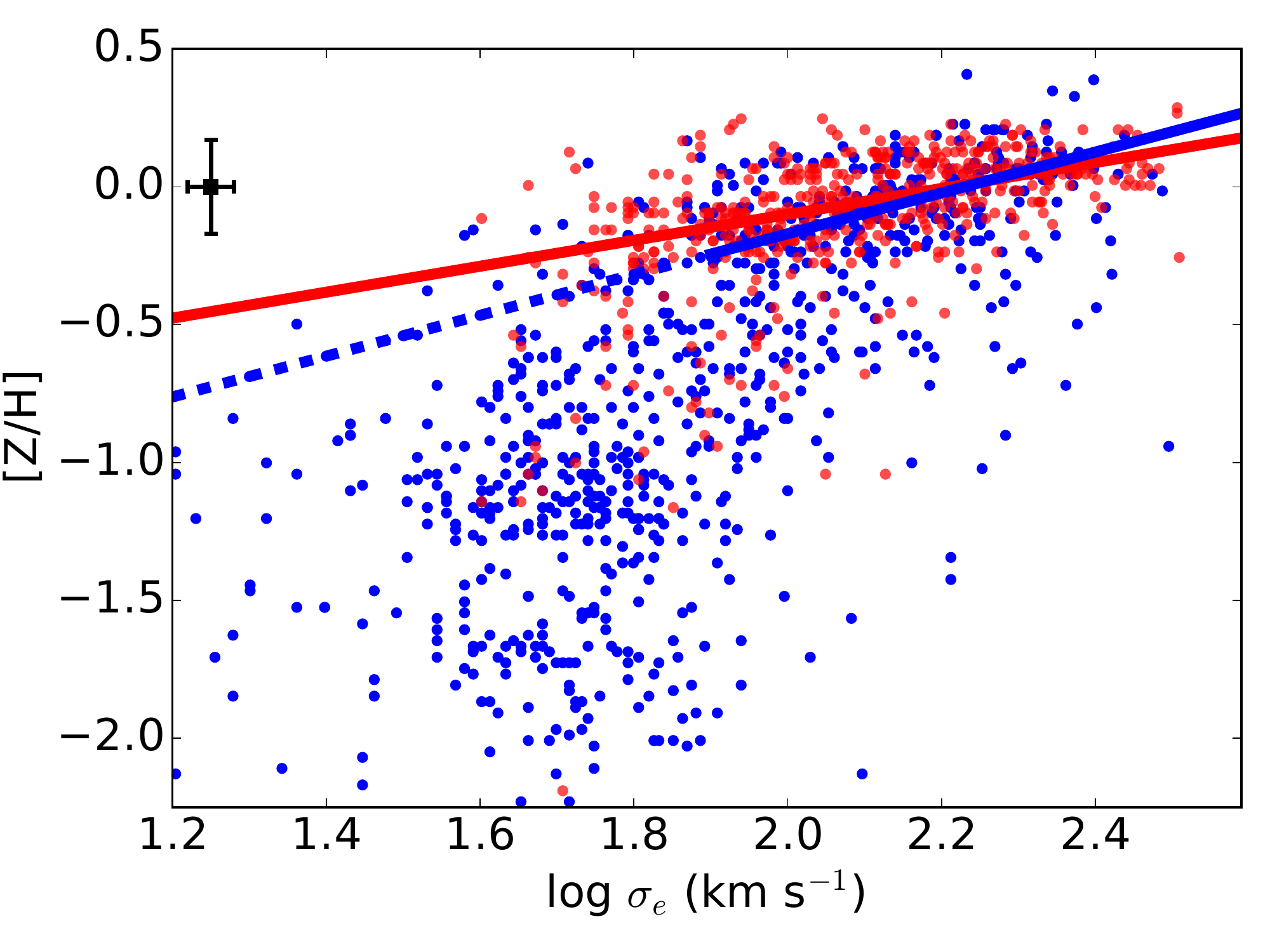}
\includegraphics[width=2.25in]{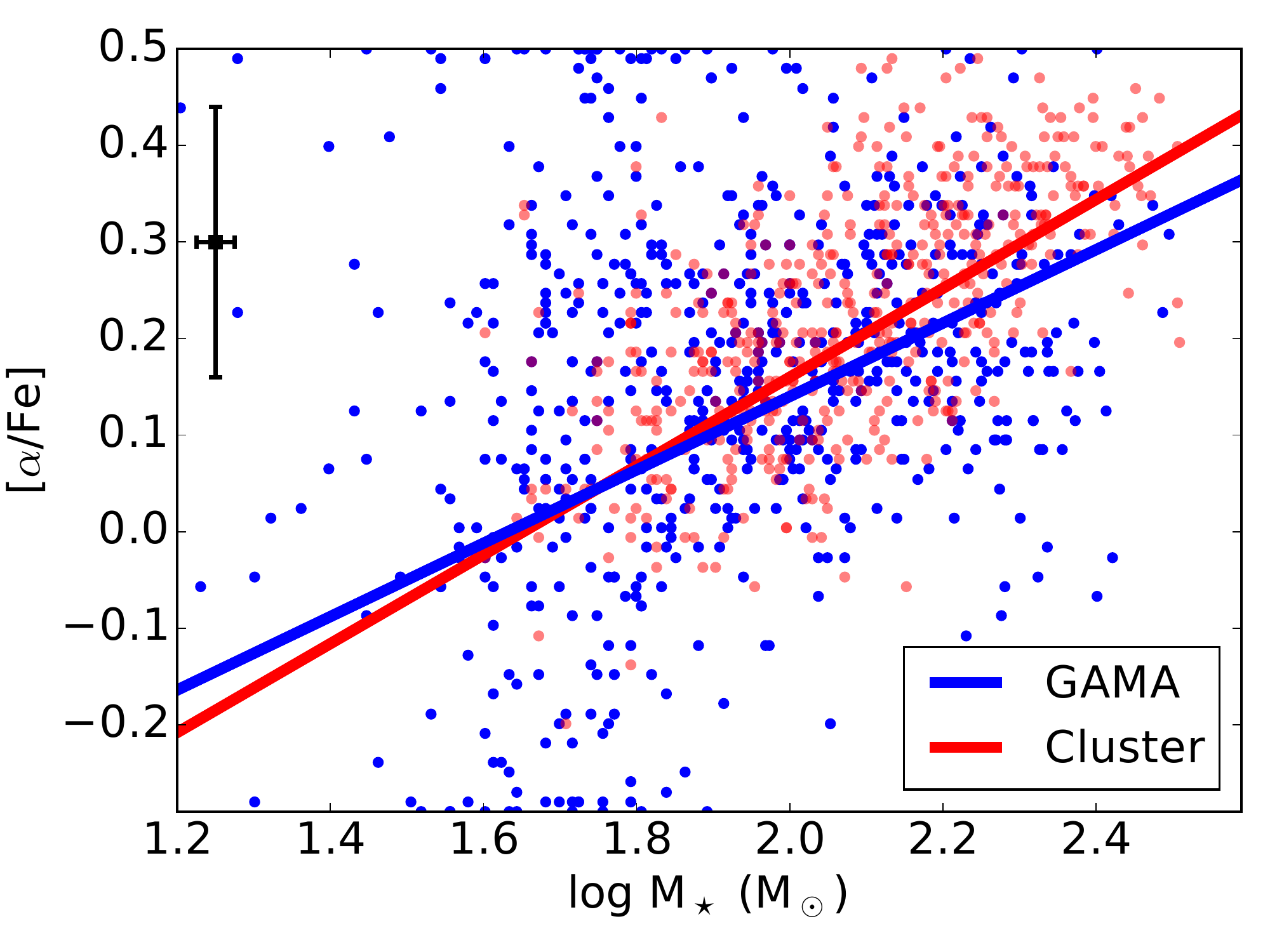}
\caption{As Figure \ref{fig:ssp_mass}, but for stellar effective velocity dispersion, $\sigma_e$. Left panel: age (in Gyrs) vs. $\sigma_e$. Centre panel: [Z/H] vs. $\sigma_e$. Right panel: $[\alpha$/Fe] vs. $\sigma_e$. Representative error bars for each quantity are shown in the upper left of each panel. The [Z/H] -- $\sigma_e$ relation shows the same bimodal behaviour as the [Z/H] -- M$_*$ relation, but there is no evidence of bimodality in the age -- $\sigma_e$ relation.}
\label{fig:ssp_disp}
\end{figure*}

In Figure \ref{fig:tmj_sch_comp} we show a comparison of the age (left panel), metallicity (centre panel) and alpha-enhancement (right panel) derived using the stellar population models for the R$_e$ aperture measurements for the sample. The two models agree for [Z/H], except at low-metallicity where the restricted range of the S07 model saturates for the lowest metallicity galaxies. For [$\alpha$/Fe], there are a number of outliers, though the scatter reduces significantly when we restrict our sample to galaxies with S/N $\geq 20$. There is also a small offset between the models, in the sense that the TMJ model predicts higher [$\alpha$/Fe] than the S07 model, with the magnitude of the offset increasing with [$\alpha$/Fe]. This increase in [$\alpha$/Fe] with respect to previous SSP models was noted by TMJ. The S07 model also shows signs of saturation, with a number of galaxies having [$\alpha$/Fe] = -0.2 or 0.5, the extent of the S07 model range. Saturation in [$\alpha$/Fe] appears minimal in the TMJ model, with few galaxies found at the edge of the model range.

\begin{table}
\caption{SSP-M$_*$ and SSP-$\sigma_e$ linear fit parameters}
\label{tab:ssp_relations}
Fits are of the form: $y = a + b\times x$
\begin{center}
\begin{tabular}{l l c c c c c}
\hline
Parameter & Sample & $a$ & err($a$)& $b$ & err($b$) & $\Delta_{int}$ \\
\hline
\multicolumn{7}{c}{versus log M$_*$} \\
\hline
{[Z/H]}$^\dagger$ & GAMA & -3.02 & 0.01 & 0.28 & 0.02 & 0.01\\
{[Z/H]} & Cluster & -1.84 & 0.01 & 0.17 & 0.01 & 0.02 \\
{[$\alpha$/Fe]} & GAMA & -1.23 & 0.01 & 0.13 & 0.01 & -- \\
{[$\alpha$/Fe]} & Cluster & -1.32 & 0.01 & 0.15 & 0.01 & -- \\
\hline
\multicolumn{7}{c}{versus log $\sigma_e$} \\
\hline
log age & Cluster & -1.05 & 0.01 & 0.83 & 0.05 & 0.18 \\
{[Z/H]}$^\dagger$ & GAMA & -1.65 & 0.01 & 0.74 & 0.06 & 0.15 \\
{[Z/H]} & Cluster & -1.04 & 0.01 & 0.47 & 0.03 & 0.02 \\
{[$\alpha$/Fe]} & GAMA & -0.62 & 0.01 & 0.38 & 0.03 & -- \\
{[$\alpha$/Fe]} & Cluster & -0.76 & 0.01 & 0.46 & 0.03 & -- \\
\hline
\end{tabular}
\end{center}
$\dagger$ The fits to [Z/H] for the GAMA sample were restricted to galaxies with M$_* > 10^{9.5}$ M$_\odot$ or log $\sigma_e > 1.9$.
\end{table}

The comparison with age shows the poorest agreement between the models, with the TMJ model predicting older ages for many galaxies, particularly for galaxies with the youngest ages from the S07 model. That age shows the largest scatter is unsurprising; the luminosity weighting of the SSP models means the mean age can be strongly affected by residual bursts of star formation \citep[e.g.][]{Kaviraj:2007}, whereas the SSP-equivalent metallicity and abundance are closer to their mass-weighted values \citep{Serra:2007}. Both models show signs of saturation at the oldest ages, though this effect may be accounted for by uncertainties biasing galaxies to old ages at the edge of the model grid \citep[see Appendix B of][]{McDermid:2015}. The disagreement in age between the two models is primarily driven by a change in behaviour of the Balmer indices with age at low metallicities.  At low [Z/H], the TMJ model predicts older ages at a given H$\beta$ than the S07 model. A significant number of galaxies with low [Z/H] have Balmer absorption line measurements that lie outside the TMJ model grid \citep[as previously found by][]{Kuntschner:2010}, indicating that the TMJ age measurements are unreliable in this low metallicity regime. 

In summary, the S07 model provides more reliable age estimates in the low-[Z/H] regime, whereas the TMJ model extends to lower [Z/H], and does not saturate in the low-[Z/H] regime. In other respects the models agree well. We therefore adopt the S07 model to derive SSP-equivalent ages. We adopt the TMJ model to derive SSP-equivalent [Z/H] and [$\alpha$/Fe]. The results of this work do not depend on this choice of different SSP models for the different parameters, outside the different behaviour in the low-[Z/H] regime discussed in this section. Neither model perfectly describes the full set of Lick indices for the diverse population of galaxies present in our sample, but that the choice made here to use the S07 model ages and TMJ model [Z/H] and [$\alpha$/Fe] represents the best compromise given these limitations. We note that more recent population models \citep[e.g.][]{Vazdekis:2015, Conroy:2014} may eventually reconcile these issues, however at present they are more limited in their parameter coverage compared to the well-established models used here and cannot yet be easily applied to large and diverse samples of galaxies.

\begin{figure}
\includegraphics[width=3.15in]{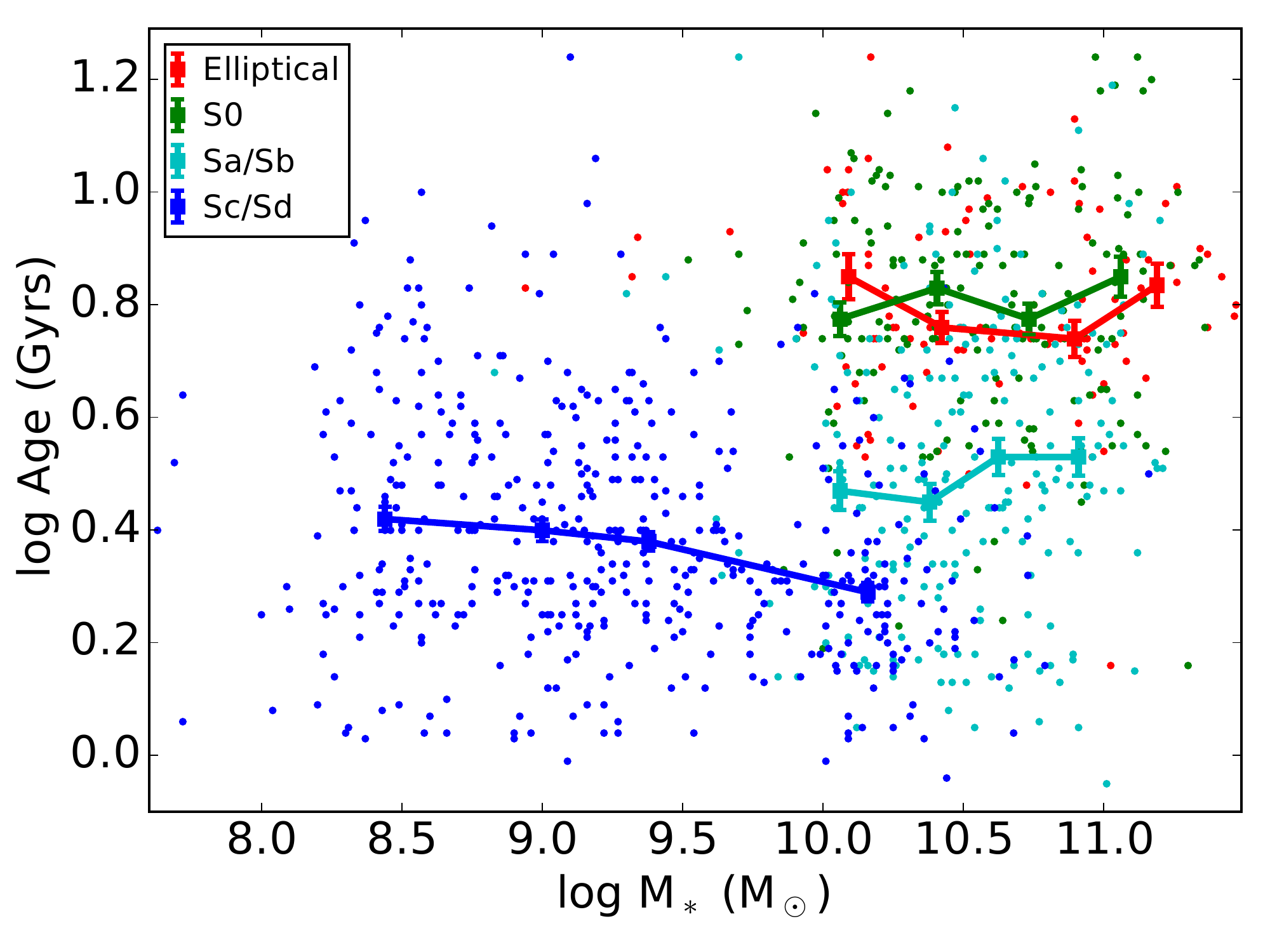}
\includegraphics[width=3.15in]{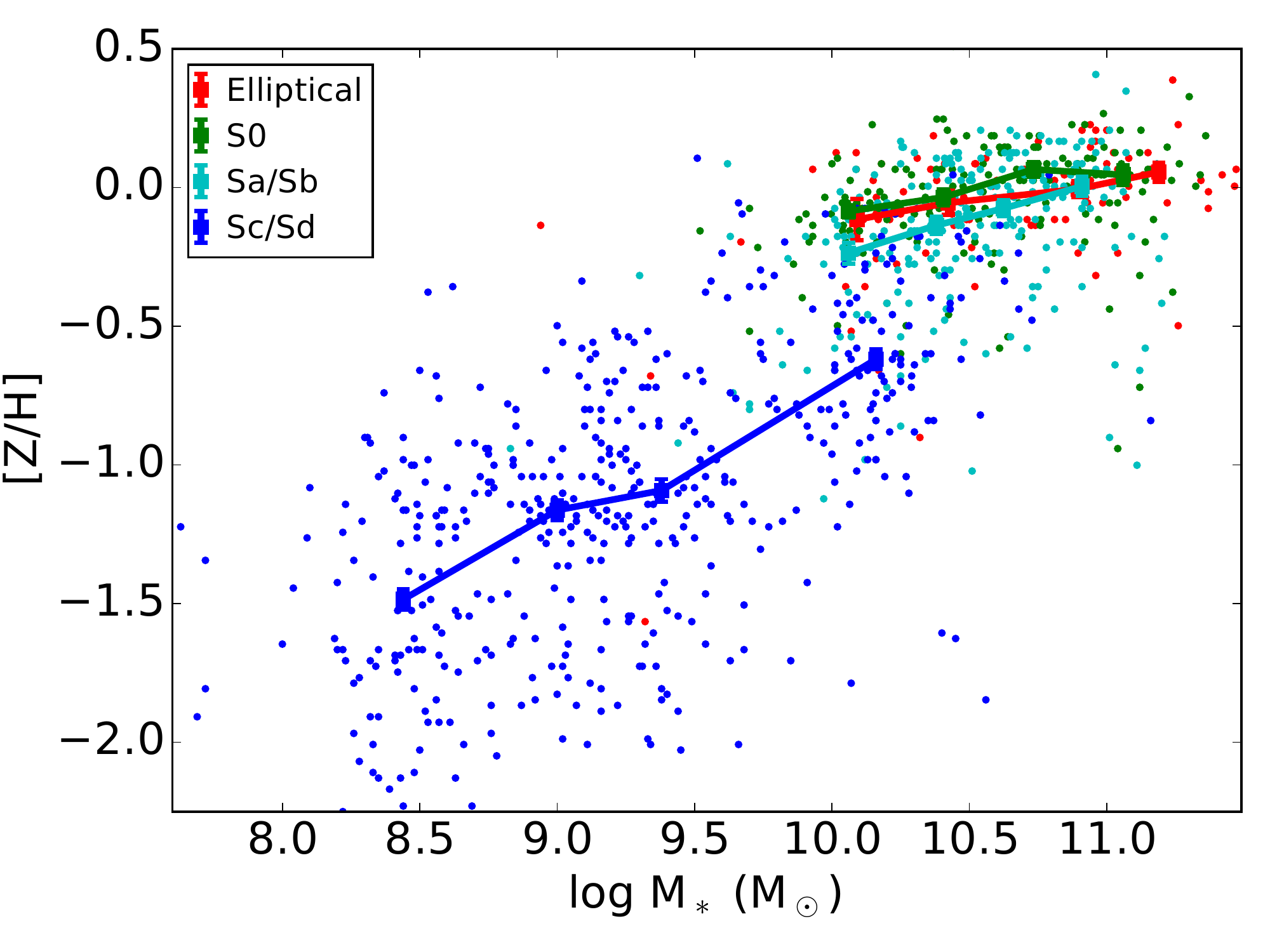}
\includegraphics[width=3.15in]{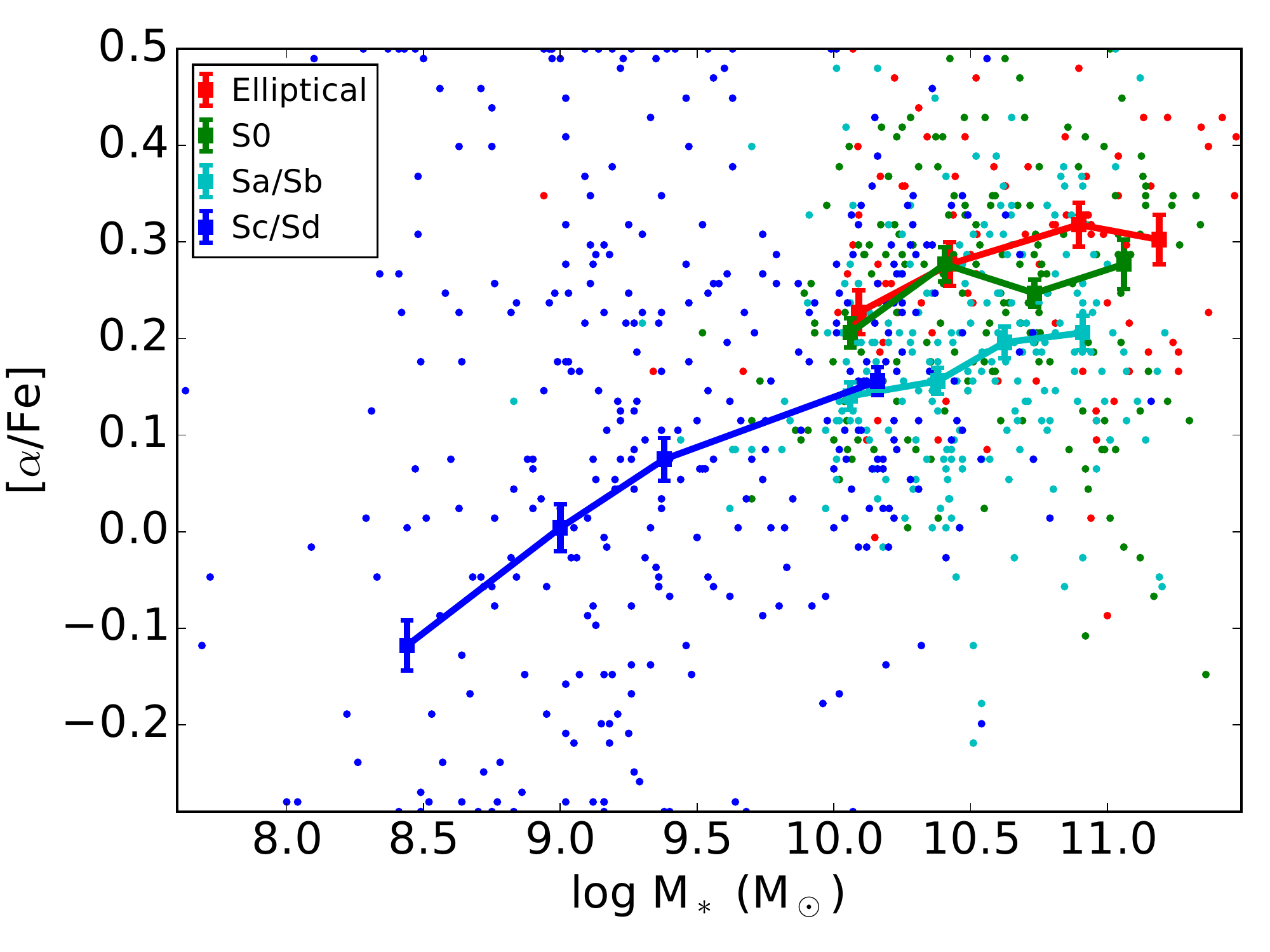}
\caption{Residual dependence of the SSP parameters on morphology. In each panel we show the SSP-M$_*$ relation, with average trends for elliptical (red), S0 (green), Sa/Sb (cyan) and Sc/Sd/Irr (blue) galaxies indicated by the coloured lines. The squares indicate the mean value for all galaxies in each bin. The uncertainty on the mean is small due to the large number of galaxies within each bin, and typically is on the order of the symbol size unless explicitly indicated by the error bars. These average trends were determined as described in the text. Unsurprisingly age shows the strongest variation with morphology. [Z/H] shows little morphology dependence, but there is a significant separation between early-type galaxies and spirals with respect to [$\alpha$/Fe] at fixed mass.}
\label{fig:ssp_mass_morph}
\end{figure}

Finally, we caution that the quantities measured here are not the {\it true} age, metallicity and alpha-enhancement of a galaxy, but the luminosity-weighted, single stellar population-equivalent value. No attempt is made to account for differing star formation histories or chemical abundance patterns, beyond variations in [$\alpha$/Fe]. In particular, luminosity-weighted SSP-equivalent ages are likely to be biased to young ages (relative to mass-weighted values) for all galaxies that have experienced recent star formation. However, the relative (rather than absolute) differences in SSP parameters between galaxies are robust and reflect real differences in stellar populations. The results presented in this work should be viewed in this light.

\section{Global stellar populations}
\label{sec:results1}

We begin by examining the dependence of age, [Z/H] and [$\alpha$/Fe] on galaxy stellar mass, M$_*$, and stellar effective velocity dispersion, $\sigma_e$. In Figures \ref{fig:ssp_mass} and \ref{fig:ssp_disp} we show the SSP--M$_*$ and SSP--$\sigma_e$ relations for GAMA (blue points) and cluster (red points) galaxies. The GAMA and cluster samples show similar behaviour so we consider them together here, with the exception that we fit to the two samples separately. All three SSP parameters show significant variation with M$_*$ and similar behaviour with $\sigma_e$. The overall picture is that massive, high $\sigma_e$ galaxies are older, metal-rich and more alpha-enhanced than less massive, lower $\sigma_e$ systems, though the details differ for each SSP parameter.

Age shows weak bimodal behaviour with mass, decreasing rapidly from high masses to M$_* \sim 10^{10}$ M$_\odot$, then flattening at low masses. There is significant scatter in age at all masses. The age -- $\sigma_e$ relation shows no strong bimodality, with a single log-linear relation describing the trend in age with $\sigma_e$. Metallicity shows consistent behaviour with both M$_*$ and $\sigma_e$: increasing [Z/H] with increasing M$_*$ and $\sigma_e$, with a tight relation above M$_* \sim 10^{10}$ M$_\odot$ ($\sigma_e \sim 100$ km s$^{-1}$) and substantial scatter below that. Abundance shows the simplest behaviour, with single log-linear relations providing a good description of the increase of [$\alpha$/Fe] with M$_*$ and $\sigma_e$.
 
The  [$\alpha$/Fe] -- $\sigma_e$, [$\alpha$/Fe] -- M$_*$, [Z/H] -- M$_*$ and [Z/H] -- $\sigma_e$ relations for massive, high dispersion galaxies are all well approximated by single log-linear relations. We do not attempt to fit the age -- M$_*$ and age -- $\sigma_e$ relations due to the large intrinsic scatter and complex behaviour (with the exception of the age -- $\sigma_e$ relation for cluster galaxies). We fit each of these regimes using the $\it lts\_linefit$ routine\footnote{Available from http://purl.org/cappellari/software}, described in \citet{Cappellari:2013a}, which finds the best-fitting linear relation by minimising the orthogonal residuals to the fit while accounting for measurement errors, intrinsic scatter and outliers. The best-fitting linear relations are shown as the solid lines in the respective panels, and the coefficients of these linear fits are given in Table \ref{tab:ssp_relations}. 

\subsection{Residual dependence of stellar populations on morphology and environment}
\label{sec:ssp_resids_mass}
 
There remains significant intrinsic scatter in the above relations, suggesting other parameters may affect these relations. The large sample and ancillary data provided by the SGS enable us to sub-divide the population into several different groupings while maintaining a sufficient number of objects per bin to determine robust statistical quantities. Here we examine the dependence of the SSP parameters on morphology, and the environmental parameters, local number density of galaxies, $\Sigma_5$, and the group or cluster halo mass, M$_\mathrm{halo}$.

 \begin{figure}
\includegraphics[width=3.1in]{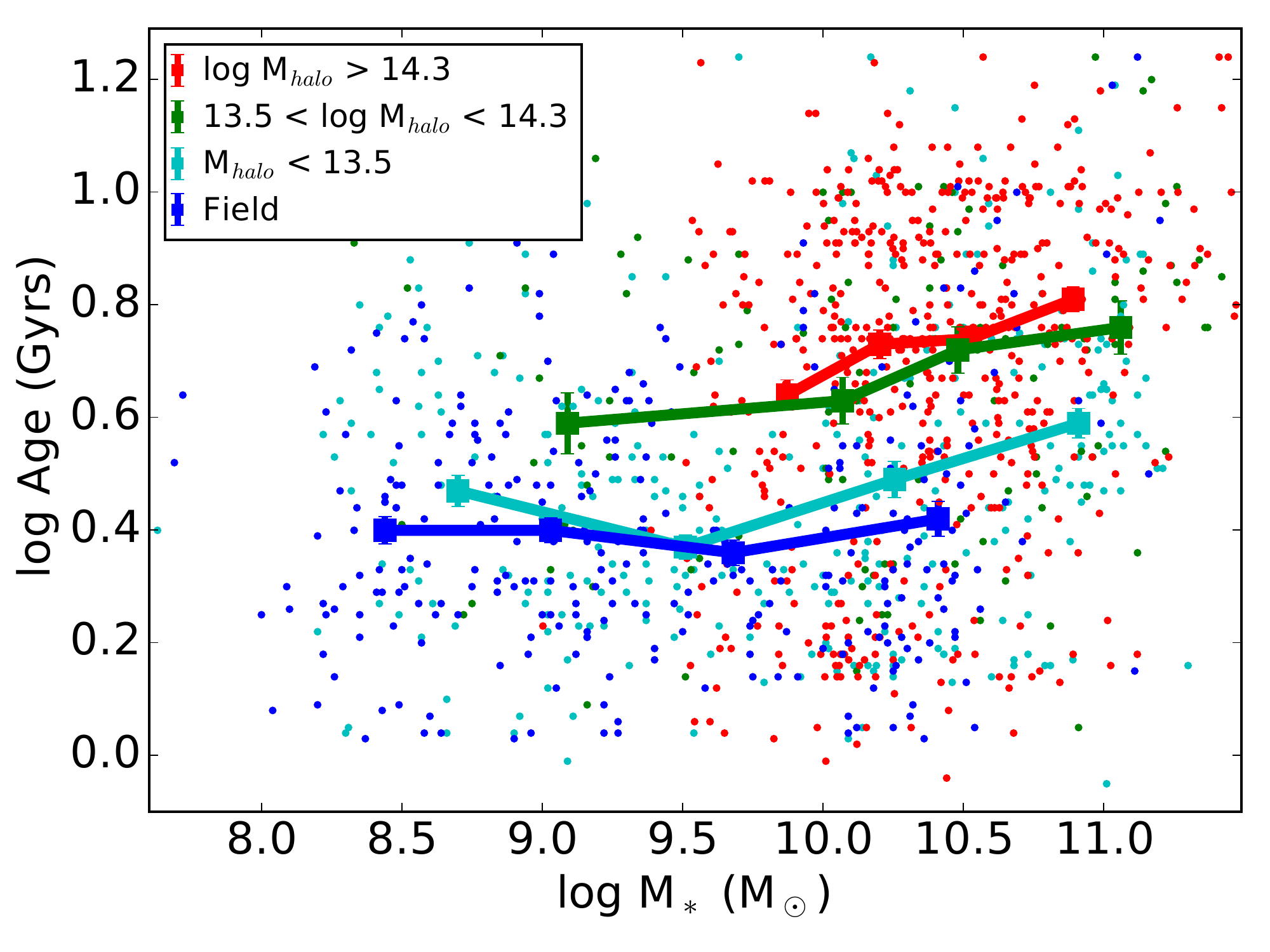}
\includegraphics[width=3.1in]{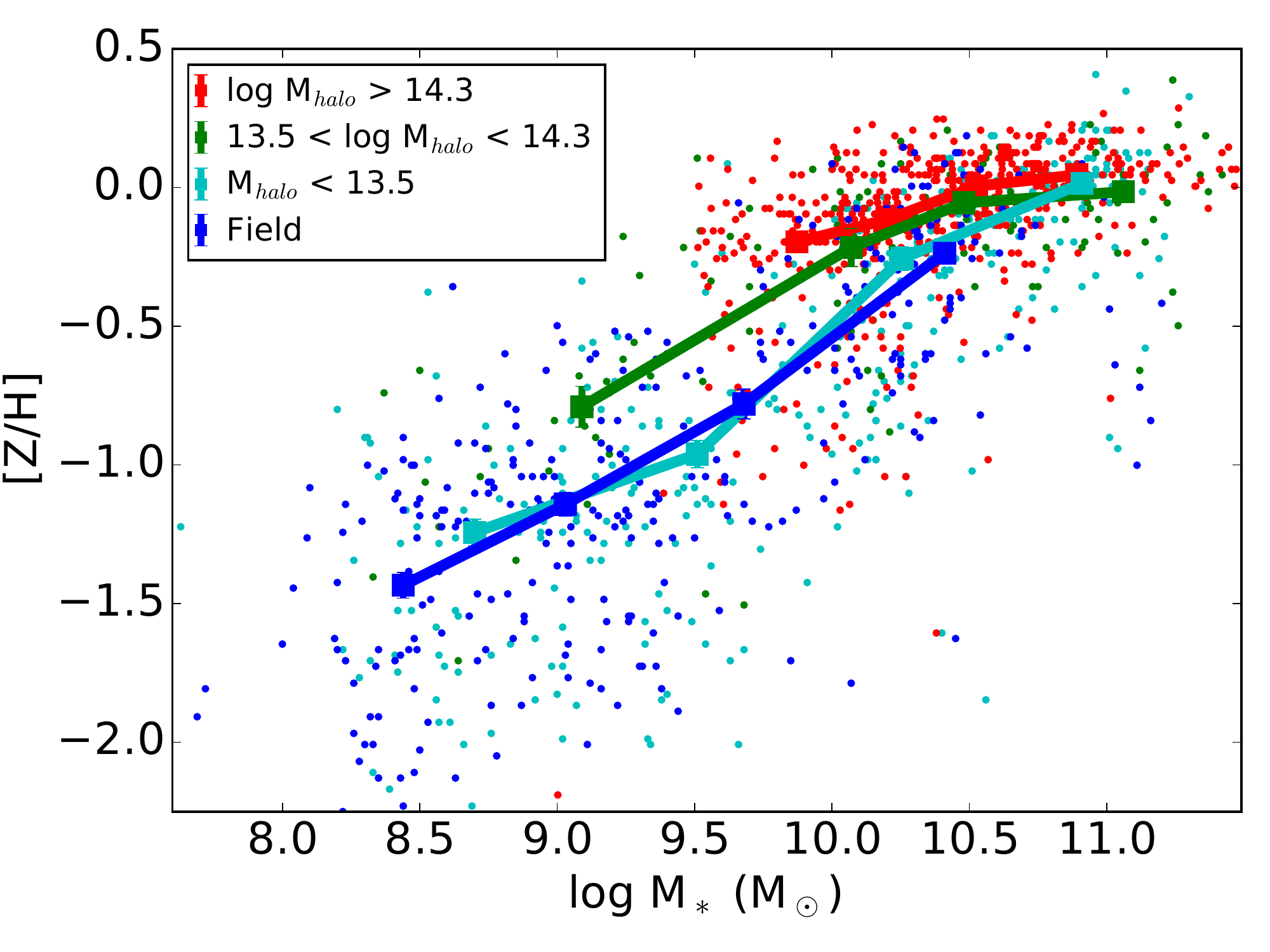}
\includegraphics[width=3.1in]{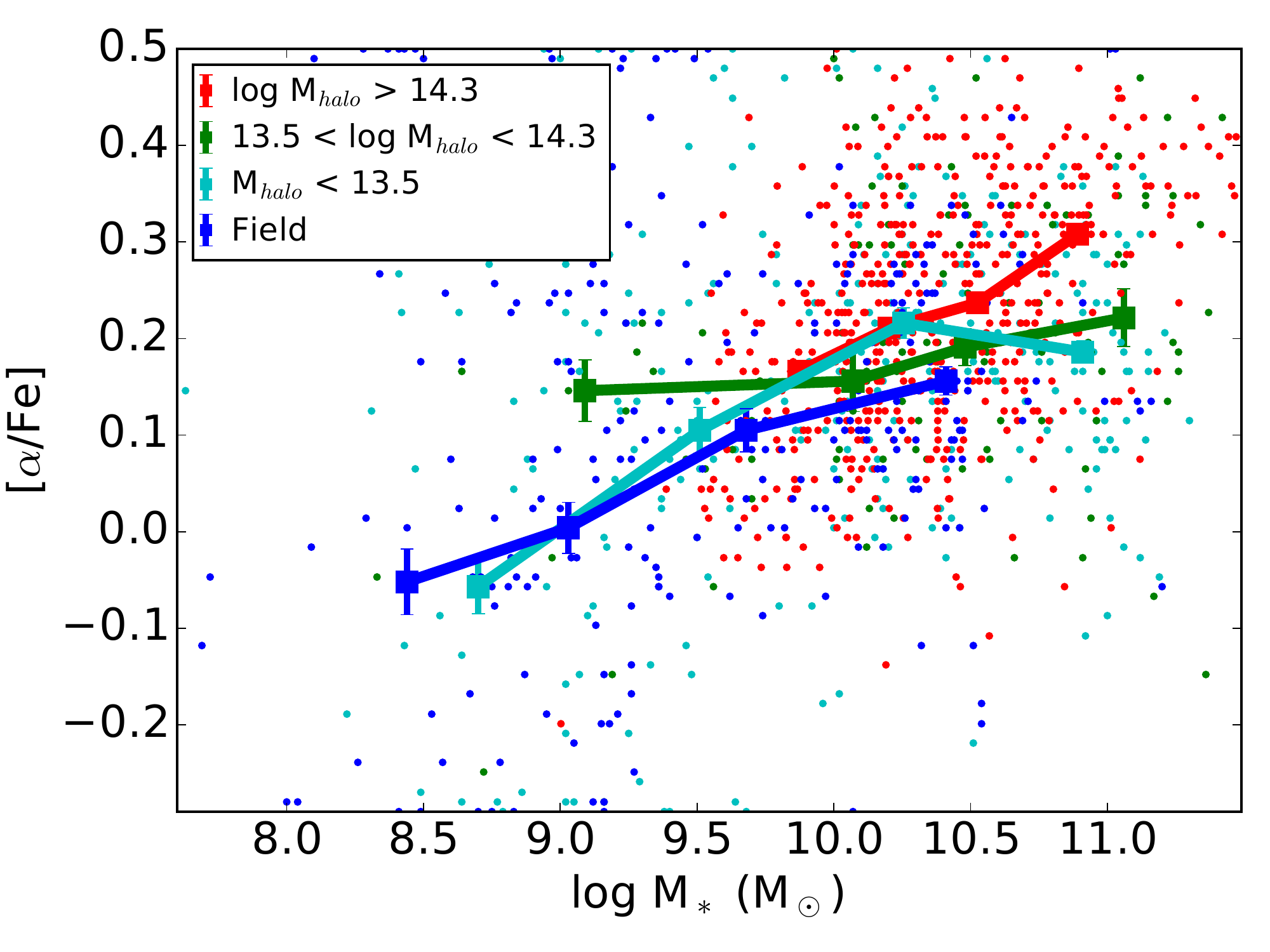}
\caption{Residual dependence of the SSP parameters on halo mass, M$_\mathrm{halo}$. We divide the sample into three bins of high (red), intermediate (green) and low (cyan) halo mass. We also include a fourth bin of isolated field galaxies, where M$_\mathrm{halo}$ could not be determined from group velocity dispersions. Average trends with M$_*$ were derived as in Figure \ref{fig:ssp_mass_morph}. Galaxies in more massive halos are, on average, older, more metal rich and alpha-enhanced compared to those in lower mass halos.}
\label{fig:ssp_mass_mhalo}
\end{figure}

\begin{figure}
\includegraphics[width=3.1in]{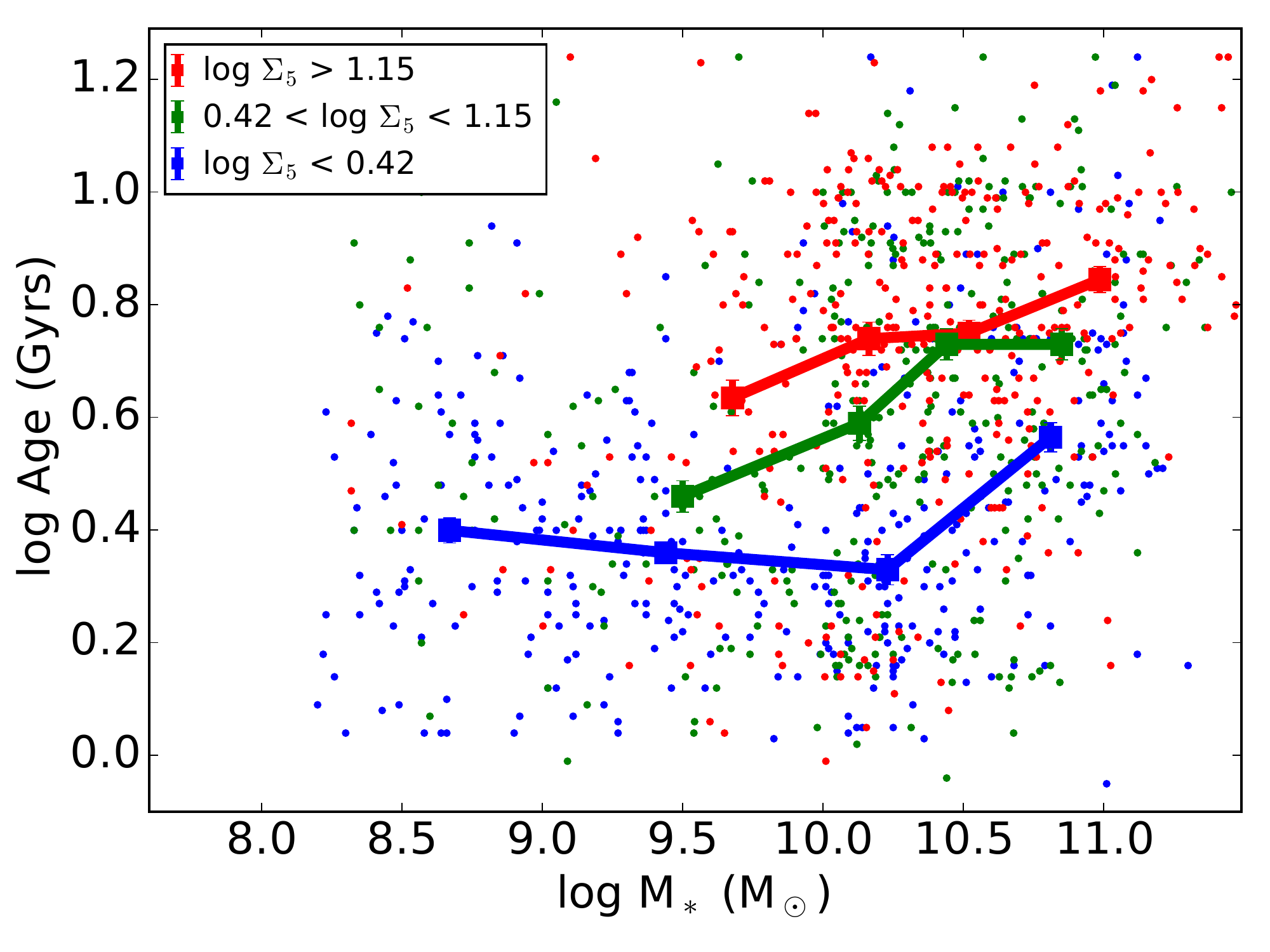}
\includegraphics[width=3.1in]{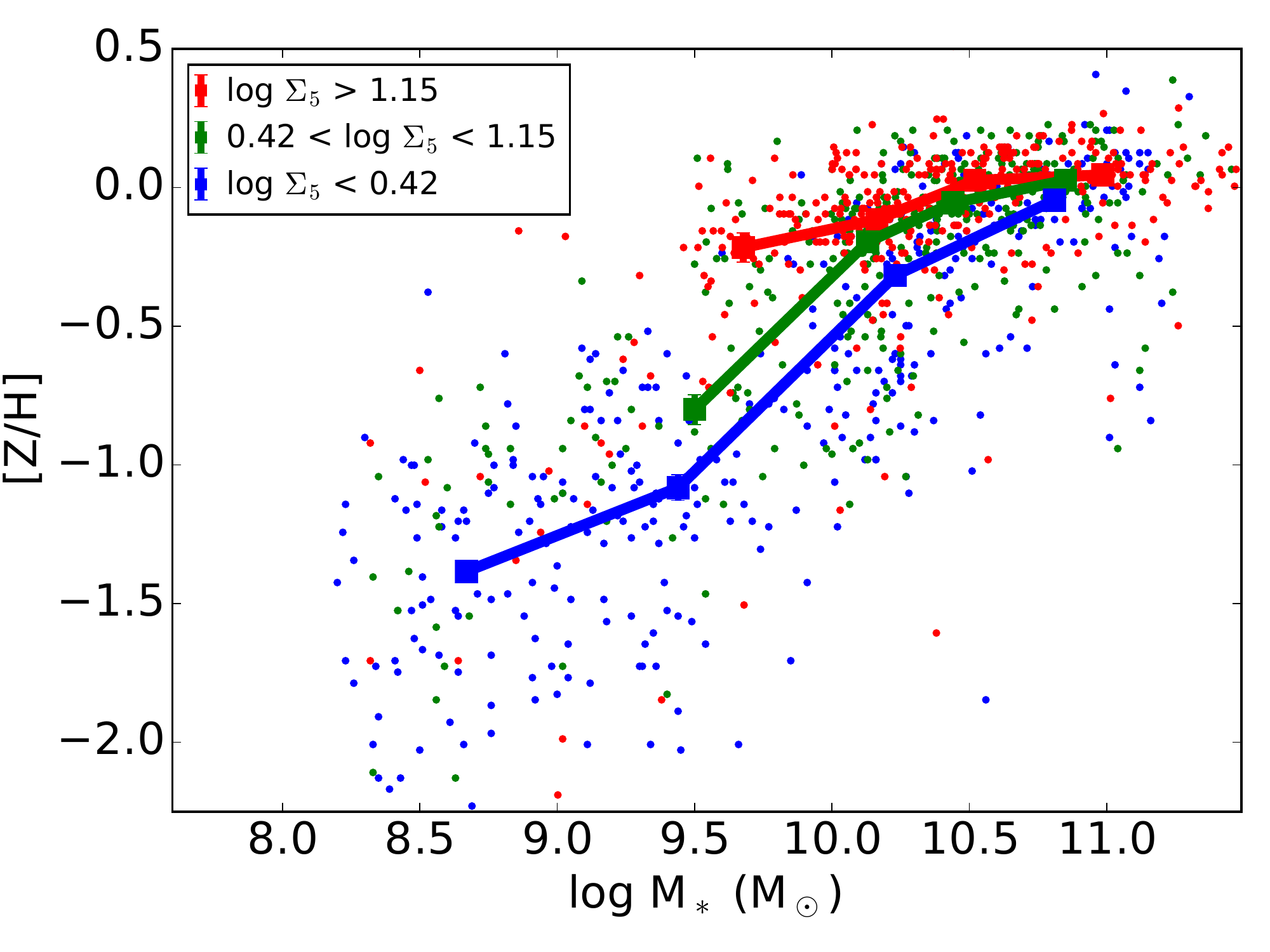}
\includegraphics[width=3.1in]{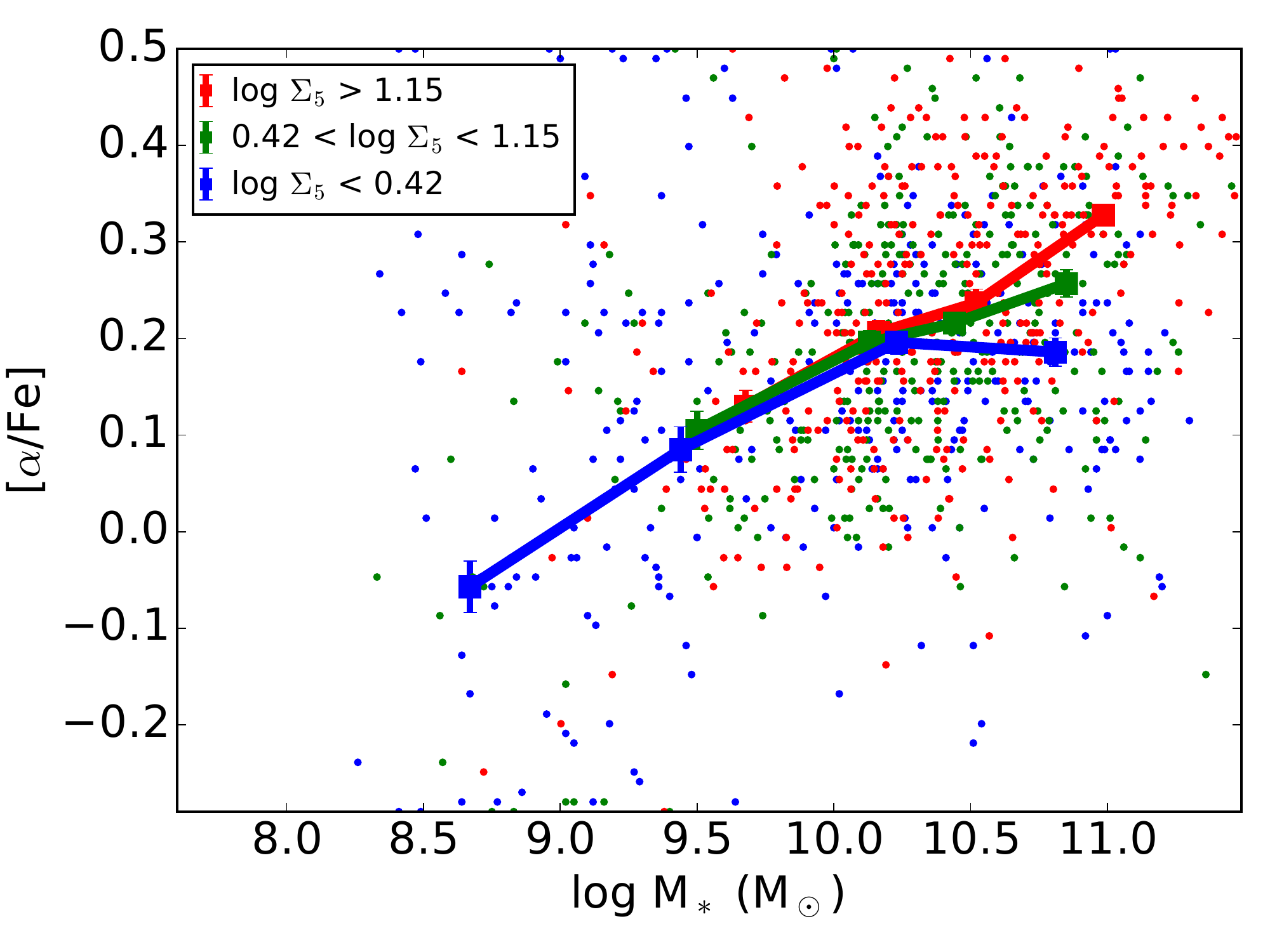}
\caption{Residual dependence of the SSP parameters on the local surface density of galaxies, $\Sigma_5$. We divide the sample into three bins of high (red), intermediate (green) and low (blue) $\Sigma_5$, then determine average trends with M$_*$ as in Figure \ref{fig:ssp_mass_morph}.}
\label{fig:ssp_mass_env}
\end{figure}

In Figure \ref{fig:ssp_mass_morph} we show the dependence of the SSP relations on morphology. Here we consider only the GAMA sample to avoid any bias introduced by the overrepresentation of cluster galaxies in the effective volume due to the cluster sample. In each panel we have binned the sample by morphology into elliptical, S0, early-type spiral (i.e. spiral galaxies with prominent bulges) and late-type spiral (i.e. spiral galaxies without a prominent bulge) galaxies. Within each morphology bin, we determine the mean SSP value in four bins of M$_*$. These values are shown by the coloured lines, with the uncertainty ion the mean indicated by the error bars. These uncertainties are small, typically on the order of the symbol size, because of the large number of galaxies within each bin.

Morphology is closely related to stellar populations, in that to have clear spiral arms, and therefore be classified as a spiral, a galaxy is likely to have experienced recent star formation, and therefore have a young light-weighted age. It is unsurprising then that we find substantial differences in SSP age at fixed M$_*$ or $\sigma_e$ with morphology: elliptical and S0 galaxies are predominantly old \citep[though some contain small mass fractions of young populations e.g.][]{Kaviraj:2007}, late-type spirals are uniformly young and early-type spirals are intermediate in age. We find no significant difference in [Z/H] for elliptical, S0, or early-type spirals; however, at fixed mass, late-type spirals show significantly lower metallicity. At M$_\star \sim 10^{10}$ M$_\odot$, late-type spirals are $0.47 \pm 0.1$ lower in [Z/H] than earlier morphology galaxies of the same mass. [$\alpha$/Fe] shows the most interesting behaviour with morphology, with the late- and early-type spirals forming a continuous sequence in [$\alpha$/Fe] versus M$_*$, and elliptical and S0 galaxies having higher [$\alpha$/Fe] at the same mass. At M$_\star \sim 10^{10}$ M$_\odot$, spiral galaxies are $0.07 \pm 0.03$ lower in [$\alpha$/Fe] than early type galaxies of the same mass. High [$\alpha$/Fe] is associated with rapid, efficient star formation, consistent with the early-type galaxies having experienced shorter, more intense star formation histories.

We carry out the same exercise with environment, but include both the GAMA and cluster galaxies because we are specifically examining the dependence on environmental metrics. We divide the sample into thirds of high ($\log \Sigma_5 > 1.15$), intermediate ($0.42 > \log \Sigma_5 > 1.15$) and low ($\log \Sigma_5 < 0.42$) local number density. For halo mass we divide galaxies into cluster (log M$_\mathrm{halo} > 14.3$), high-mass group ($ 13.5 < $ M$_\mathrm{halo} < 14.3$) and low-mass group samples (M$_\mathrm{halo} < 13.5$) , with the addition of a field subsample where halo masses could not be determined from group velocity dispersion measurements. Massive field galaxies may have halo masses that overlap with those of low-mass groups and so cannot strictly be considered a low halo-mass sample. We note that the two most massive groups from the GAMA regions have halo masses in the same range as the clusters and are therefore included in the cluster bin. The resulting trends are shown in Figures \ref{fig:ssp_mass_mhalo} and \ref{fig:ssp_mass_env}. Low-mass galaxies in high-density environments are missing from the sample due to the SGS sample selection \citep{Bryant:2015}. High-mass galaxies in low-density environments are also missing, but due to their intrinsic rareness rather than the sample bias.

As both $\Sigma_5$ and M$_\mathrm{halo}$ are measures of environment, the SSP parameters are expected to show similar behaviour with both indicators. In general, galaxies living in denser environments or more massive halos are older, metal rich and $\alpha$-enhanced relative to galaxies in lower density environments or less massive halos at fixed mass. The largest separation between environments is seen with age, where, at all masses, the galaxies in denser environments are uniformly older. However, this may simply be a consequence of the well-known morphology--density relation \citep{Dressler:1980}. With the current sample size we have insufficient galaxies to simultaneously separate the effect of both morphology and environment at fixed mass. At high M$_*$, metallicity shows no dependence on environment, but at M$_* <\sim10^{10.5}$ a difference of $\Delta$ [Z/H] $= 0.20 \pm 0.04$ is found between the highest and lowest density environments. At lower masses this difference increases still further to $\Delta$[Z/H] $= 0.87 \pm 0.07$. In contrast, at high M$_*$, we find [$\alpha$/Fe] increases from low density environments to high density environments by $\Delta$[$\alpha$/Fe] $= 0.14 \pm 0.02$. At intermediate masses, M$_* \sim 10^{10}$, there is no significant dependence of [$\alpha$/Fe] on environment. 

\section{Stellar populations on the size-mass plane}
\label{sec:results2}

\begin{figure*}
\includegraphics[width=3.15in]{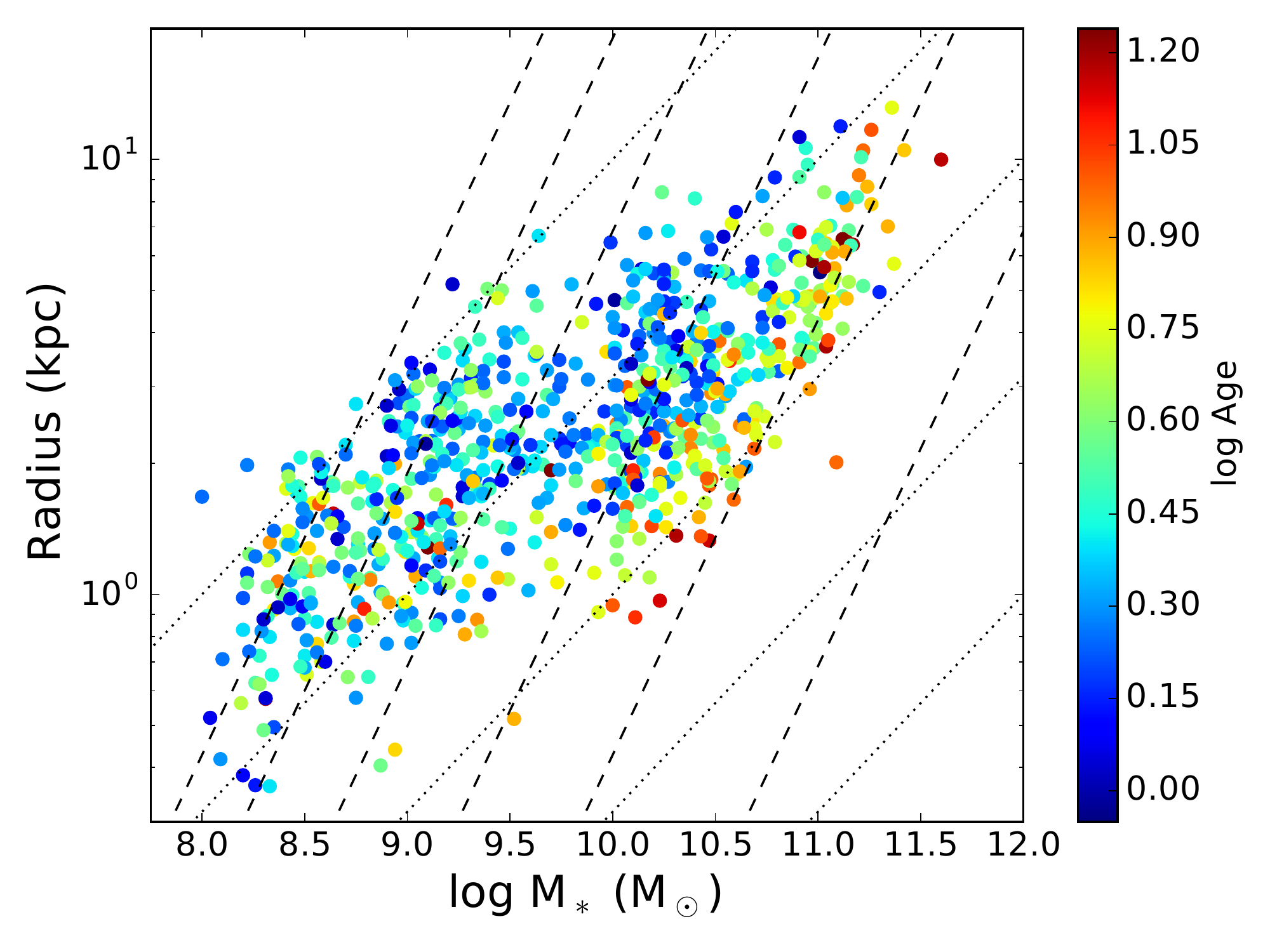}
\includegraphics[width=3.15in]{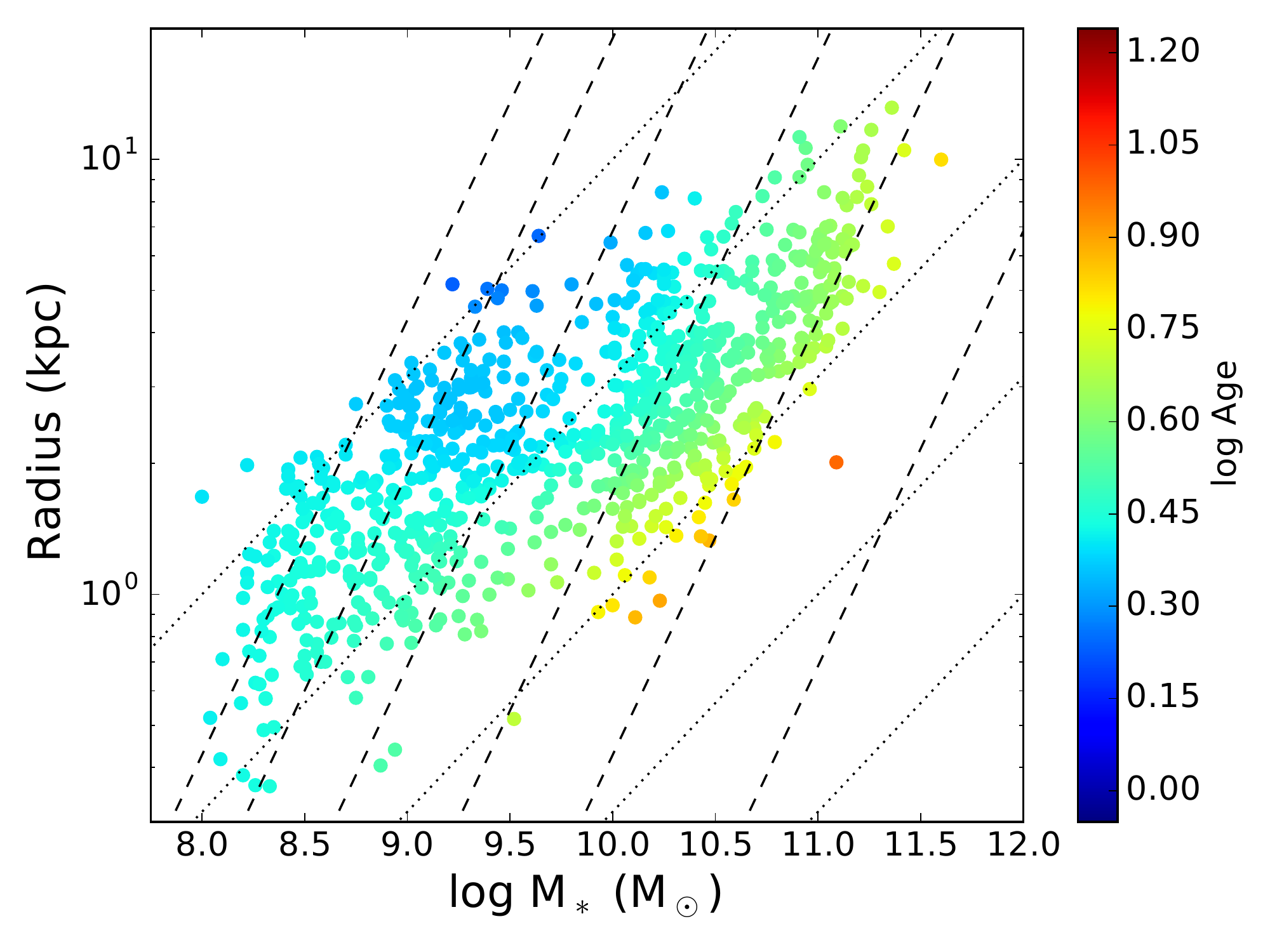}
\includegraphics[width=3.15in]{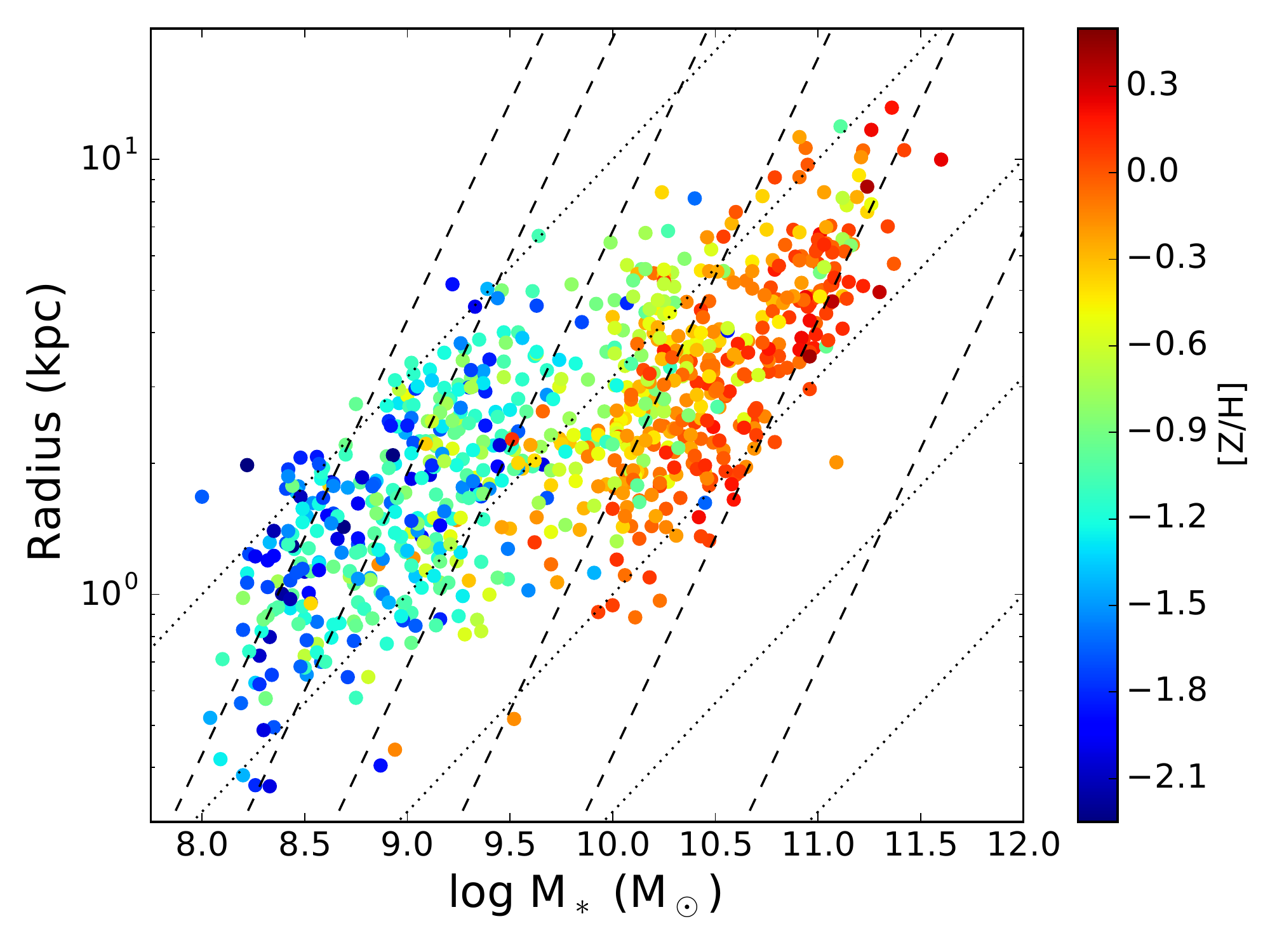}
\includegraphics[width=3.15in]{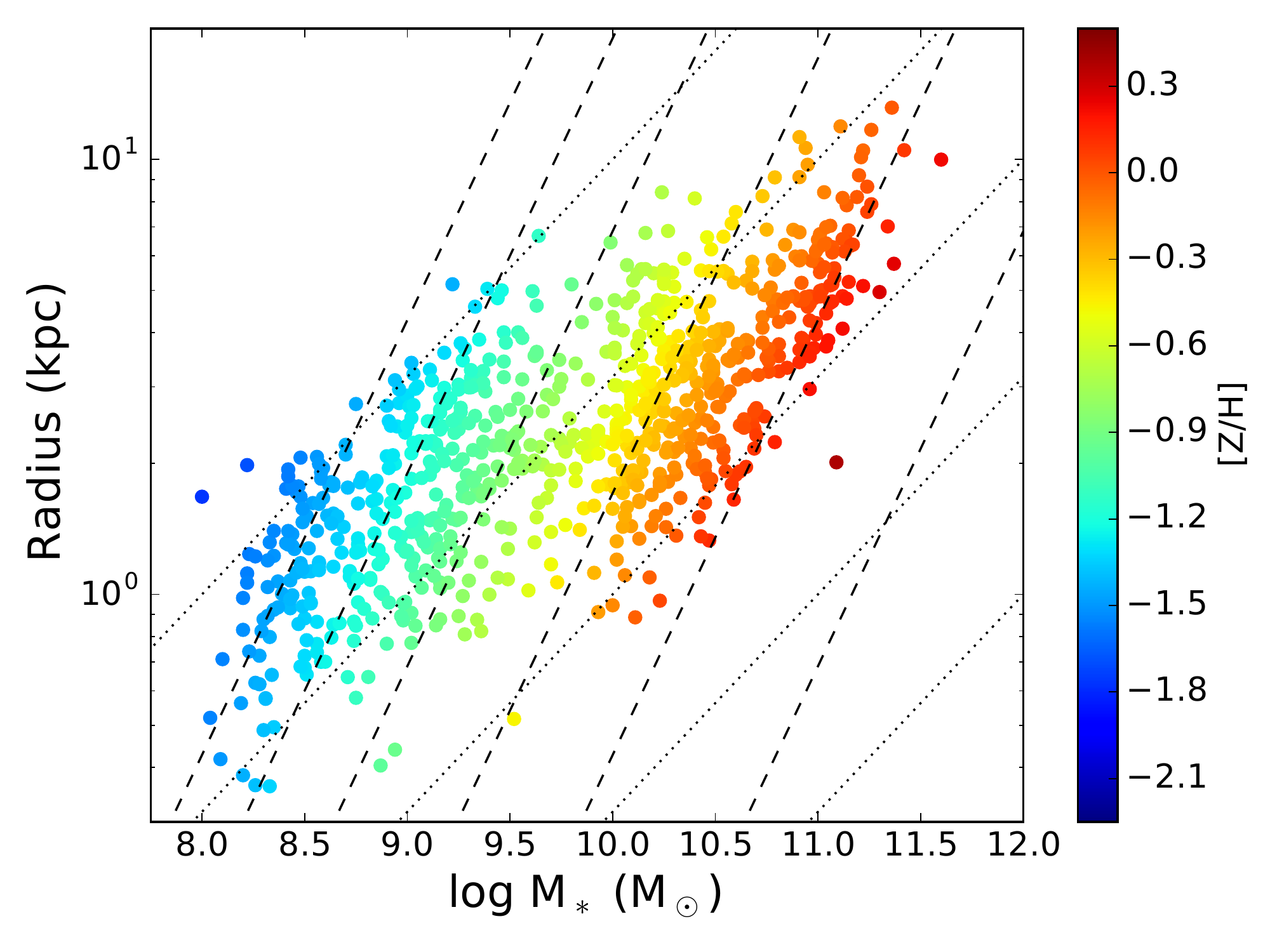}
\includegraphics[width=3.15in]{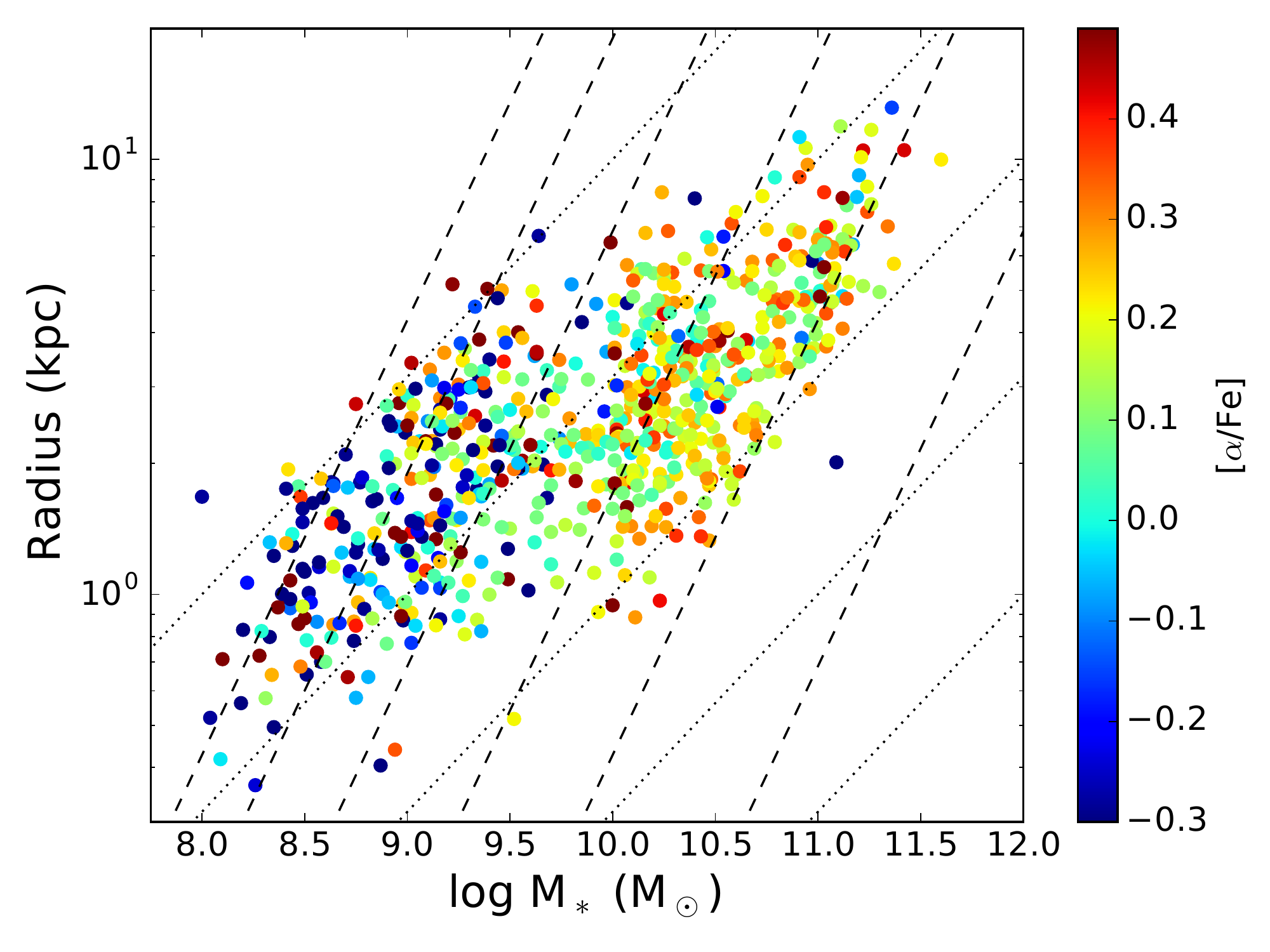}
\includegraphics[width=3.15in]{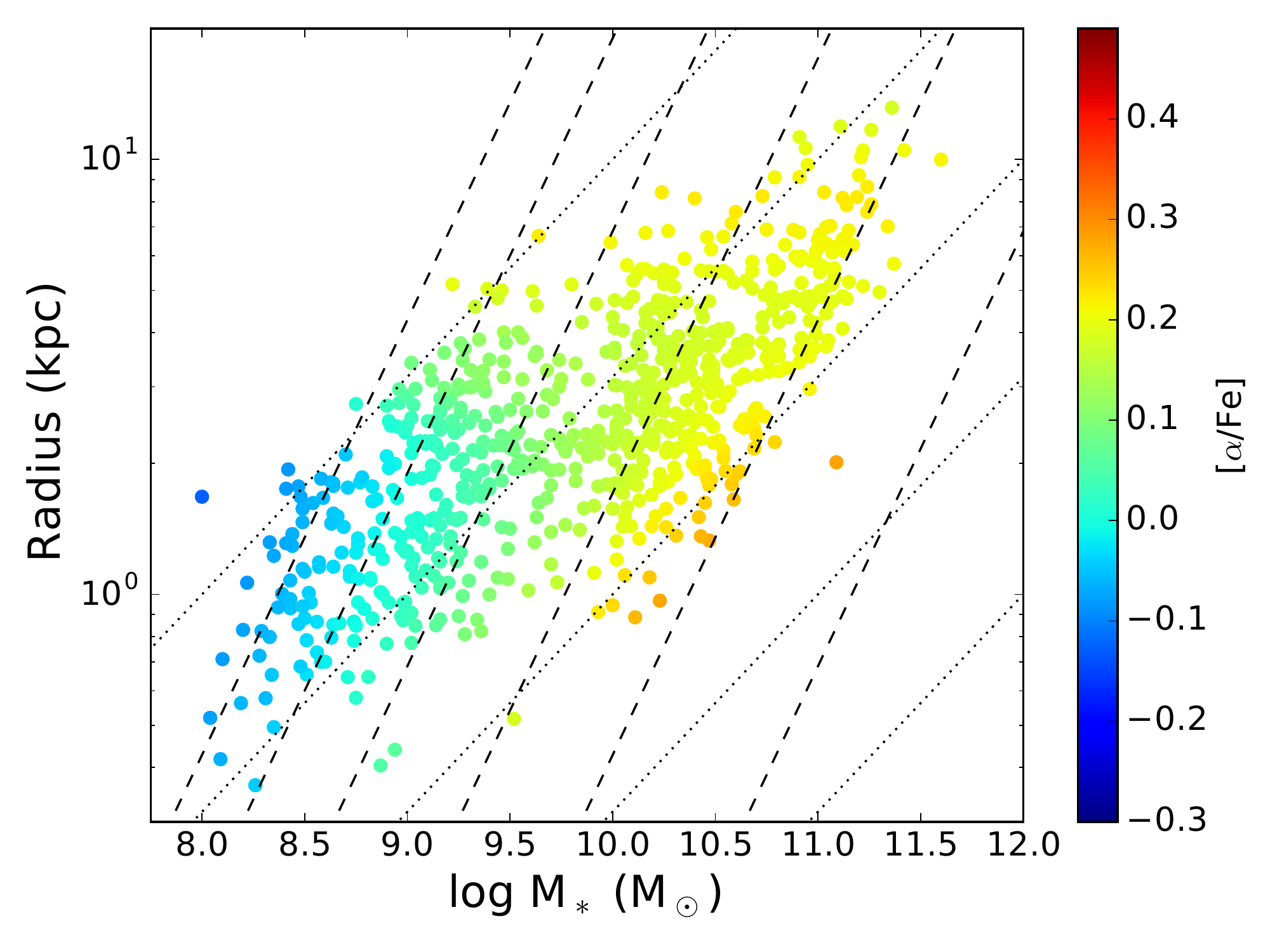}
\caption{Distribution of the SSP parameters in the mass-size plane, for the GAMA sample only. The colour scale in each panel indicates the value of the corresponding SSP parameter, age (top panels), [Z/H] (centre panesl) and [$\alpha$/Fe] (bottom panels). The left column shows the SSP values for individual galaxies, the right column shows the LOESS-smoothed locally averaged value for each data point. Dashed lines indicate lines of constant velocity dispersion (20, 30, 50, 100, 200, 500 km s$^{-1}$ from left to right), assuming all galaxies are Virialised and have a constant dynamical to stellar mass ratio. The dotted lines indicate lines of constant surface mass density, increasing from left/top to right/bottom.}
\label{fig:ssp_mass_size_gama}
\end{figure*}

\begin{figure*}
\includegraphics[width=3.15in]{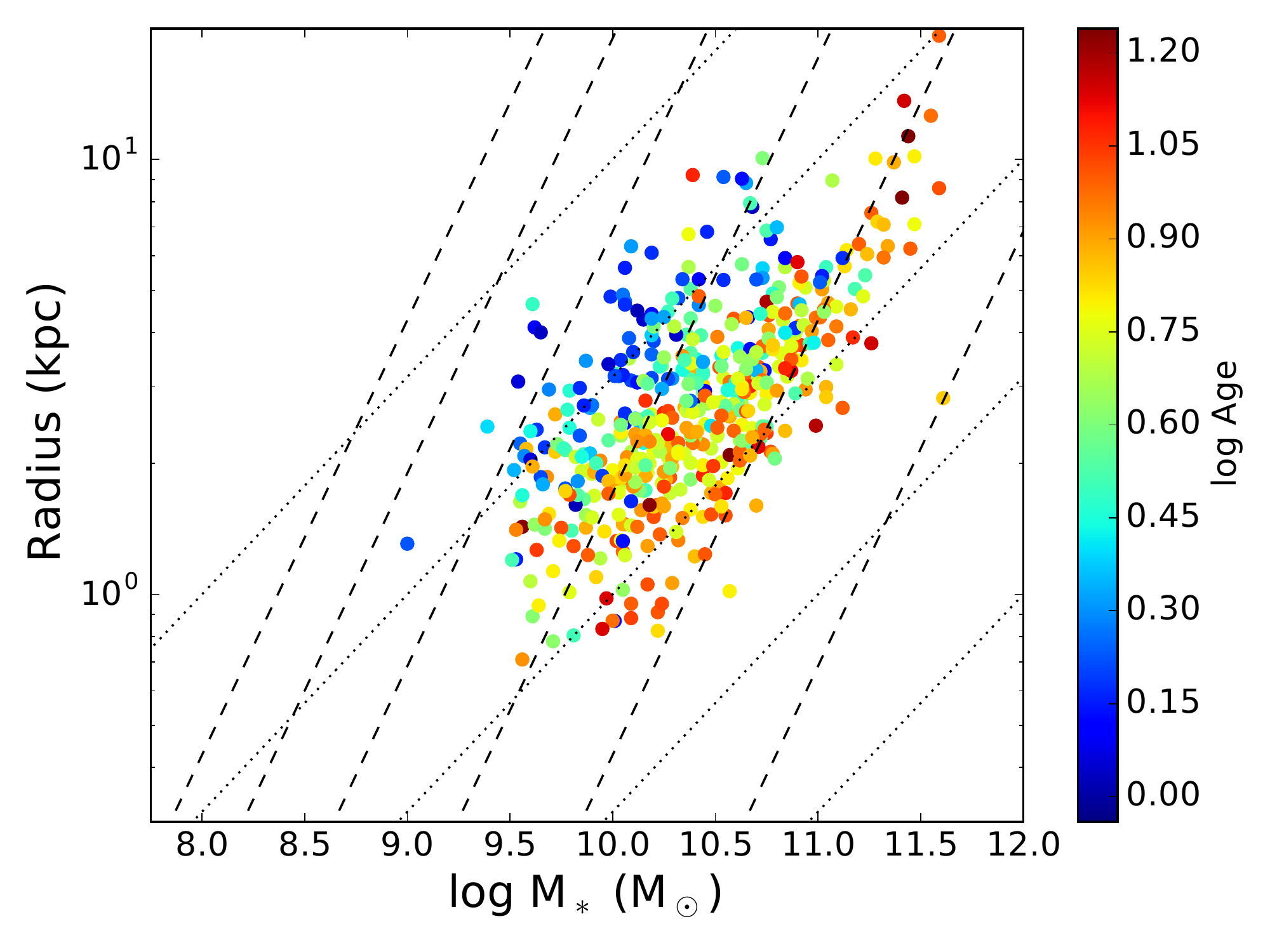}
\includegraphics[width=3.15in]{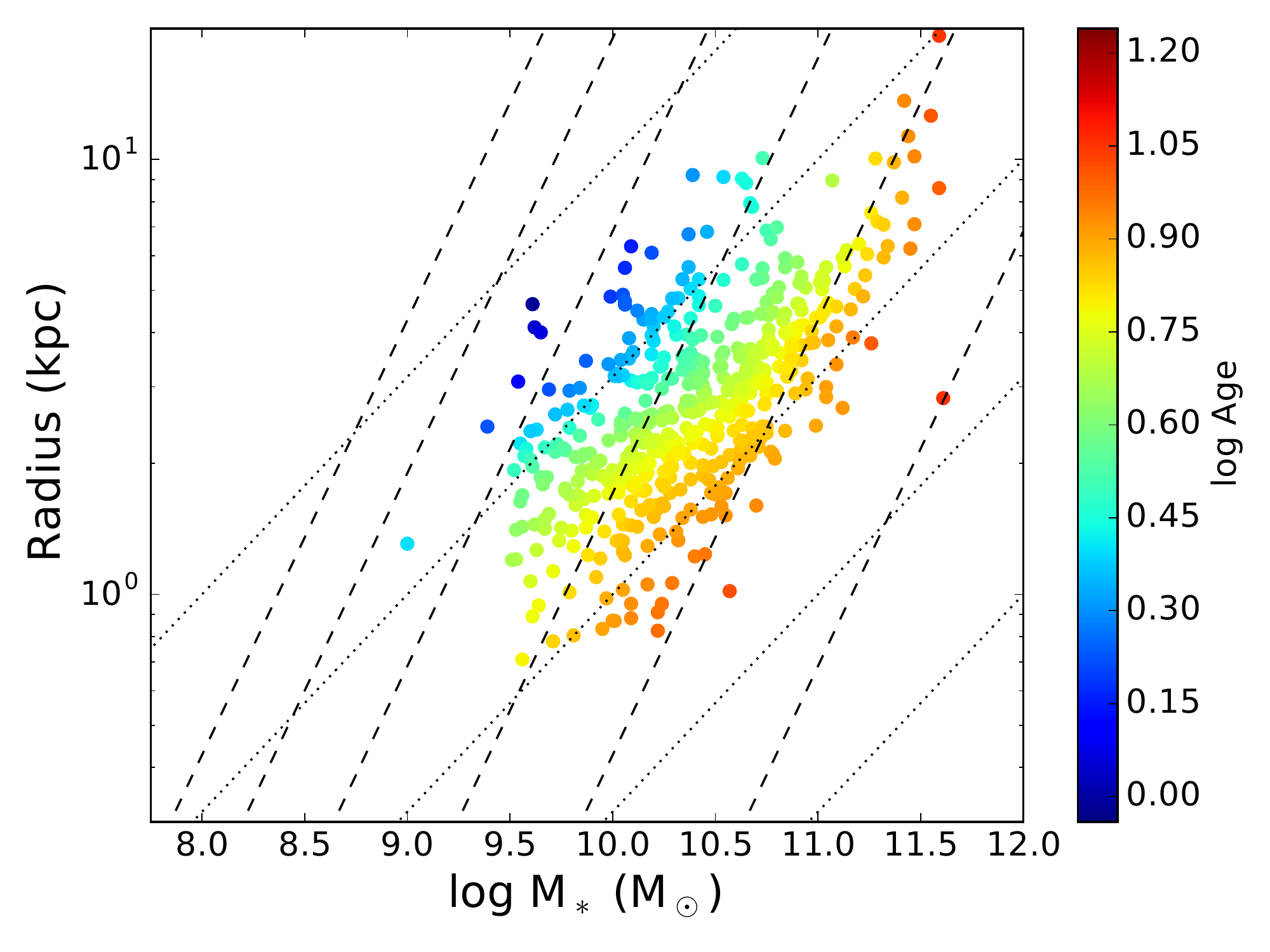}
\includegraphics[width=3.15in]{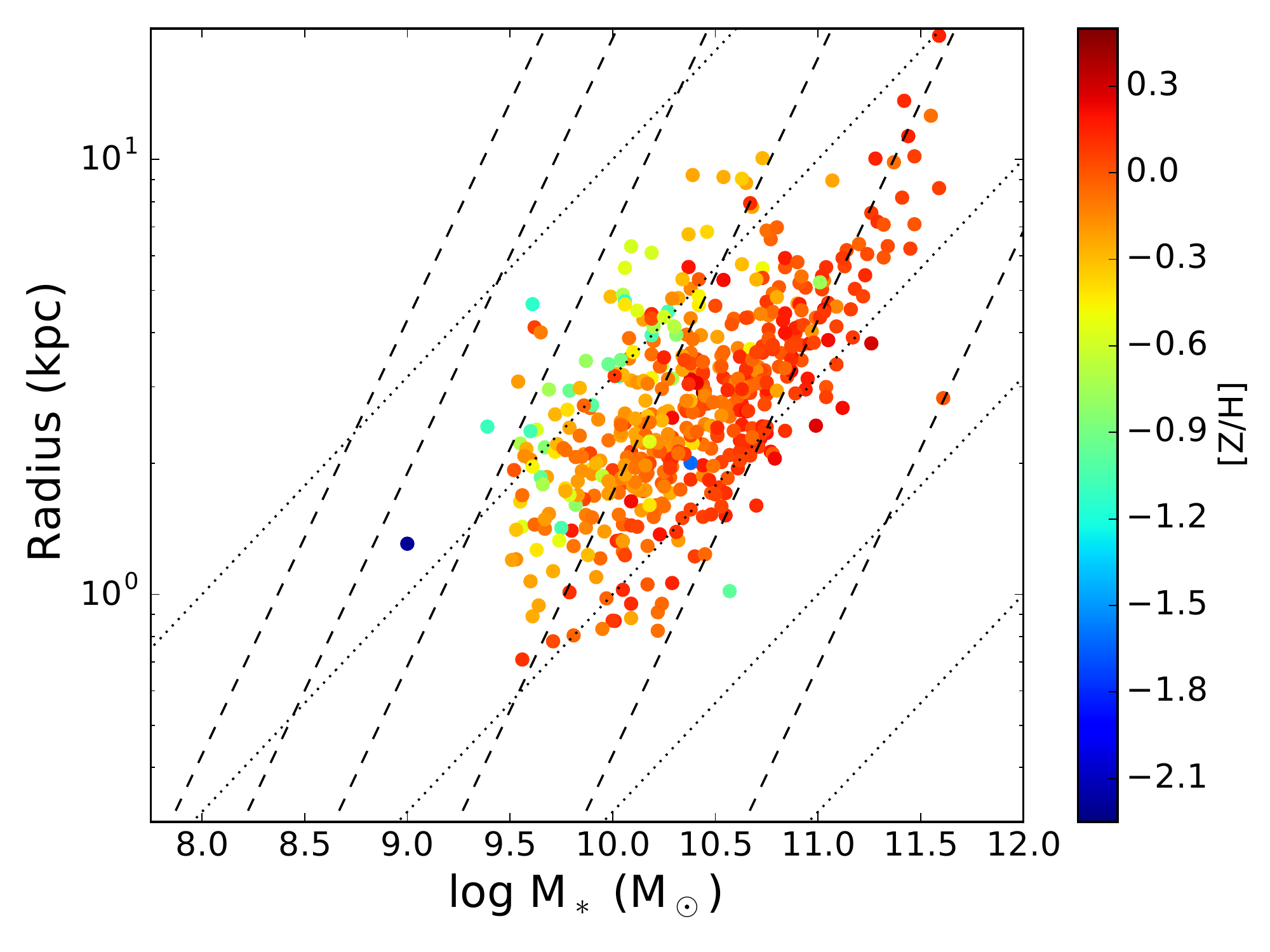}
\includegraphics[width=3.15in]{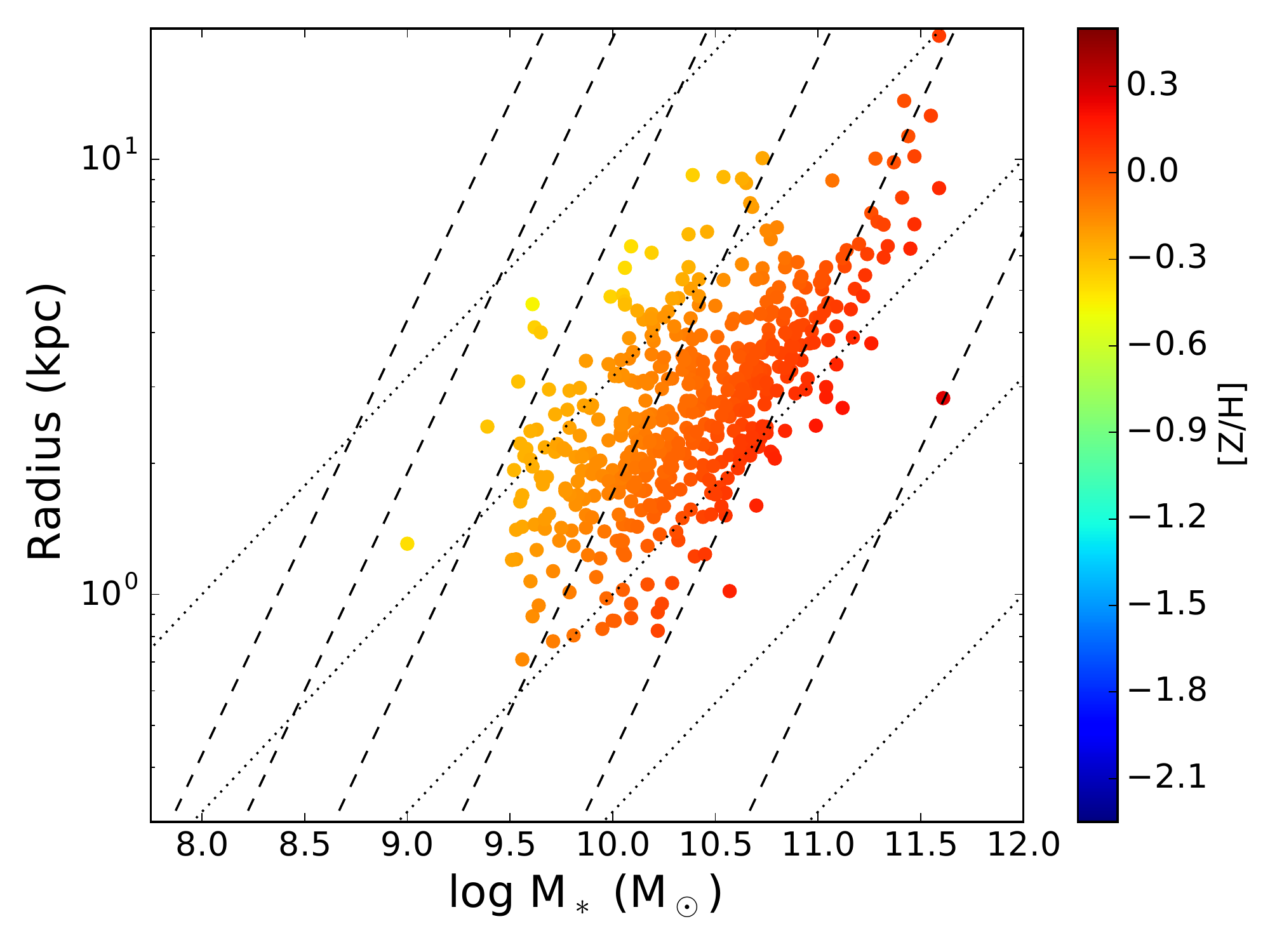}
\includegraphics[width=3.15in]{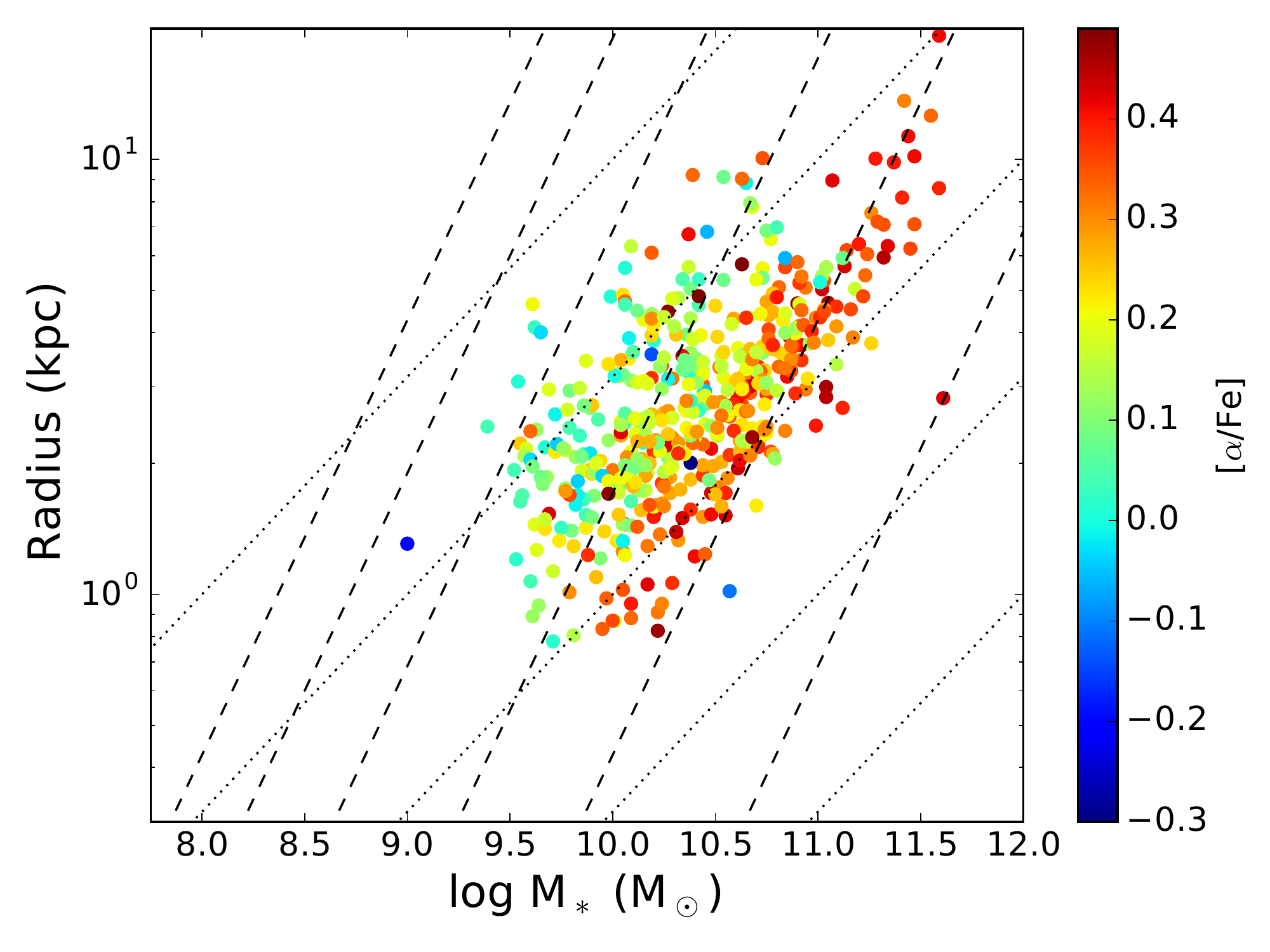}
\includegraphics[width=3.15in]{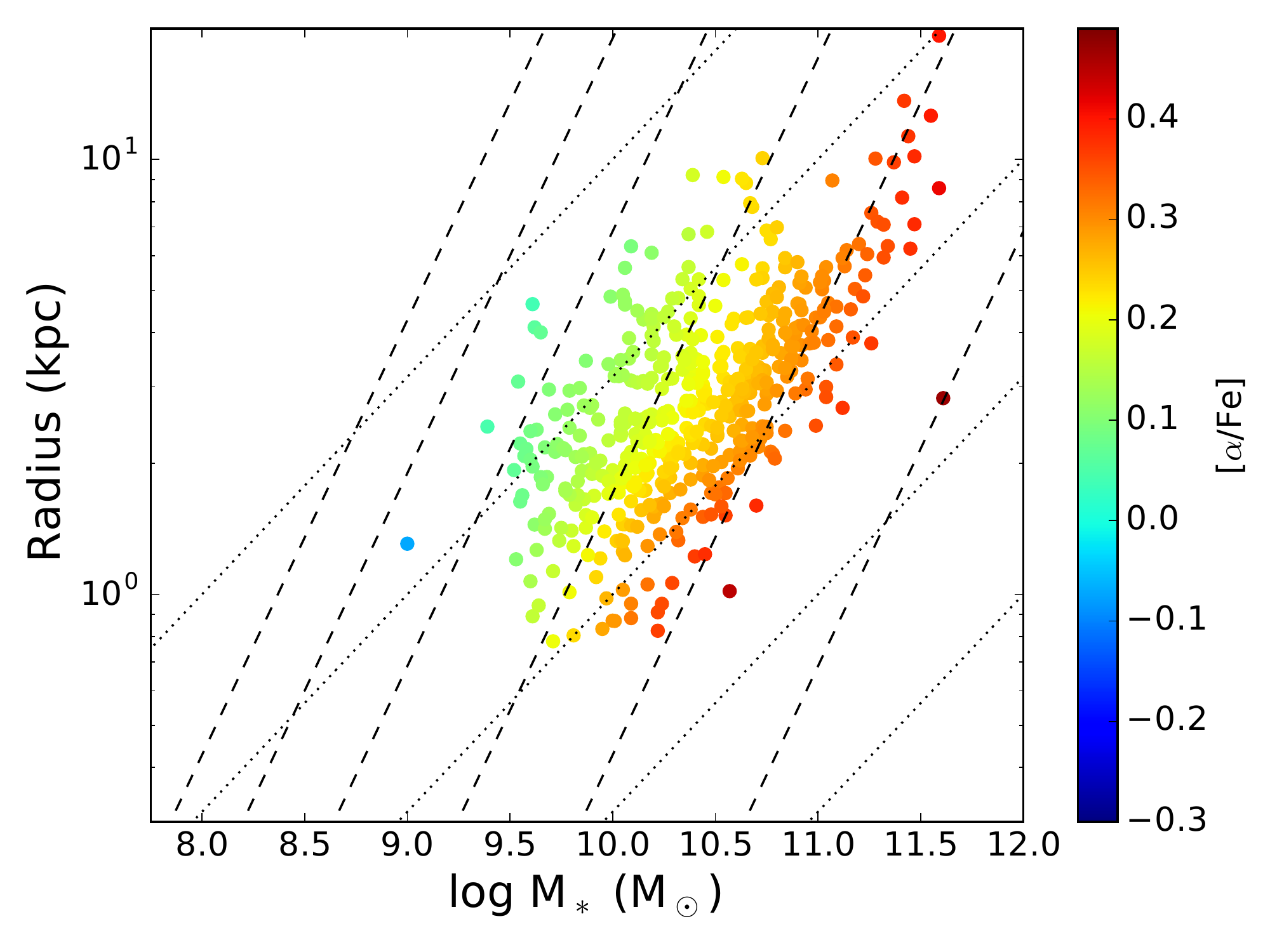}
\caption{As Figure \ref{fig:ssp_mass_size_gama}, but for the cluster galaxies only.}
\label{fig:ssp_mass_size_cluster}
\end{figure*}

In recent years the size--mass plane has become a popular tool for visualising galaxy populations, and in particular their evolution with redshift \citep[e.g.][]{Daddi:2005a,Trujillo:2007,Toft:2007,van-Dokkum:2008,vdWel:2014}. As the stellar population of a galaxy must bear the record of any redshift evolution, such evolution should be reflected in a variation of galaxy stellar populations across the size--mass plane. \citet{McDermid:2015} found that, for early-type galaxies, the more compact galaxies were older, more metal-rich and alpha-enhanced relative to more extended galaxies of the same mass. We extend that analysis to galaxies of all morphological types and with lower masses using the larger SGS sample. Physical effective radii (in kiloparsecs) were derived from the apparent angular size and the distance determined from the GAMA flow-corrected redshift. For galaxies in clusters we used the cluster redshift to determine all galaxy sizes rather than the individual galaxy redshifts as the relationship between cluster redshift and distance is less affected by galaxy peculiar velocities than the individual galaxy redshifts. As with the previous section, all results here are based on stellar populations measured within a 1 R$_e$ aperture. Any strong radial variation, or evidence of significant differences between an inner core and outer envelope \cite[as in e.g.][]{Oser:2010} will be washed out in these global measurements.

In Figures \ref{fig:ssp_mass_size_gama} and \ref{fig:ssp_mass_size_cluster} we show the size--mass plane for each of the three SSP parameters for the GAMA and cluster samples respectively. In the left column we show the values for individual galaxies, where the colour of each point indicates its SSP value. In the right column we show galaxies in the same position in the size--mass plane, but their SSP value has instead been replaced with a locally smoothed value, determined using the Locally Weighted Regression (LOESS) robust technique of \citet{Cappellari:2013b}\footnote{available from http://purl.org/cappellari/software}. We find both representations valuable; the LOESS smoothed values better illustrate the underlying trends across the plane, while the raw values preserve the local scatter at any given point in the plane.

Age, metallicity and alpha abundance all increase with increasing mass or decreasing size, consistent with the early-type galaxy picture presented in \citet{McDermid:2015}. The variation of SSP parameters with size at fixed mass accounts for much of the scatter seen in the SSP--mass correlations of Figure \ref{fig:ssp_mass}. The spread in size at fixed mass reaches its largest extent at M$_* \sim 10^{10}$ M$_\odot$. At this stellar mass, the LOESS-smoothed age ranges from 10 Gyrs at $R_e = 1$\ kpc to 2 Gyrs at $R_e = 10$\ kpc, essentially the full range in age spanned by this sample. At the same mass, the variation in [Z/H] from small to large sizes is 0.8, again accounting for the majority of the scatter seen in the [Z/H] -- M$_*$ relation. The variation of [$\alpha$/Fe] with size at fixed mass is 0.15-0.2, compared to an observed scatter in the [$\alpha$/Fe] -- M$_*$ relation of $\sim 0.4$ at M$_* \sim 10^{10}$ M$_\odot$; size variation does not account for the majority of scatter observed in the [$\alpha$/Fe] -- M$_*$ relation. 

Contours of constant metallicity (i.e. constant colour in the [Z/H] size--mass plane) closely follow lines of constant velocity dispersion (dashed lines in all panels), as previously noted by \citet{McDermid:2015}. This result accounts for the tight [Z/H]--$\sigma_e$ relation in Figure \ref{fig:ssp_disp}. The increased scatter in the [Z/H]--$\sigma_e$ relation at low $\sigma_e$ appears to be due to increased intrinsic scatter, and not due to a change in behaviour of [Z/H] in the size--mass plane. At high mass, contours of constant age are shallower, more closely tracing lines of constant surface mass density (dotted lines in all panels). Below M$_* \sim 10^{10}$ M$_\odot$, contours of constant age flatten further, becoming almost flat with mass while still showing the same variation with size. The result is that the youngest galaxies in the sample are not the lowest mass galaxies, but the galaxies with the lowest surface mass density, i.e. those tracing the upper left envelope of the size--mass distribution. This is surprising, in that we might expect stellar populations to be related to the volume mass density, however it is consistent with previous results from the CALIFA survey\citep{Gonzalez-Delgado:2015}. Contours of constant [$\alpha$/Fe] appear shallower than lines of constant $\sigma_e$ but steeper than lines of constant surface mass density. In Barone et al. (in prep.) we quantify which single parameters best describe the variation of age and [Z/H] and examine the connection to galaxy formation mechanisms.

\subsection{Residual correlations from the size-mass plane}

In Section \ref{sec:ssp_resids_mass} we showed that, at fixed mass, the SSP parameters have a significant residual dependence on morphology and environment. However, both morphology and environment vary across the size--mass plane, in the sense that galaxies towards the lower-right boundary of the population tend to have earlier type morphologies and to be found in denser environments \citep[e.g.][]{Maltby:2010,Cappellari:2013c,Cebrian:2014,Lange:2015}. We now ask whether this residual dependency on morphology and environment can be entirely explained by variations with size, or whether there remains a correlation at fixed mass {\it and} size. In sections \ref{subsec:morph} and \ref{subsec:env} we qualitatively examine residual trends of the stellar population parameters within the size -- mass plane before attempting to quantify which parameters have a significant residual effect in section \ref{subsec:significance}.

\subsubsection{Visual and kinematic morphology}
\label{subsec:morph}

\begin{figure*}
\includegraphics[width=7in,clip,trim = 20 40 10 70]{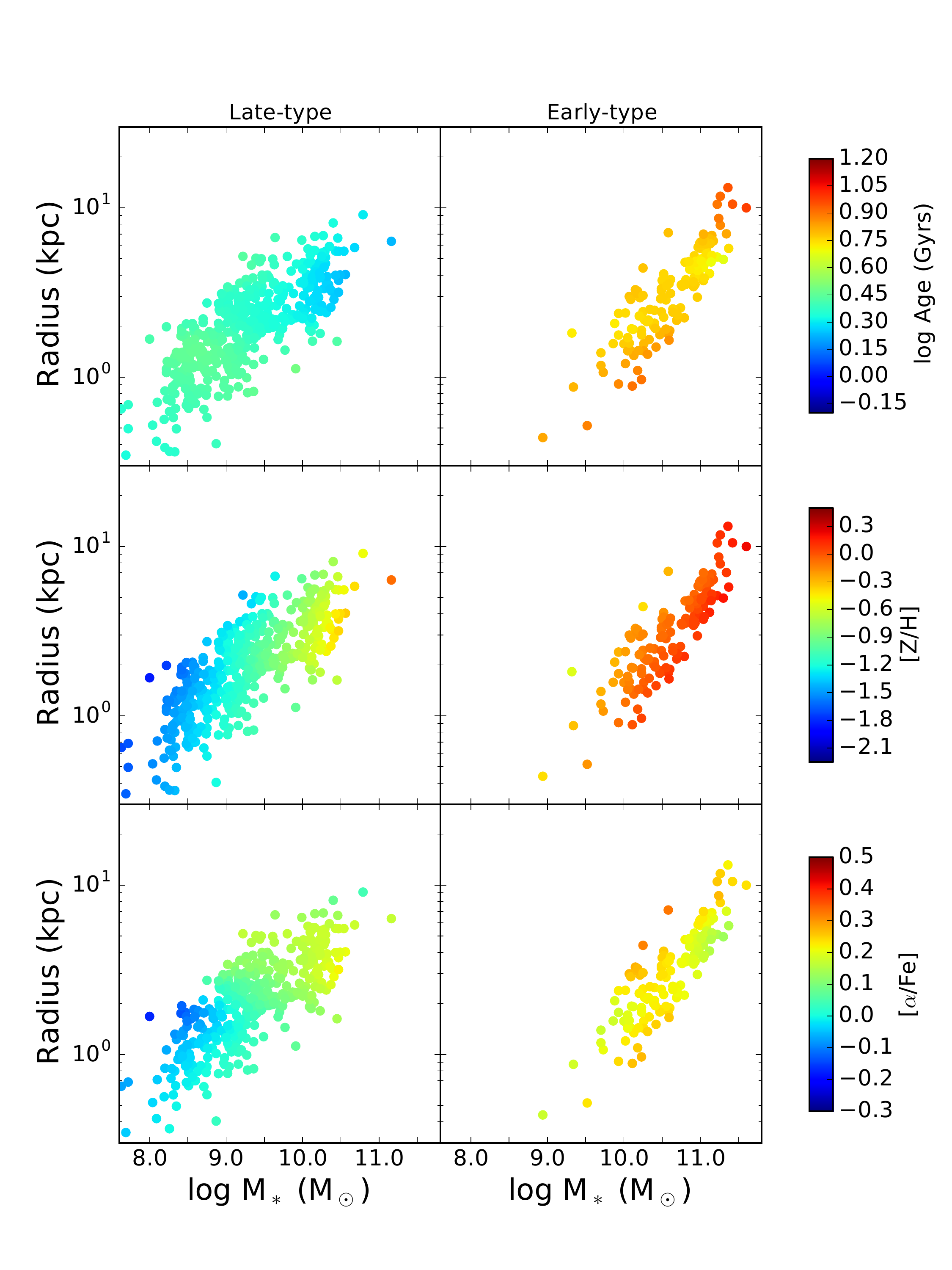}
\caption{The LOESS-smoothed size-mass plane for the GAMA sample, separated into morphological late-type (left column) and early-type (right column) galaxies. The colour scale indicates age (top row), [Z/H] (middle row) and [$\alpha$/Fe] (bottow row) respectively. The clear trend of age across the size-mass plane seen for both the GAMA and cluster samples is no longer evident at fixed morphological type. [Z/H] and [$\alpha$/Fe] both show the same general trends as for the full sample.}
\label{fig:ssp_mass_size_morph}
\end{figure*}

\begin{figure*}
\includegraphics[width=7in,clip,trim = 20 40 10 70]{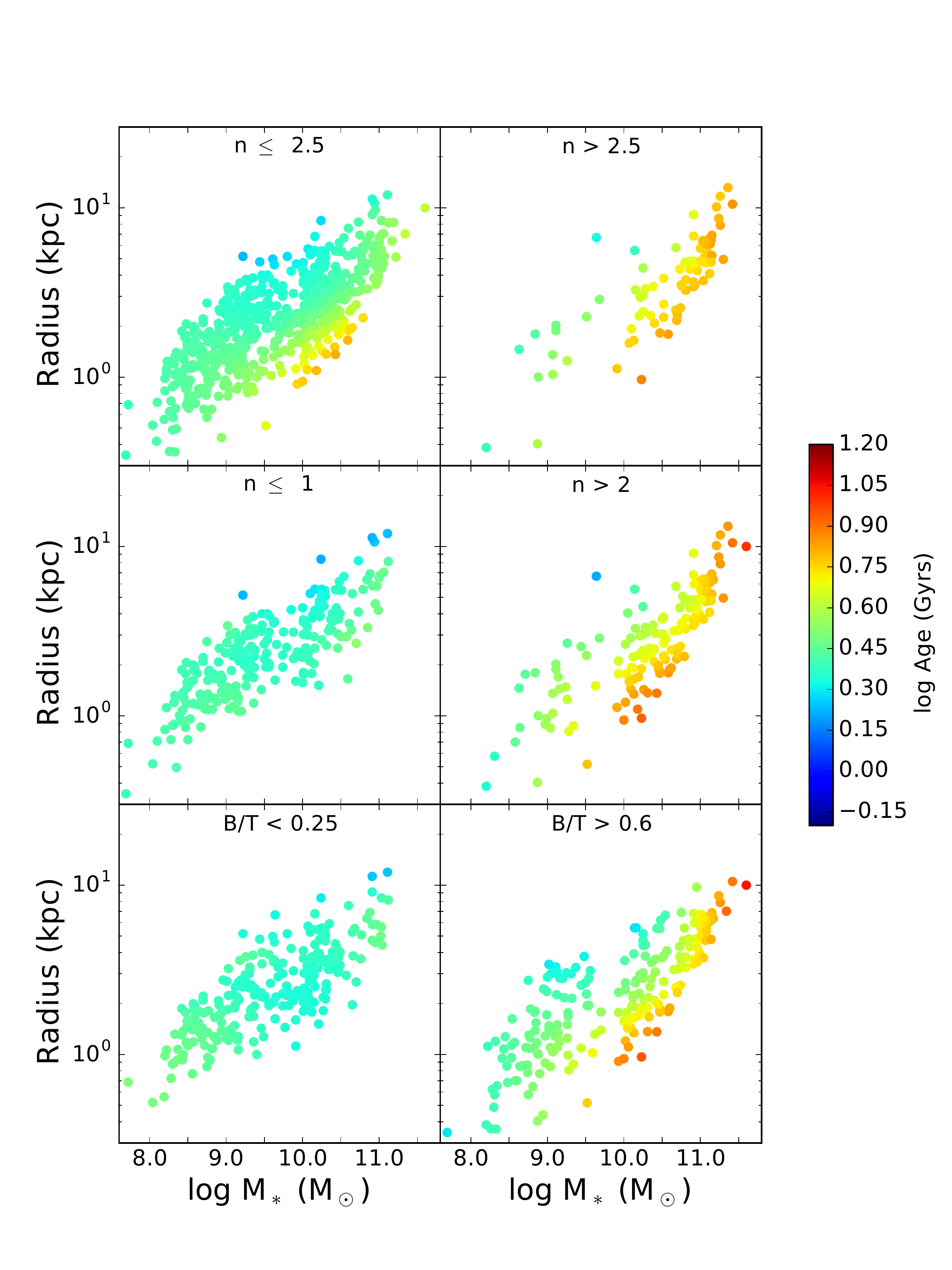}
\caption{Variation of age within the size-mass plane for the GAMA sample for i) low ($n < 2.5$) and high ($n > 2.5$) S\'{e}rsic index galaxies, ii) pure exponential ($n \leq 1$) galaxies, and those with a significant non-disk component ($n > 2$), and iii) galaxies with small (B/T < 0.25) and large (B/T > 0.6) bulges in the left and right columns respectively. The pure exponential and small-bulge systems show the same weak or non-existent variation of age within the size-mass plane as the late-type separation (upper left panel of Figure \ref{fig:ssp_mass_size_morph}), whereas the traditional separation of galaxies into structurally early-type or late-type galaxies at $n = 2.5$ still shows a strong variation of age within the plane.}
\label{fig:age_mass_size_planes_struct}
\end{figure*}

\begin{figure*}
\includegraphics[width=7in,clip,trim = 20 10 10 50]{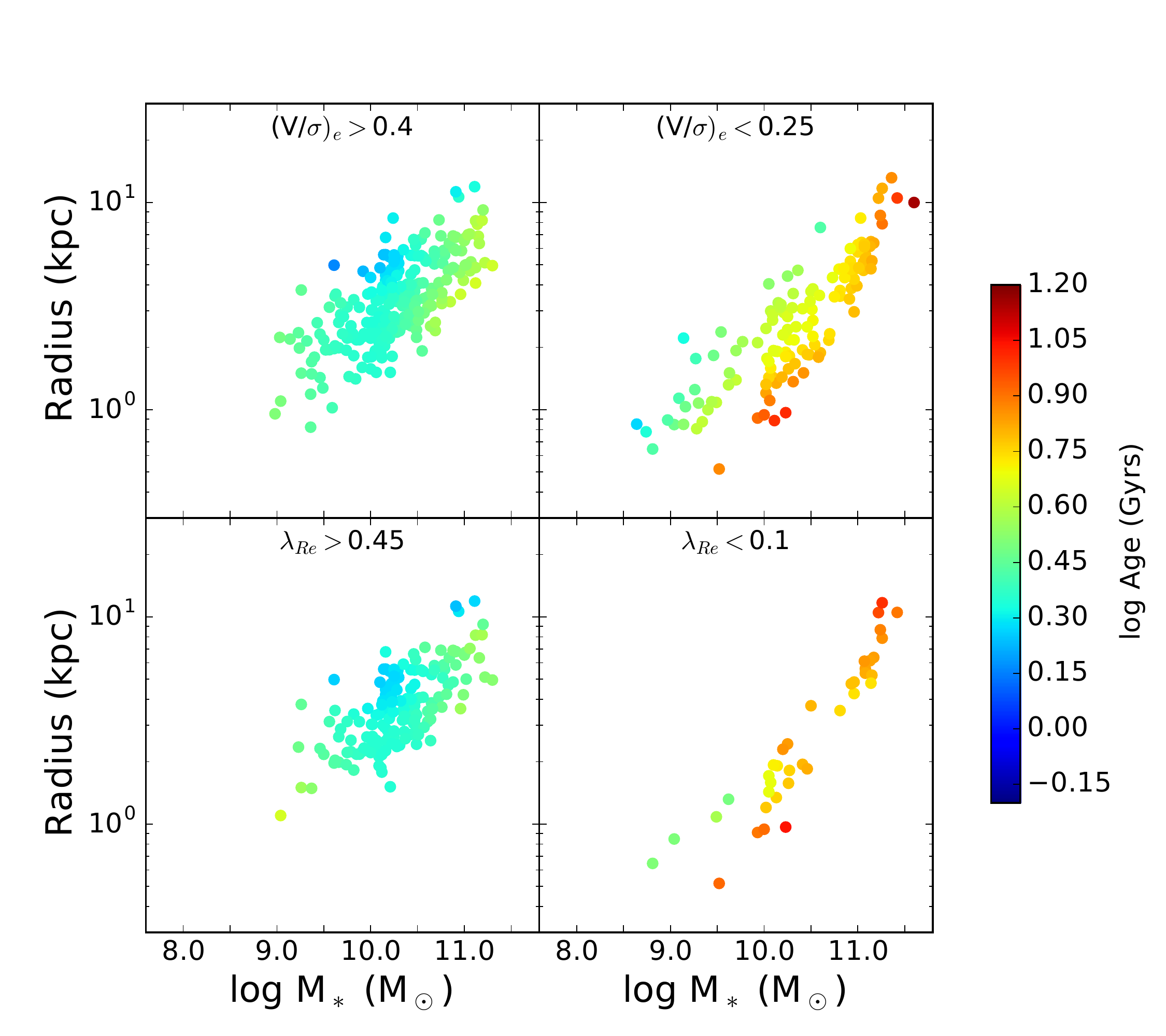}
\caption{As Figure \ref{fig:age_mass_size_planes_struct}, but separating galaxies based on their kinematic morphology. In the upper row galaxies are separated into low (V/$\sigma$ < 0.25) and high (V/$\sigma$ > 0.4) rotational support subsamples. In the lower row galaxies are separated by $\lambda_{Re}$, a proxy for specific stellar angular momentum, into slow rotators ($\lambda_{Re} < 0.1$) and high ($\lambda_{Re}$ > 0.45) specific angular momentum systems, roughly consistent with late-type galaxies. The values of V/$\sigma$ were chosen to best reproduce the stricter Sersic index selection in Figure \ref{fig:age_mass_size_planes_struct}. Unlike the visual and structural morphology separations shown in Figures \ref{fig:ssp_mass_size_morph} and \ref{fig:age_mass_size_planes_struct}, the kinematic morphology subsamples do show strong variations of age within the size-mass planes, implying a weaker correspondence between age and kinematic morphology than between age and visual or structural morphology.}
\label{fig:age_size_mass_planes_kin}
\end{figure*}

In Figure \ref{fig:ssp_mass_size_morph} we show the size--mass planes for each SSP parameter for late-type (left column) and early-type (right column) galaxies from the GAMA regions only. For these regions we have homogeneous imaging data, ensuring a consistent morphological classification. In the upper row, the strong trend of decreasing age from massive, compact galaxies to lower mass extended objects (i.e. bottom right to top left) seen in Figures \ref{fig:ssp_mass_size_gama} and \ref{fig:ssp_mass_size_cluster} is much less evident, implying that the age of galaxies within the size--mass plane is driven largely by the different loci of morphological types within this plane. This lack of mass dependence can also be seen in Figure \ref{fig:ssp_mass_morph}, where the age-mass relations for individual morphological types are shallow or even flat with mass. At fixed morphological type, [Z/H] and [$\alpha$/Fe] still show clear trends within the size--mass plane, suggesting a dependence other than morphology. 

Our visual morphological classifications may be strongly influenced by the stellar population of the galaxy, rather than the underlying structure. To test this, in Figure \ref{fig:age_mass_size_planes_struct} we examine the age dependence within the size--mass plane for galaxies separated based on their S\'{e}rsic $n$ value and bulge-to-total ratio (B/T, lower row). High S\'{e}rsic index ($n > 2$) and high B/T (B/T > 0.6) galaxies can be considered `structural' early-type galaxies. We define those with $n \leq 1$ or B/T $< 0.25$ as `structural' late-type galaxies. Traditionally, galaxies have been separated into early and late-type at $n = 2.5$ \citep[e.g.][]{vdWel:2008} and we show this division in the upper row. As the majority (92 per cent) of our sample have $n \leq 2.5$, little is revealed by the traditional separation --- the low-n sample simply reproduces the trend of the full GAMA sample discussed earlier. However, when we apply the stricter criteria of $n \leq 1$ or B/T $< 0.25$, we find a similar picture to the morphological late/early-type classification. The age trend apparent for the full GAMA sample is much weaker for samples consisting of only morphological late-type galaxies, galaxies with small bulges (B/T $< 0.25$), or structurally pure-disk ($n \leq 1$) systems. 

\begin{figure*}
\includegraphics[width=6.9in]{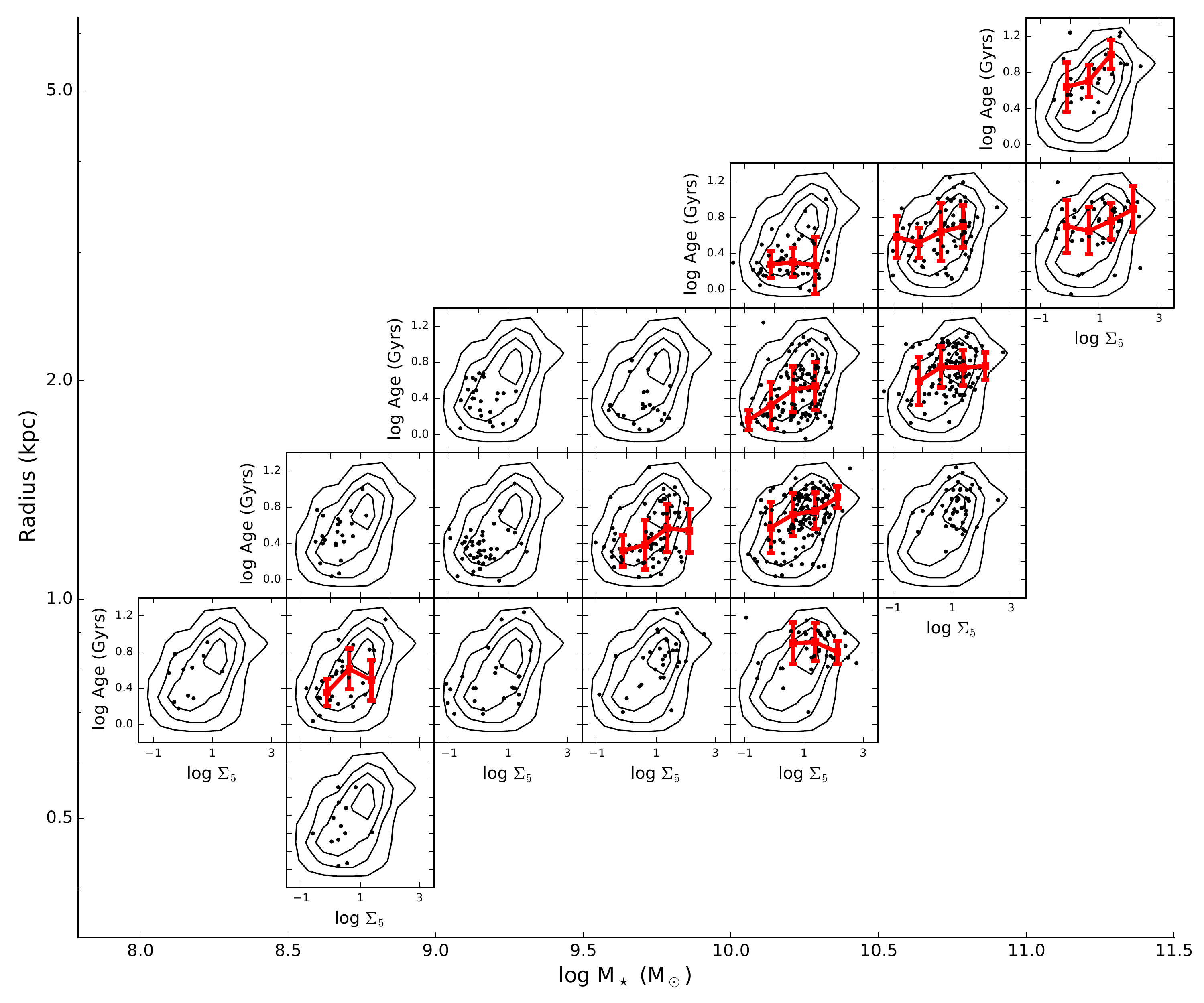}
\caption{The age--$\Sigma_5$ relation in bins of fixed size and mass. Each subplot shows the relation within a bin of 0.5 dex in mass and 0.2 dex in size. Mass increases from left to right, from $10^{8}$ M$_\odot$ to $10^{11.5}$ M$_\odot$, with size increasing from bottom to top, from 0.63 kpc to 6.3 kpc, as indicated by the outer axes. The black points indicate values for only galaxies that lie within a given bin, while the contours show the global distribution of the full sample. The red line and error bars indicate the median and 1-$\sigma$ scatter about the median in bins of $\Sigma_5$.}
\label{fig:age_mass_size_sig5}
\end{figure*}

An alternative way of classifying galaxies uses their stellar kinematics to separate the population into rotation and dispersion dominated types, or alternatively those with high or low specific stellar angular momentum. This is sometimes referred to as a kinematic morphology classification. While the angular momentum itself is challenging to measure, largely due to projection effects, the ratio of the velocity to the velocity dispersion has acted as a direct estimate of the degree of rotational support in a system \citep[e.g.][]{Illingworth:1977,Binney:1978}. More recently, the SAURON and ATLAS$^\mathrm{3D}$ surveys have promoted the related quantity, $\lambda_{R}$ \citep{Emsellem:2007,Cappellari:2007} as a proxy for the specific stellar angular momentum. In Figure \ref{fig:age_size_mass_planes_kin} we separate our sample into populations with low (V/$\sigma$ < 0.25, $\lambda_{Re} < 0.225$) and high (V/$\sigma$ > 0.4, $\lambda_{Re} > 0.35$) subsamples. These divisions approximately correspond to the $n \leq 1$, $n > 2$ separation in Figure \ref{fig:age_mass_size_planes_struct}. Despite this, we still see strong age variations across the size-mass plane for the two subsamples, unlike in the corresponding S\'{e}rsic selected samples (middle row of Figure \ref{fig:age_mass_size_planes_struct}),  indicating that kinematic morphology may be less related to mean stellar age than the visual and structural morphology measures discussed previously. However, we caution that both $V/\sigma$ and $\lambda_{Re}$ are only proxies for the specific stellar angular momentum, with both being strong functions of inclination at fixed angular momentum. Correcting for inclination or utilising dynamical models to directly infer the specific stellar angular momentum may still reveal the same strong connection between age and kinematic morphology as found for both visual and structural morphology indicators.

\subsubsection{Environment}
\label{subsec:env}

\begin{figure*}
\includegraphics[width=6.9in]{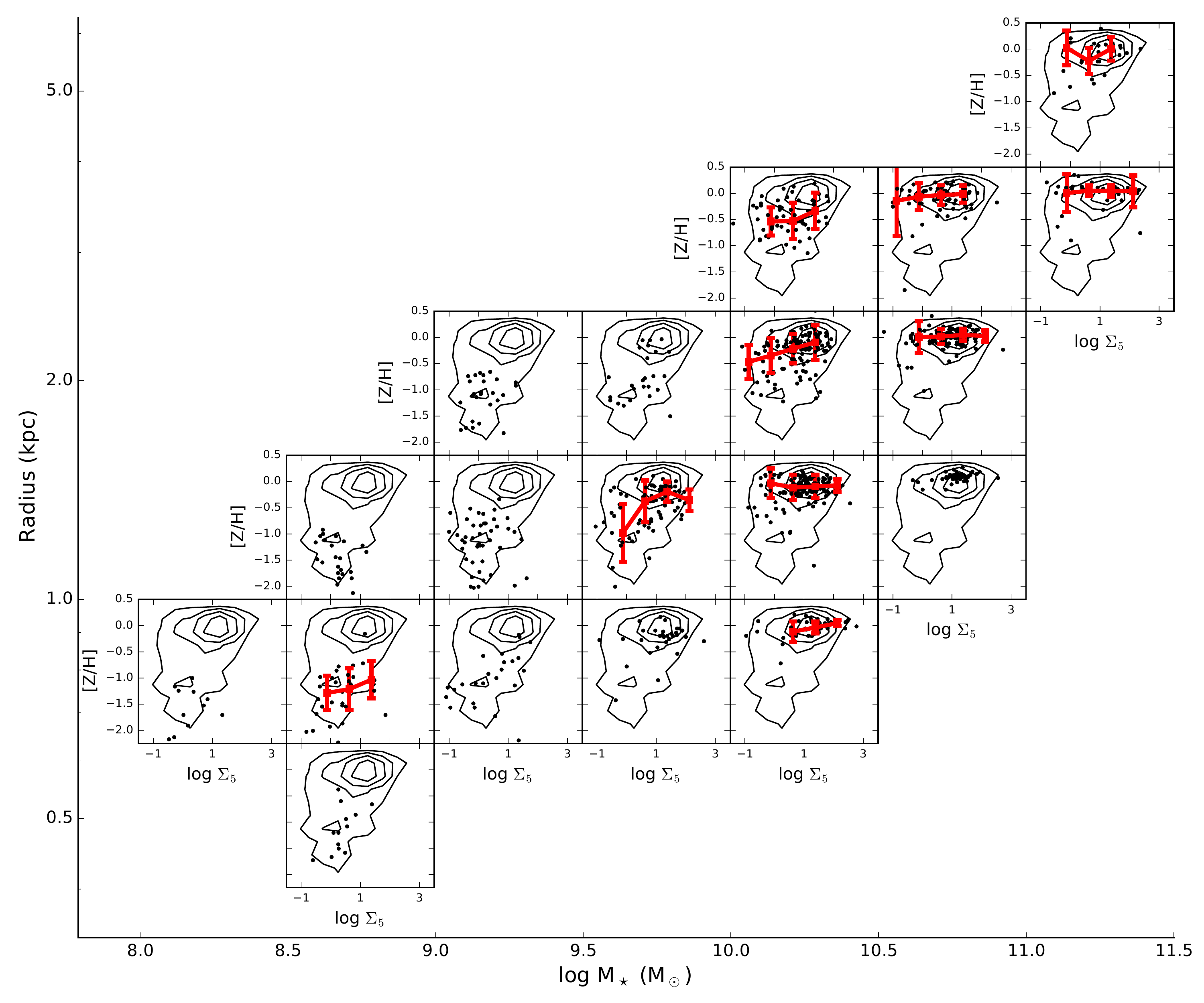}
\caption{As Figure \ref{fig:age_mass_size_sig5}, but for [Z/H]. At fixed mass, the scatter in [Z/H] increases substantially with size, though there is little residual correlation with environment.}
\label{fig:z_mass_size_sig5}
\end{figure*}

One disadvantage of the LOESS smoothing technique is that it hides the scatter in the SSP parameters at any particular location within the size-mass plane. We can examine this scatter by dividing the size--mass plane into small regions of approximately constant size and mass, then examining the relation between the SSP parameters and other independent variables within each of these sub-regions. We divide the plane into bins of 0.5 dex in mass and 0.2 dex in size. We consider only bins with at least 20 galaxies, which yields 19 regions spanning the range $10^{8}$ to $10^{11.5}$ M$_\odot$ in M$_*$ and 0.4 to 6.3 kpc in radius. In Figures \ref{fig:age_mass_size_sig5} and \ref{fig:z_mass_size_sig5} we show the age and [Z/H] -- $\Sigma_5$ relations within each of these bins for individual galaxies. Each sub-plot shows the relation for a single bin spanning 0.5 dex in mass and 0.2 dex in size and is located in the corresponding location of the size--mass plane. The right-hand column shows the most massive galaxies, with size increasing from the bottom to the top of each column. The bottom row shows the most compact galaxies, with mass increasing from the left to the right of each row. The distribution of the full SGS sample is shown in each plot as contours of number density. The red points show the median value in bins of $\Sigma_5$, with the error bars indicating the $1-\sigma$ dispersion within each bin. The median trends are only shown where the data span at least three bins in the independent variable, with at least 3 galaxies in each bin.

\begin{figure*}
\includegraphics[width=6.9in]{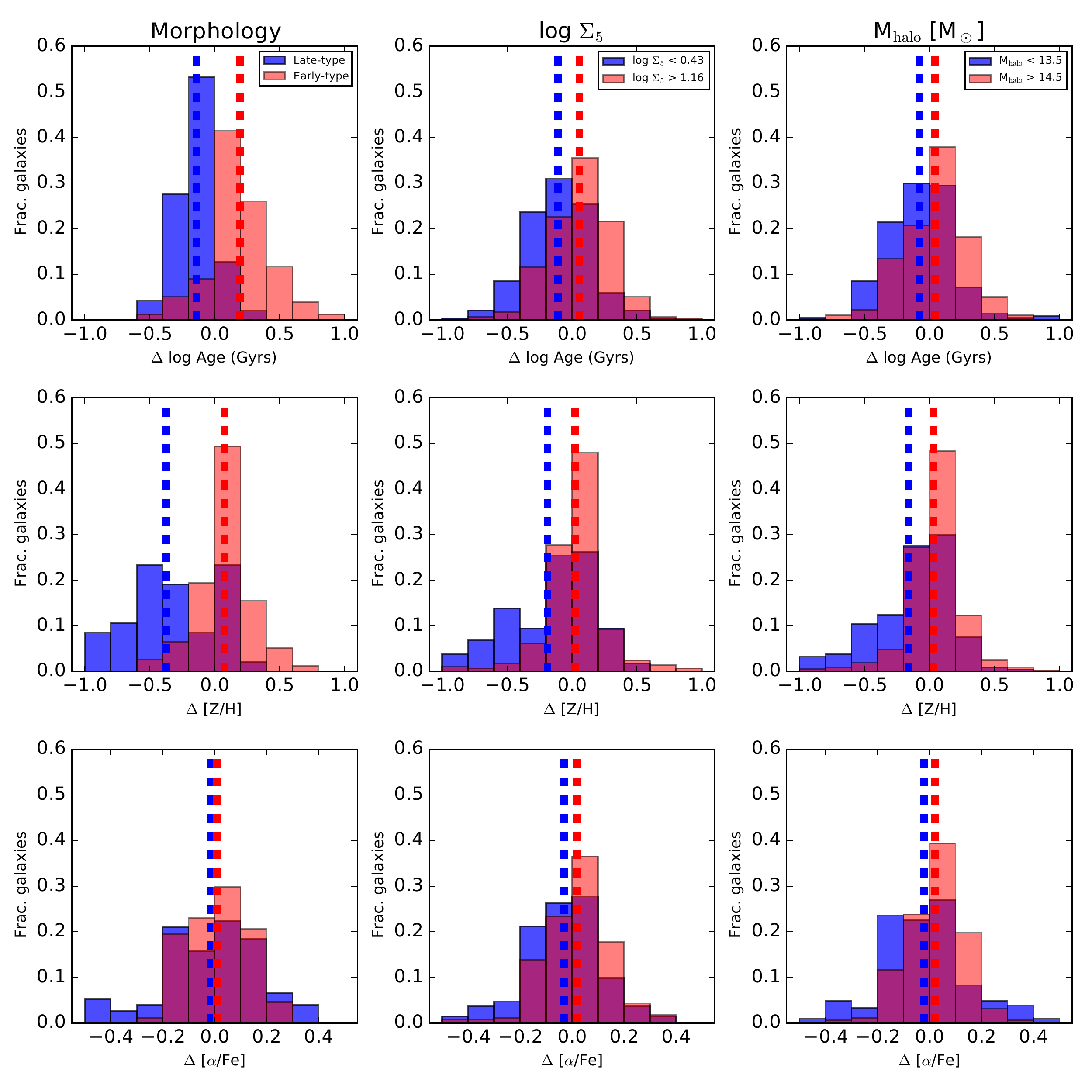}
\caption{Histograms of residuals $\Delta$ from the SSP--size--mass planes for different subsamples. For details of how the residuals were determined see the text. The rows indicate, from top to bottom, the log age, [Z/H] and [$\alpha$/Fe] residual. Each column shows the residuals for separate subsamples, separated into early- and late-type galaxies (left column), high and low local environmental density (centre column) and high and low halo mass (right column). The dashed vertical lines indicate the median residual of each subsample. Typical uncertainties on the median residuals are 0.02--0.04, though refer to the text for details.}
\label{fig:size_mass_histograms}
\end{figure*}

Two clear results stand out from these plots. Firstly, in the majority of bins that span a broad range in environment, there is a trend of age (at fixed mass and size) with $\Sigma_5$ (and similarly for M$_\mathrm{halo}$), in the sense that galaxies in lower density environments or lower mass halos tend to be younger than those in denser environments or more massive halos. This trend is less clear for the most extended (upper left envelope) or the most compact (lower right envelope) galaxies --- these galaxies are either predominantly young or old respectively, and so any residual correlation is suppressed by the effect of the primary age--surface density relation. The other clear trend is that, in massive galaxies (M$_* > 10^{10} $M$_\odot$, there exists a tight [Z/H] sequence which shows little, if any, residual dependence on environment. However, the scatter about this sequence increases significantly with increasing galaxy size, in the sense that the most compact, highest $\sigma_e$ galaxies (lower right envelope) show the smallest range in [Z/H] (at fixed mass and size), with the range of [Z/H] increasing rapidly as galaxy size increases (moving up in each column of plots). No residual trends are apparent in the [$\alpha$/Fe] -- size -- mass plane, though there is significant scatter in [$\alpha$/Fe] at fixed size and mass. This scatter is likely due to the higher uncertainty on our [$\alpha$/Fe] measurements compared to those of [Z/H] and log age; however, it is possible that an unknown additional parameter is responsible for this variation in [$\alpha$/Fe].

\subsubsection{Significance of residual trends}
\label{subsec:significance}

In each bin within the size--mass plane we determined the local median of age, [Z/H] and [$\alpha$/Fe]. We then determined the residuals $\Delta$ from the size--mass plane by subtracting this local median from the observed SSP parameters for each galaxy, i.e. the residual accounting for the observed trends in size and mass found within the sample. We consider only bins with a broad range in the independent variable. In Figure \ref{fig:size_mass_histograms} we show these residuals, separating the full sample in three different ways: late-type (morphological classification of early- and late-type spiral) galaxies and early-type (elliptical and lenticular) galaxies (left column), high ($\log \Sigma_5 > 1.16$) and low ($\log \Sigma_5 < 0.43$) local environmental density samples (central column) and high ($\log M_\mathrm{halo} > 14.5$) and low ($\log  M_\mathrm{halo} < 13.5$) halo mass environments (right column). The environmental selections divide the sample approximately into thirds. The blue histograms show the late-type, low local density or low halo mass samples, with the red histograms showing the early-type, high local density or high halo mass samples. The vertical dashed lines show the median residual for the corresponding sample. The upper row shows residuals with age, $\Delta$ log age, the middle row $\Delta $[Z/H] and the bottom row $\Delta $[$\alpha$/Fe].

The residuals in log age and [Z/H] show the strongest dependence on morphology. The morphology dependence is consistent with the previous discussion, reflecting the close link between visual morphology and age. However, we again note that the consistent behaviour found with the visual and structural morphology measures suggests there is a genuine physical dependence, rather than a simple classification bias for young, blue galaxies to be classified as later visual types. The [Z/H] dependence on morphology is consistent with the results of Figure \ref{fig:ssp_mass_morph}, where we found a large difference in [Z/H] between Sc/Sd galaxies and all earlier morphological types, and only small variations with the E/S0/Sa/Sb populations. At fixed size and mass, late-type galaxies are, on average, $0.34 \pm 0.03$ dex younger and $0.44 \pm 0.06$ dex more metal poor than early-type galaxies. The difference in [$\alpha$/Fe] between late- and early-type galaxies at fixed mass and size, $0.02 \pm 0.02$, is consistent with zero, however, we again note the large observational uncertainty in [$\alpha$/Fe] for low mass galaxies.

The observed dependence on environment is smaller than that seen with morphology in all three SSP parameters. We find a residual dependence, in the sense that galaxies in denser environments or more massive halos are older and more metal rich compared to galaxies in lower mass halos of the same galaxy mass and size. For age, this appears to be a global offset in the distributions, but for [Z/H] this offset is driven by a low-metallicity tail that is absent in the more massive halos. We find no significant residual dependence of [$\alpha$/Fe] on either morphology or environment, unsurprising given the larger observational uncertainty on that measurement. The dependence of age on $\Sigma_5$ ($\Delta$ log  age$ = 0.17 \pm 0.02$) is larger than for M$_\mathrm{halo}$ ($\Delta$ log  age$ = 0.12 \pm 0.02$), whereas the difference in [Z/H] is similar for $\log \Sigma_5$ and M$_\mathrm{halo}$ ($\Delta$ [Z/H]$ = 0.21 \pm 0.03$ and $0.19 \pm 0.03$ respectively). Galaxies in lower density environments or lower mass halos are, on average, 0.2 dex more metal poor than galaxies of the same size and mass in high density environments or more massive halos.

To summarise this section: for late-type galaxies, galaxies with $B/T < 0.25$ and galaxies with $n \leq 1$ we find little variation of age with size and mass. Variations of morphology with size and mass explain essentially all of the age variation within the plane, whereas [Z/H] and [$\alpha$/Fe] show additional variation beyond that explained by morphology. At fixed size and mass, galaxies in denser local environments or higher mass halos are older than galaxies in lower density environments or low mass halos. Galaxies with the highest surface density exhibit a very narrow range in [Z/H], whereas, at fixed mass, more extended galaxies show a much larger spread in [Z/H]. Overall, morphology has a stronger influence on age and [Z/H] than environment. We find a slightly larger dependence of age on local environmental density than on halo mass, though the difference is small. We find no residual dependence of [$\alpha$/Fe] on either morphology or environment, though smaller observational uncertainties may still reveal such a dependence.

\section{Discussion}
\label{sec:discussion}

\subsection{Comparison to previous work}

Comparisons between stellar population measurements can be challenging to interpret, because the detailed results are sensitive to the choice of stellar population measurement technique and the particulars of the data. However, while absolute comparisons should be treated with caution, the relative trends are more robust.

\begin{figure}
\includegraphics[width=3.15in,clip,trim = 0 15 10 10]{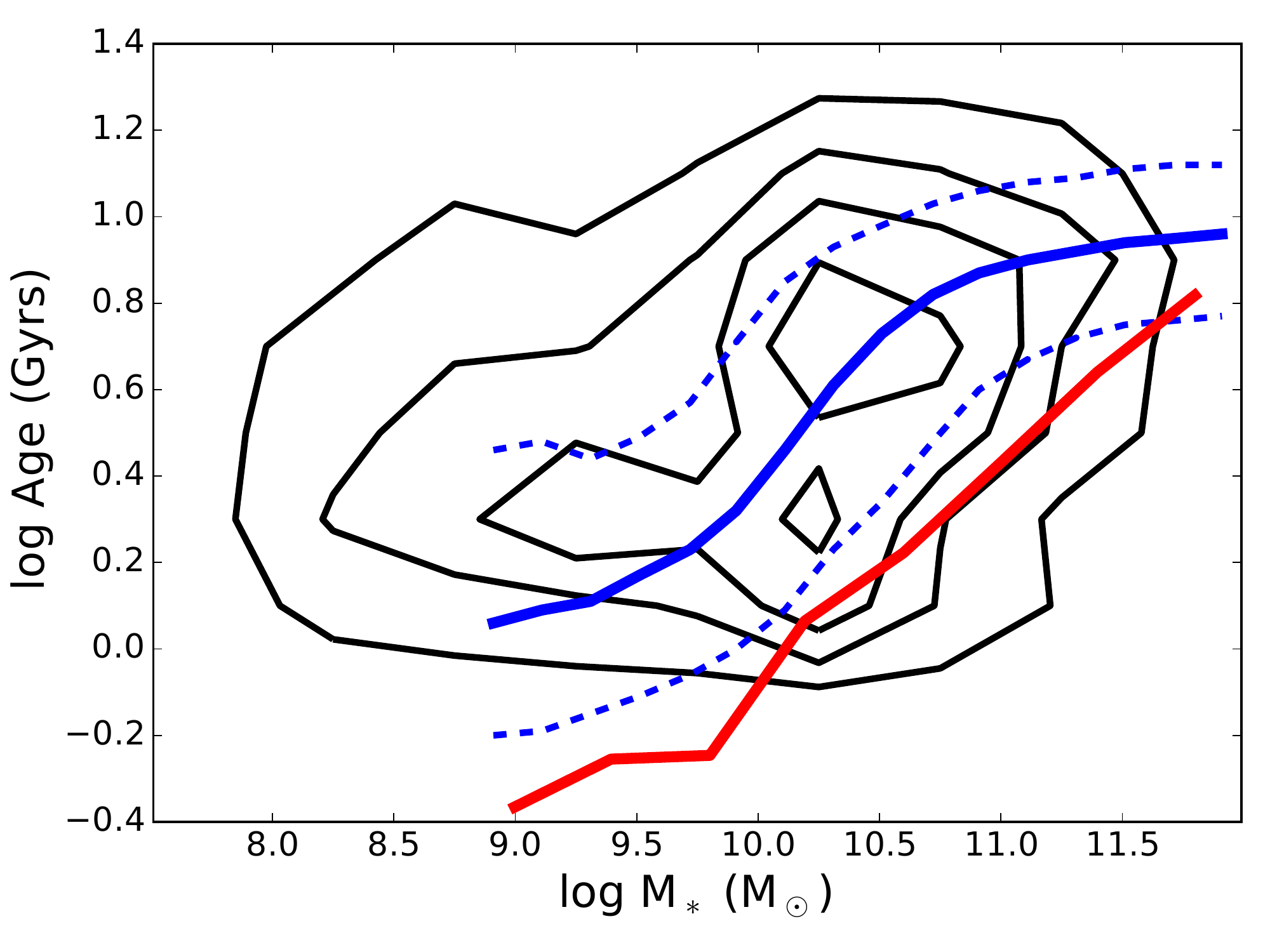}
\includegraphics[width=3.15in,clip,trim = 0 15 10 10]{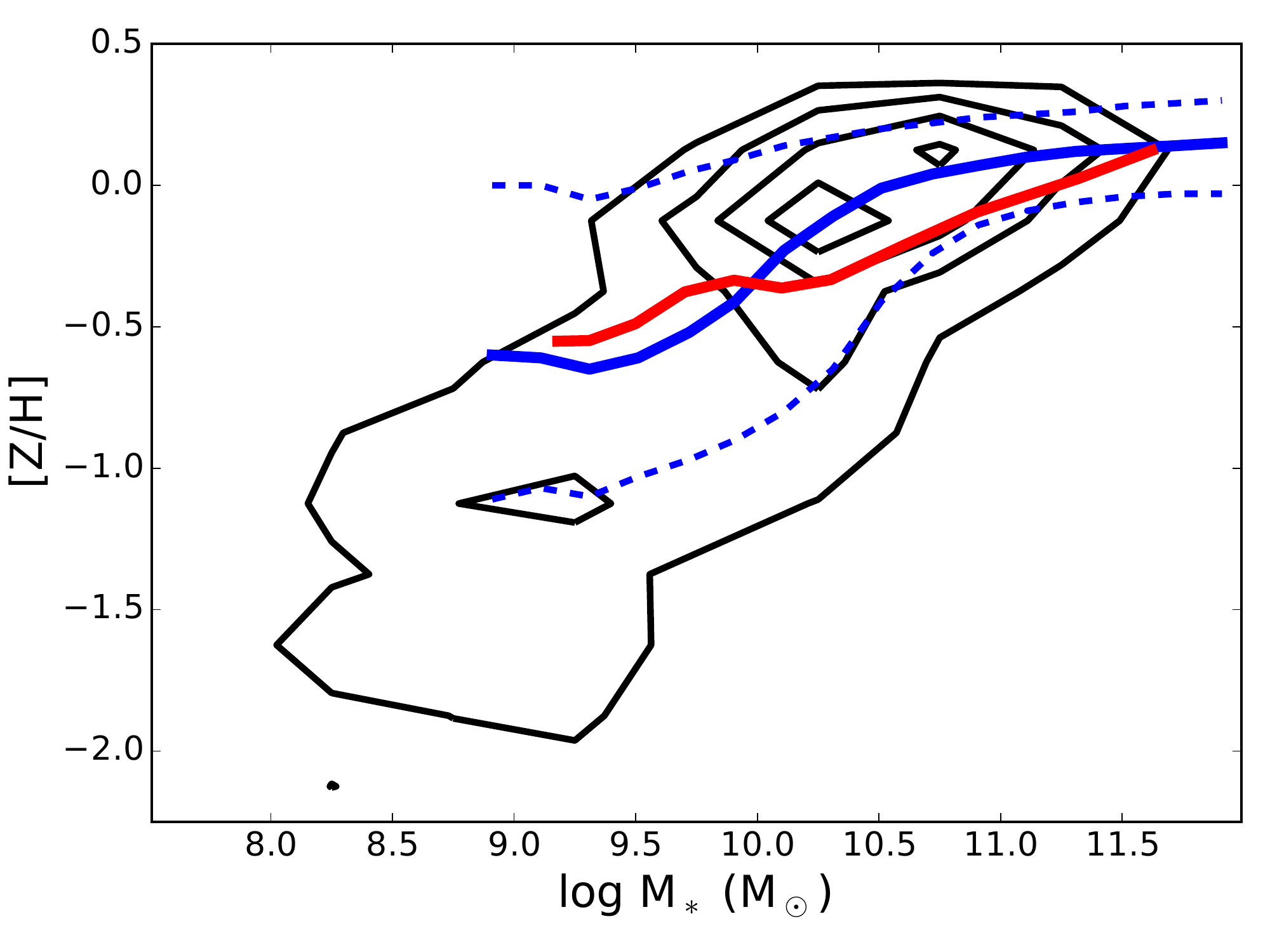}
\caption{Upper panel: Age -- M$_*$ relations from \citet[blue lines, mean and 16$^\mathrm{th}$ and 84$^\mathrm{th}$ percentiles]{Gallazzi:2005} and \citet[red line]{Gonzalez-Delgado:2015}. Lower panel: as upper panel but for [Z/H]. The contours show the distribution of the full (GAMA and cluster combined) sample from this work, enclosing 5, 20, 40 and 70 per cent of the data respectively.}
\label{fig:ssp_mass_comp}
\end{figure}

While no single sample combines the advantages of size, parameter space and integral field spectroscopy provided by the SGS, other works do provide a useful comparison to some of the global trends presented here. The works of \citet{Gallazzi:2005} and \citet{Gonzalez-Delgado:2015} both explored the global stellar population --- stellar mass relations for a sample of galaxies spanning a broad range in mass and morphology. In Figure \ref{fig:ssp_mass_comp} we show the mean trends from \citet[blue lines]{Gallazzi:2005} and \citet[red line]{Gonzalez-Delgado:2015} over the data from this work (black contours). At high mass the agreement between our work and \citet{Gallazzi:2005} is excellent, but at masses below 10$^{10}$ M$_\odot$ we find lower metallicities and older ages. However, the overall trend is consistent; a tight high-metallicity sequence and older ages at high mass, with a sharp transition to younger ages and lower metallicities below M$_* \sim 10^{10}$ M$_\odot$. The discrepancies at low mass likely arise from i) the narrower set of spectral features and ii) the more complex star formation histories utilised by \citet{Gallazzi:2005}. We also note that our lower metallicities are consistent with [Z/H] measurements for local group dwarf galaxies with masses in the range M$_* = 10^8$ -- $10^9$ M$_\odot$ from \citet{Kirby:2013}. \citet{Gonzalez-Delgado:2015} find younger ages at all masses, with lower metallicities at high mass and higher metallicities at low mass \citep[consistent with][below M$_* < 10^{10}$ M$_\odot$]{Gallazzi:2005}. This is expected as \citet{Gonzalez-Delgado:2015} utilise mass-weighted properties, whereas the light-weighted SSP parameters presented in this work are more sensitive to young populations that dominate in luminosity while contributing proportionally less to the total stellar mass.

\begin{figure}
\includegraphics[width=3.15in,clip,trim = 0 15 10 10]{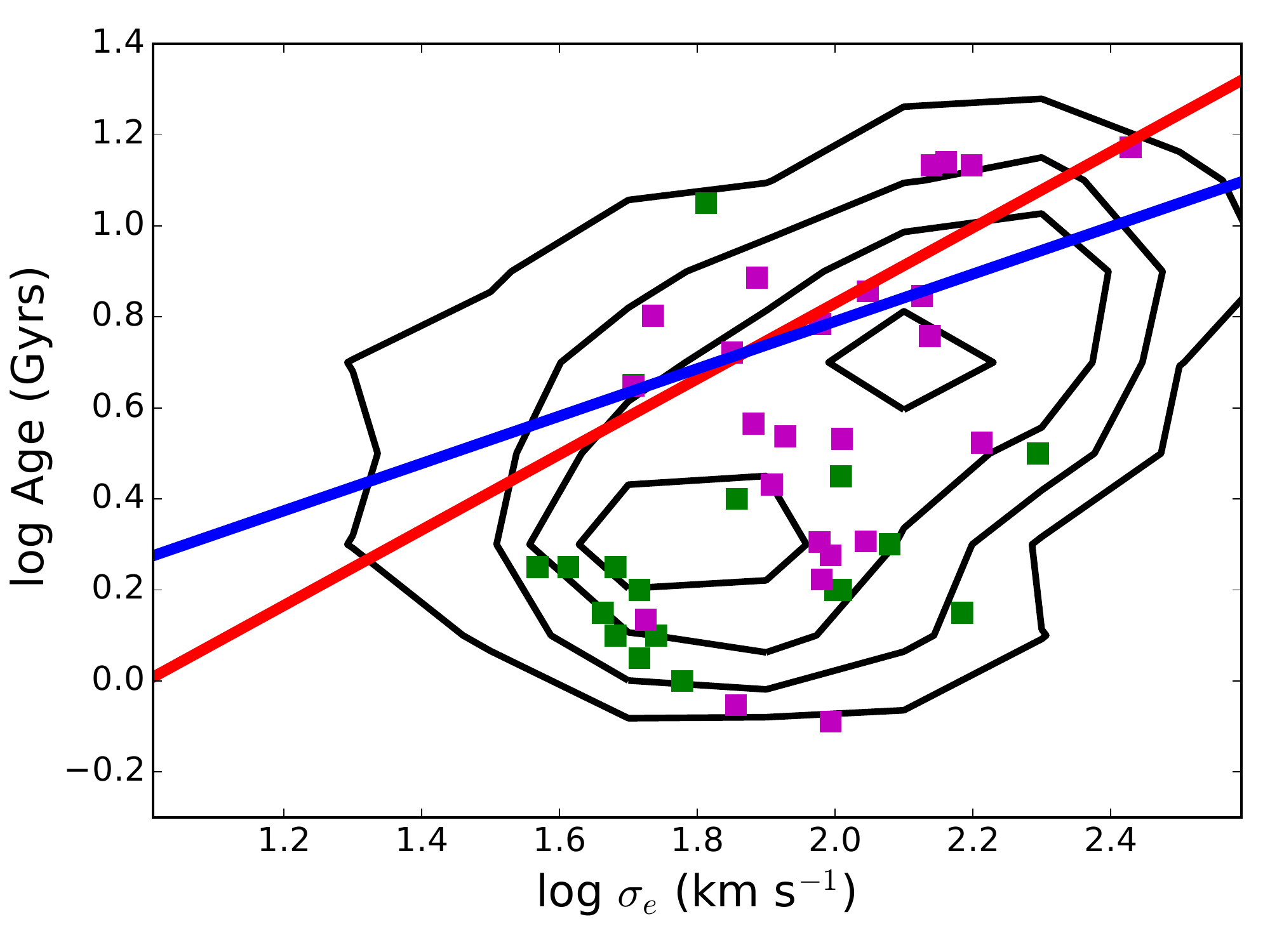}
\includegraphics[width=3.15in,clip,trim = 0 15 10 10]{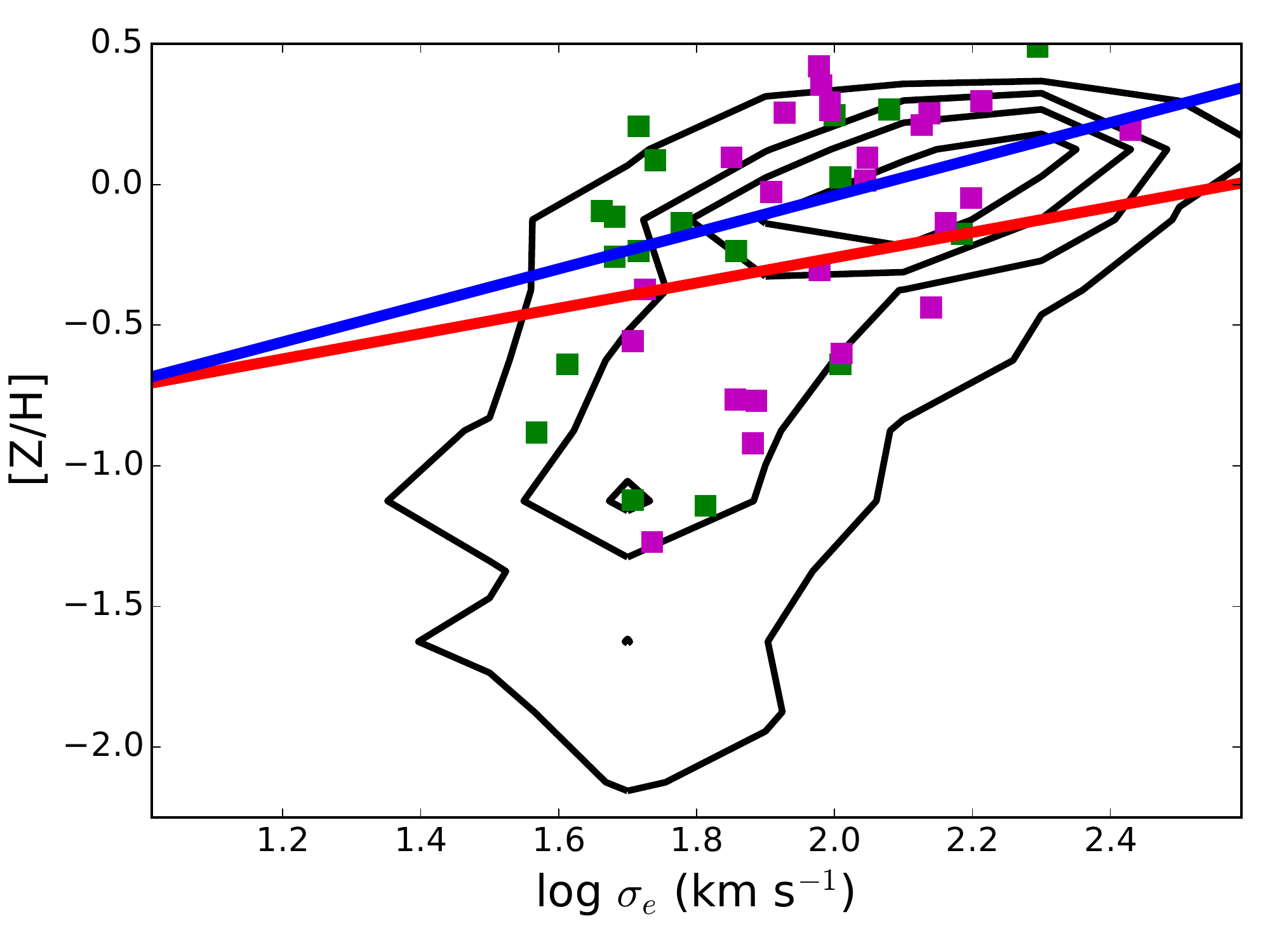}
\includegraphics[width=3.15in,clip,trim = 0 15 10 10]{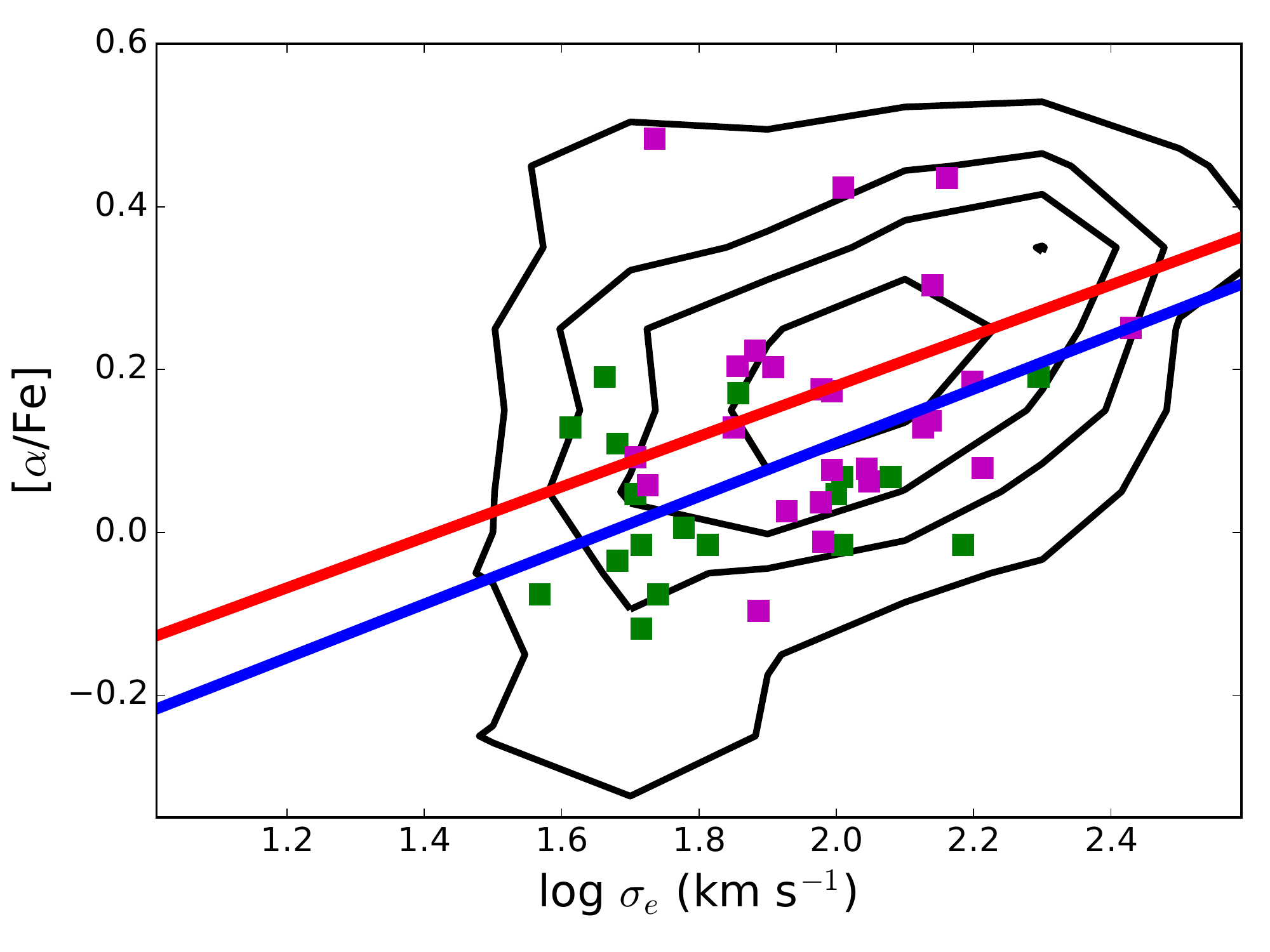}
\caption{Upper panel: Age -- $\sigma$ relations from \citet[red line]{McDermid:2015} and \citet[blue line]{Thomas:2010} and data for spiral galaxies from \citet[magenta squares]{Peletier:2007} and \citet[green squares]{Ganda:2007}. Middle panel and lower panels: as upper panel but for the [Z/H] -- $\sigma$ and [$\alpha$/Fe] -- $\sigma$ relations respectively. The contours show the distribution of the  full (GAMA and cluster combined) sample from this work, enclosing 5, 20, 40 and 70 per cent of the data respectively.}
\label{fig:ssp_disp_comp}
\end{figure}

Other works have utilised smaller samples or focused on particular morphological types. In Figure \ref{fig:ssp_disp_comp} we compare the mean SSP relations with $\sigma_e$ found by \citet[blue lines]{Thomas:2010} and \citet[red lines]{McDermid:2015}, whose samples were restricted to early-type (i.e. E and S0) galaxies only. We also show data from \citet[magenta points]{Peletier:2007} and \citet[green points]{Ganda:2007}, who examined early-type and late-type spirals respectively. The distribution of spiral galaxies is consistent with our work in all three SSP parameters --- in particular we find good agreement in luminosity-weighted SSP-equivalent age between our work and \citet{Peletier:2007} and \citet{Ganda:2007}. Some of their spiral galaxies have higher metallicities than found in our work (and the others examined here), but this may be due to the very limited range of spectral indices used in these studies.

\citet{Thomas:2010} and \citet{McDermid:2015} both examined early-type galaxies with $\sigma > 100$ km s$^{-1}$ and $\sigma > 60$ km s$^{-1}$ respectively, though we extrapolate their reported best-fitting relations to lower values of $\sigma$ in Figure \ref{fig:ssp_disp_comp}. The other key difference between our work and \citet{Thomas:2010} is that that work used a fixed angular aperture for both their SSP and $\sigma$ measurements, whereas both this work and \citet{McDermid:2015} use an R$_e$ aperture. Our best-fitting [Z/H]--$\sigma_e$ relation is intermediate between that from \citet{Thomas:2010} and \citet{McDermid:2015}, agreeing well with the relation \citet{McDermid:2015} found using an R$_e/2$ aperture. For the [$\alpha$/Fe] comparison the agreement is reasonable, though we find slightly higher values of [$\alpha$/Fe] in early-type galaxies and a steeper general trend with $\sigma$, driven by our inclusion of all morphological types. A similar effect is evident in the age--$\sigma$ relation, where we find a broadly consistent relation for high-$\sigma$, early-type galaxies but a steeper global relation due to the late-type galaxies in our sample.

\citet{Thomas:2010} and \citet{McDermid:2015} also examined the residual dependence of the SSP parameters on environment at fixed mass (fixed stellar mass and dynamical mass respectively). \citet{Thomas:2010} conclude that for old early-type galaxies there is no residual dependence of the SSP parameters on environment, but that the relative proportion of young to old early-type galaxies does increase with decreasing environmental density \citep[however, see][for a contrasting view]{Rogers:2010,LaBarbera:2014}. \citet{McDermid:2015} find that in the highest density environments in their sample (corresponding to galaxies in cluster-mass halos) galaxies are older and alpha-enhanced, with little metallicity dependence. Outside clusters, they find no significant environmental dependence of the SSP parameters, suggesting that the [Z/H] variation we find between different environments, even at fixed mass and size, is likely driven by the changing morphological mix of our sample, which was not present in the pure early-type samples discussed in \citet{Thomas:2010} and \citet{McDermid:2015}. Our present sample is too small to simultaneously distinguish the effects of mass, morphology and environment, however the final SGS sample of $\sim 3600$ galaxies will allow us to robustly make this separation.

The apparent increase in [$\alpha$/Fe] with environment identified by \citet{McDermid:2015} is consistent with the smaller average sizes of galaxies in higher density environments. However, whether the environment causes galaxies to be more compact, which in turn accelerates their star-formation timescales (as implied by higher [$\alpha$/Fe]), or whether environment directly increases the efficiency of star formation, which indirectly causes the galaxies to become more compact through e.g. a centrally concentrated starburst cannot be distinguished by this study alone.

\subsection{Comparison to simulation predictions}

Simulations and semi-analytic models of galaxy formation have traditionally shied away from predictions of galaxy stellar populations. This is mainly because the stellar populations, and the chemical enrichment in particular, are sensitive to the details of the feedback implementation, which is poorly constrained in most simulations. In recent years a number of authors \citep{Lu:2014,Henriques:2015,Ma:2016} have predicted stellar ages and metallicities from their models with some success. Here we focus on the predictions of \citet{Hirschmann:2016} and \citet{Tonini:2016} as broadly representative of the state-of-the-art in stellar age and metallicity relations from simulations.

\citet{Hirschmann:2016} examine the age-- and [Z/H]--M$_*$ relations for several different feedback models implemented in the same cosmological framework. They find tight relations for galaxies with M$_* > 10^{10.5}$ M$_\odot$, consistent with those presented here and in previous studies. Below this mass the scatter increases both for individual galaxies within a given feedback implementation, but also between different implementations, confirming that the observed stellar metallicity and age are highly sensitive to the chosen feedback prescription. Their preferred models, that they characterise as `strong feedback', while successfully predicting the high-mass [Z/H] relation and its evolution with redshift, overestimate the metallicity of galaxies with M$_* < 10^{10}$ M$_{\odot}$, relative to both this work and \citet{Gallazzi:2005}. In contrast, models with weaker feedback do match the observed metallicities of lower mass galaxies, but fail to match a number of other key observables, in particular the mean age of low-mass galaxies. Despite these inconsistencies, the analysis of \citet{Hirschmann:2016} does capture some key behaviour of the stellar population--M$*$ relations: i) the tight and relatively flat age and metallicity relations above M$_* = 10^{10.5}$ M$_\odot$, ii) the sharp downturn and increased scatter in both age and metallicity below this mass, and iii) the flat, or slightly rising age--M$_*$ relation towards the lowest masses.

The work of \citet{Tonini:2016} is of particular interest because they consider not only the dependency of age and metallicity on mass, but also the simultaneous dependence on galaxy size that we have shown is critical to understanding the variation of stellar populations with galaxy properties. They find that, at fixed mass, more compact galaxies are older and more metal rich, consistent with our findings in Figure \ref{fig:ssp_mass_size_gama}. They also find that the youngest galaxies are not the lowest mass, but the most diffuse, and for low-mass galaxies the age--M$*$ relation becomes essentially flat. They connect these trends to galaxy structure, distinguishing between systems with merger-driven spheroids, those with instability-driven spheroids and disk-dominated systems, and argue that angular momentum drives the variation in stellar populations at fixed mass, in the sense that lower angular momentum galaxies have more evolved stellar populations (i.e. older and more metal rich). This age -- angular momentum relation is consistent with the finding here, that even at fixed mass and size, early-type galaxies (i.e. those with a larger spheroidal component and therefore lower angular momentum) are older (Figure \ref{fig:age_mass_size_planes_struct}).

\subsection{Implications for galaxy formation scenarios}

A galaxy's location in the size-mass plane depends to some degree on the processes that have dominated its evolutionary history. \citet{Cappellari:2016}, based on previous ideas from \citet{Cappellari:2013b} and \citet{van-Dokkum:2015}, put forward a view where galaxies change location in the plane through three predominant mechanisms. Disk galaxies grow through regular star formation, rapidly increasing their mass with a modest increase in size. Disk instabilities can result in an episode of bulge growth, driving gas to the centre where it forms stars, resulting in a significant increase in the central density and an apparent reduction in size (evolving towards the lower right corner of the plane). The most massive galaxies experience dry merging, growing proportionally in both size and mass, evolving along lines of constant velocity dispersion. This evolutionary path dominates the high-mass edge of the plane.

With respect to the stellar populations, the dry merging track is the simplest to understand. Because dry merging involves no new star formation we expect little change in the stellar populations of galaxies as they evolve along this path in the size-mass plane. The redistribution of stars in a merger, or the growth of the effective radius including younger or more metal-poor stars from the outskirts of a galaxy within a 1 R$_e$ aperture can modify the observed stellar populations, but this is a more modest effect than that caused by star formation. While the [Z/H] and [$\alpha$/Fe] distributions in the plane are consistent with this picture, the age distribution is less consistent -- there is some variation of age as radius and mass increase along the high-mass edge of the plane. This age variation can be understood if the mergers are not perfectly dry and some new stars form as part of the merger, or if the merger adds young stars from a lower-mass, younger progenitor. An alternate view \citep[argued by][amongst others]{Valentinuzzi:2010,Carollo:2013,Poggianti:2013,Fagioli:2016,Yano:2016}, is that progenitor bias sets the present day distribution of massive galaxies in the size--mass plane. In this picture, the mass and size of a quiescent galaxy is set by the typical mass and size of a star-forming galaxy at the epoch that the galaxy was quenched. As the typical star-forming galaxy becomes less dense with redshift \citep{vdWel:2014,vDokkum:2014}, they argue that galaxies that quench earlier, i.e. have older stellar populations and shorter star-formation duration, are more compact. This scenario is consistent with the variation in stellar populations of the most massive galaxies in our sample as shown in Figure \ref{fig:ssp_mass_size_cluster}. The youngest galaxies being the most diffuse, rather than simply the lowest mass, is also consistent with this picture.

The other tracks are harder to interpret in terms of their expected effects on stellar populations. We note that the change in contours of constant age, shifting from a shallow dependence on mass below M$_* \sim 10^{10}$ to a steeper dependence at higher masses, occurs approximately at the transition from disk-dominated growth to bulge growth or dry merging. At fixed mass and size we find a residual dependence of age on morphology, in the sense that earlier type galaxies are older than later type galaxies. There is a residual effect on a galaxy's stellar population, beyond the simple structural dependence. This morphology dependence is inconsistent with a scenario where bulges are built up by a central starburst at later times --- in that case we would expect galaxies with earlier type morphologies (i.e. larger bulges) to be younger at fixed size and mass. If the central star formation only adds a modest amount of mass, its effect on the mean stellar age may not be detectable, but this would also limit the increase in central mass density. This residual morphology dependence agrees well with the picture of \citet{Tonini:2016}, who find intermediate ages for their instability driven bulges.

After controlling for size and mass we find a residual dependence of both age and metallicity on environment. The age dependence, combined with the lack of [$\alpha$/Fe] dependence on environment (at fixed size and mass), suggests that star formation is truncated at earlier times, rather than being more efficient at higher local environmental densities (which would lead to an alpha-enhanced population). We also find a tail of low-metallicity galaxies in lower density environments that is not present in the higher density environments. This low-metallicity tail indicates that some galaxies in lower density environments may have accreted relatively pristine gas, whereas this is less likely to occur in the higher density environments. Massive halos are effective at preventing the cooling of newly accreted gas, therefore the lack of low metallicity galaxies in these environments is expected. We find some evidence (at the 2.5 $\sigma$ level) that luminosity-weighted age is more sensitive to local galaxy number density than to host halo mass, suggesting that local environment plays a stronger role in determining the star formation history of a galaxy than the halo mass. As luminosity-weighted age is sensitive to recent star formation, and local environment is likely to be more closely correlated with recent merger or accretion events than global halo environment this is consistent. Mass-weighted ages would provide a more robust test of this correlation -- if the stronger correlation between age and local environmental density remains with mass-weighted ages this would imply a physical connection as opposed to the similar timescale argument consistent with the luminosity-weighted correlation.

\section{Conclusions}
\label{sec:conclusion}
In this work we have examined the dependence of the single stellar population equivalent age, metallicity ([Z/H]) and alpha-abundance ([$\alpha$/Fe]) of galaxies from the SAMI Galaxy Survey on stellar mass, stellar velocity dispersion, size, morphology (both kinematic and visual) and environment. The key advances of this work over previous studies are the combination of i) high S/N spectra free from aperture bias due to the use of IFS with ii) a large sample of galaxies spanning a broad range in mass, morphology and environment, and iii) a wealth of supporting data, in particular detailed environmental metrics for the local environmental density and host halo mass. These three advantages enable us to probe an unprecedented range in mass, and to examine residual dependences on morphology and environment while controlling for structural parameters. We find a general picture consistent with previous work in which high mass galaxies are old, more-metal rich and alpha-enhanced compared to low mass galaxies being young, metal-poor and alpha-deficient, albeit with significant scatter. Our conclusions are based on light-weighted stellar population parameters, which can be challenging to interpret, as we describe in Section \ref{sec:ssp_model_choice}. However, our key results depend only on the relative ages and metallicities of galaxies (which are consistent between mass- and light-weighted galaxies for all but the most extreme star formation histories), and can therefore be compared to galaxy formation scenarios.

Beyond this simple picture, our key results are:
\begin{itemize}
\item{At fixed mass, compact galaxies are older, more metal rich and more alpha-enhanced compared to more extended galaxies. This trend holds for galaxies of all masses.}
\item{Age is most closely correlated with surface mass density. Contours of constant age in the size--mass plane flatten at low masses to show almost no mass dependence. After controlling for mass {\it and} size, we find a residual dependency of age on both morphology (later type galaxies are younger) and environment (galaxies in low-density regions or in low mass halos are younger).}
\item{Visual morphology and its quantitative proxies, bulge-to-total ratio and S\'{e}rsic $n$ are tightly coupled to stellar age, in the sense that when these parameters are fixed there is little residual age variation in the size--mass plane. In contrast, when kinematic morphology is fixed, significant variation remains of age with size and mass. Visual and structural morphology are better predictors of mean stellar age than kinematic morphology.}
\item{[Z/H] is closely correlated with $\sigma_e$, particularly at high galaxy M$_*$ or $\sigma_e$. We find that late-type spiral galaxies have lower [Z/H] than expected for galaxies of their mass and size. The distribution of [Z/H] for galaxies in low density environments or low-mass halos shows a tail of values to low metallicities.}
\item{Environment has a weak effect on galaxy stellar populations at fixed size and mass. Galaxies in higher mass halos or in regions of higher galaxy number density are older and more metal-rich than galaxies of the same size and mass in lower density environments. Local galaxy number density has a stronger effect on stellar population age than host halo mass does, whereas [Z/H] shows similar variation with the two environmental metrics.}
\item{We can associate a galaxy's location in the size--mass plane with physical processes that are likely to have influenced its evolution. Stellar populations at fixed size and mass provide further evidence for which processes are more important in different locations within the plane.}
\end{itemize}

\section*{Acknowledgements}
NS thanks Richard McDermid for helpful discussions and the anonymous referee for their constructive suggestions about the manuscript.

The SAMI Galaxy Survey is based on observations made at the Anglo-Australian Telescope. The Sydney-AAO Multi-object Integral field spectrograph (SAMI) was developed jointly by the University of Sydney and the Australian Astronomical Observatory. The SAMI input catalogue is based on data taken from the Sloan Digital Sky Survey, the GAMA Survey and the VST ATLAS Survey. The SAMI Galaxy Survey is funded in part by the Australian Research Council Centre of Excellence for All-sky Astrophysics (CAASTRO), through project number CE110001020, and other participating institutions. The SAMI Galaxy Survey website is http://sami-survey.org/

NS acknowledges the support of a University of Sydney Postdoctoral Research Fellowship.

SB acknowledges the funding support from the Australian Research Council through a Future Fellowship (FT140101166). SMC acknowledges the support of an Australian Research Council Future Fellowship (FT100100457). RLD acknowledges travel and computer grants from Christ Church, Oxford and support from the Oxford Centre for Astrophysical Surveys which is funded by the Hintze Family Charitable Foundation. JvdS is funded under Bland-Hawthorn's ARC Laureate Fellowship (FL140100278). JTA acknowledges the award of a SIEF John Stocker Fellowship. M.S.O. acknowledges the funding support from the Australian Research Council through a Future Fellowship (FT140100255). C.F. gratefully acknowledges funding provided by the Australian Research Council's Discovery Projects (grants DP150104329 and DP170100603). Support for AMM is provided by NASA through Hubble Fellowship grant \#HST-HF2-51377 awarded by the Space Telescope Science Institute, which is operated by the Association of Universities for Research in Astronomy, Inc., for NASA, under contract NAS5-26555. AJM gratefully acknowledges funding provided by the Australian Research Council's Discovery Projects (grant DP130103505). SKY acknowledges support from the Korean National Research Foundation (NRF-2017R1A2A1A05001116).

This work was supported by the UK Science and Technology Facilities Council through the `Astrophysics at Oxford' grant ST/K00106X/1.

 This research made use of Astropy, a community-developed core Python package for Astronomy (Astropy Collaboration, 2013, http://www.astropy.org).

\bibliographystyle{mnras}
\bibliography{sami_stellar_pops}


\label{lastpage}

\bsp
\label{lastpage}
\end{document}